\documentclass[10pt]{jfm}

\usepackage{graphicx}
\usepackage{epstopdf, epsfig}

\usepackage{amssymb,amsbsy, graphicx, booktabs, array, bbold, color,fancyhdr, empheq, natbib,subcaption, calligra} %bbold,natbib  bbm, amsmath,
\usepackage[most]{tcolorbox}
\newcommand{\bv}[1]{\mathbf{#1}} 
  
\newcommand{\dd}{\mathrm{d}}

\newcommand{\curlyA}{\mathcal{A}}

\newcommand{\activity}{\mathcal{A}}
\newcommand{\mobility}{\mathcal{M}}
\newcommand{\vslip}{\bv{v}_{slip}}

\newcommand{\no}{\bv{n}_f}
\newcommand{\idmat}{\bv{1}}
\newcommand{\x}{\bv{x}}

\newcommand{\Greendiff}{\mathcal{G}}
\newcommand{\R}{\bv{R}}
\newcommand{\xo}{\tilde{\bv{x}}}
\newcommand{\length}{l}
\newcommand{\rc}{\bv{r}}
\newcommand{\surf}{\bv{S}}
\newcommand{\erho}{\hat{\bv{e}}_{\rho}}
\newcommand{\etheta}{\hat{\bv{e}}_{\theta}}
\newcommand{\tanhat}{\hat{\bv{t}}}
\newcommand{\norhat}{\hat{\bv{n}}}
\newcommand{\binorhat}{\hat{\bv{b}}}
\newcommand{\thetators}{\theta_i}
\newcommand{\crossradius}{\rho}
\newcommand{\maxcrossradius}{r_f}
\newcommand{\torsion}{\tau}
\newcommand{\curvature}{\kappa}
\newcommand{\dds}[1]{\frac{\partial {#1}}{\partial s}}
\newcommand{\ddth}[1]{\frac{\partial {#1}}{\partial \theta}}
\newcommand{\ddsprime}[1]{\frac{\partial {#1}}{\partial \sdum}}
\newcommand{\ddthprime}[1]{\frac{\partial {#1}}{\partial \thetadum}}
\newcommand{\dtheta}{\theta_m(s,\theta)}

\newcommand{\dthetaprime}{\theta_m(\sdum,\thetadum)}

\newcommand{\bvR}{\bv{R}}
\newcommand{\bvRo}{\bv{R}_0}
\newcommand{\epsslend}{\epsilon}  %{{\epsilon}}

\newcommand{\bvRione}{\hat{\bv{R}}^{(1)}_{(i)}}
\newcommand{\bvRitwo}{\hat{\bv{R}}^{(2)}_{(i)}}

\newcommand{\kerone}{K_{1}}
\newcommand{\kertwo}{K_{2}}
\newcommand{\outerexp}[1]{{#1}^{(o)}}
\newcommand{\innerexp}[1]{{#1}^{(i)}}
\newcommand{\expouterkernelininner}[1]{{#1}^{(o)\in(i)}}
\newcommand{\expinnerkernelinouter}[1]{{#1}^{(i)\in(o)}}
\newcommand{\sign}[1]{\mathrm{sign}\left({#1}\right)}

\newcommand{\lambdaf}{\lambda}
\newcommand{\zerothorder}[1]{{#1}^{(0)}}
\newcommand{\firstorder}[1]{{#1}^{(1)}}
\newcommand{\Deltaerhorho}{\boldsymbol{\mathcal{D}}^{(s,\theta)}_{(\sdum,\thetadum)}}
\newcommand{\thetamode}{\theta_m(s,\theta)}
\newcommand{\coeffsinmode}{A_s}
\newcommand{\coeffcosmode}{A_c}
\newcommand{\Ucoeffcosmode}[1]{\bv{U}_{c,{#1}}}
\newcommand{\Ucoeffsinmode}[1]{\bv{U}_{s,{#1}}}
\newcommand{\fcoeffcosmode}[1]{\bv{f}_{c,{#1}}}
\newcommand{\fcoeffsinmode}[1]{\bv{f}_{s,{#1}}}
\newcommand{\bvUo}{\bv{U}_0}
\newcommand{\bvfo}{\bv{f}_0}
\newcommand{\Uswim}{\bv{U}_{sw}}
\newcommand{\Omegaswim}{\boldsymbol{\Omega}_{sw}}
\newcommand{\Eelt}[1]{E_{#1}}
\newcommand{\Nelts}{N_{elts}}
\newcommand{\bvfoelt}[1]{\bv{f}_{0}[#1]}
\newcommand{\lelt}[1]{l_{#1}}
\newcommand{\leltconst}{l_{elt}}
\newcommand{\seltmidpt}[1]{s_{#1}}
\newcommand{\Mmatrix}{\bv{M}}
\newcommand{\Gtensor}{\bv{G}}
\newcommand{\Jtensor}{\bv{J}}
\newcommand{\Ltensor}{\bv{L}}
\newcommand{\uophor}{\bv{u}_{0}^{phor}}

\newcommand{\angleampl}{\alpha}

\newcommand{\Rhelix}{R}
\newcommand{\ddrhods}{\frac{\dd \crossradius(s)}{\dd s}}
\newcommand{\ddrhodsprime}{\frac{\dd \crossradius(\sdum)}{\dd \sdum}}
\newcommand{\sdum}{\tilde{s}}
\newcommand{\thetadum}{\tilde{\theta}}
\begin{document}
\shorttitle{Slender Phoretic Theory}
\shortauthor{P. Katsamba, S. Michelin and T. D. Montenegro-Johnson}
	\title{Slender Phoretic Theory of chemically active filaments}
%	\author{Panayiota Katsamba}
%	\email{p.a.katsamba@bham.ac.uk}
%	\affiliation{School of Mathematics, University of Birmingham, Edgbaston, Birmingham, UK, B15 2TT}
%	\author{S{\'e}bastien Michelin}
%	\email{sebastien.michelin@ladhyx.polytechnique.fr}
%	\affiliation{LadHyX, D{\'e}partement de M{\'e}canique, Ecole Polytechnique, CNRS,
%	91128 Palaiseau, France.}
%	\author{Thomas D. Montenegro-Johnson}
%	\email{T.D.Johnson@bham.ac.uk}
%	\affiliation{School of Mathematics, University of Birmingham, Edgbaston, Birmingham, UK, B15 2TT}

\author{Panayiota Katsamba\aff{1}
  \corresp{\email{p.a.katsamba@bham.ac.uk}},
S{\'e}bastien Michelin\aff{2}
 \and Thomas D. Montenegro-Johnson\aff{1}}

\affiliation{\aff{1}School of Mathematics, University of Birmingham, Edgbaston, Birmingham, UK, B15 2TT
\aff{2}	LadHyX -- D{\'e}partement de M{\'e}canique, CNRS -- Ecole Polytechnique, Institut Polytechnique de Paris, 91128 Palaiseau, France.	}

	\date{\today}
	
	\maketitle
    %\tableofcontents
	
	\begin{abstract}
	Artificial microswimmers, or ``microbots'' have the potential to revolutionise non-invasive medicine and microfluidics. Microbots that are powered by self-phoretic mechanisms, such as Janus particles, often harness a solute fuel in their environment. {Traditionally, self-phoretic particles are point-like, but slender phoretic rods have become an increasingly prevalent design.  While there has been substantial interest in creating efficient asymptotic theories for slender phoretic rods, hitherto such theories have been restricted to straight rods with axisymmetric patterning. However, modern manufacturing methods will soon allow fabrication of slender phoretic filaments with complex three-dimensional shape.} In this paper, we develop a slender body theory for the solute of self-diffusiophoretic filaments of arbitrary three-dimensional shape and patterning. We {demonstrate analytically that,} unlike other slender body theories, first-order azimuthal variations arising from curvature and confinement can have a leading order contribution to the swimming kinematics. 
	\end{abstract}
	
    %\tableofcontents
% * <p.a.katsamba@bham.ac.uk> 2018-09-10T10:52:49.485Z:
% ^.
  %  
    
    %Driven by advances in manufacturing techniques \citep{walther2008janus}, 	the theory of biological locomotion \citep{lauga2009hydrodynamics, katuri2016artificial}
    
	\section{Introduction}

	{Artificial microscale swimmers (microbots) are a novel technology with promising applications in medicine~\citep{nelson2010microrobots} and microfluidics~\citep{maggi2016self}.} Microbots can be broadly classified {by whether their propulsion is }externally-actuated, or fuel-based. Externally-actuated microbots are typically magnetised and actuated by a {periodic} magnetic field, {for example} rigid helical filaments attached to a magnetised head \citep{Zhang2009a,Zhang2009b,GhoshFischer2009, Gaonanowire2010}, and  ``sperm-like'' microbots, that move a flexible tail \citep{dreyfus2005microscopic}. Other externally-actuated microbots are powered by bubbles driven to oscillate via applied ultrasound \citep{bertin2015propulsion}.  
	In contrast, fuel-based microbots harvest fuel from their surroundings to self-propel  \citep{williams2014self, paxton2004}. {One class of such fuel-based microbots are self-diffusiophoretic (autophoretic) particles. In autophoresis,} surface patterning of a particle with a catalyst gives rise to differential surface reaction, allowing the particle to self-generate solute concentration gradients which drive a propulsive slip flow \citep{paxton2004}.

	{Typically autophoretic particles are spheroids, disks, rods, and recently tori \citep{baker2019shape}, partially coated in catalyst. Nature at the microscale is, however, proliferated by flexible active filaments. Such phoretic filaments could exhibit exciting dynamic behaviours, such as spontaneous buckling and periodic oscillations, and even be selectively controlled via targeted shape change. However, dynamic simulations of slender objects can be computationally costly, owing to the need to accurately resolve multiple length scales. As such, there has been a significant drive to develop a slender body theory for autophoretic particles. Such theories not only have the benefit of numerical efficiency, but are also able to provide analytical insight into dynamic behaviours.

The development of slender body theories (SBT) of the dynamics of filaments in viscous fluids represents a magnum opus in low Reynolds number research, spanning nearly 70 years. SBT has provided the basis for numerous insights in bioactive flows, for instance cilia-driven symmetry-breaking flow in vertebrates \citep{smith2019symmetry}, mucociliary clearance~\citep{smith2008modelling}, and sperm motility \citep{gaffney2011mammalian}, amongst others. Indeed, SBT continues to be an invaluable tool for dynamic fluid-filament interaction simulations \citep{hall2019efficient, walker2019filament, schoeller2019methods} where boundary element methods would prove prohibitively costly.

SBT was pioneered by \citet{hancock1953self}, who modelled the beating tails of microorganisms via line distributions of Stokes flow singularities, from which Resistive Force Theory was soon after derived \citep{gray1955propulsion}. This work was later formalised into a framework of matched asymptotic expansions \citep{cox1970motion}, with the inner problem representing flow past a 2D cylinder, and the outer problem a distribution of singularities. Improved, algebraically accurate, SBTs were then developed (e.g. \citet{Johnson1979}), often using Chwang and Wu's exact singularity distribution for a prolate spheroid \citep{chwang1975hydromechanics}. For a more detailed overview, see \citet{lauga2009hydrodynamics}. More recently, \citet{KoensLauga2018} showed that the boundary integral representation of Stokes flow contains SBT, via asymptotic expansion of the integral kernels. 

	For phoretic particles, \citet{yariv2008slender} developed an SBT for the electrophoretic motion of slender straight rods with a varying cross-section. A similar approach was used later by \citet{schnitzer2015osmotic} for studying slender self-diffusiophoretic particles with axisymmetric chemical activity, and \citet{Ibrahim_2017} used a matched-asymptotic expansion to examine how end-shape and cross-sectional profile of straight slender catalytic rods affects swimming speed. Recently, \citet{yariv2019self} extended the work of \citet{schnitzer2015osmotic} to more complex reaction kinetics that depend on the Damk\"{o}hler number.
	%,schnitzer2015osmotic}

	In this paper, we extend this previous work to phoretic filaments of arbitrary 3D centreline, axisymmetric but varying cross-section, and arbitrary chemical patterning, by exploiting a matched asymptotic expansion from a boundary integral representation of Laplace's equation, as developed by \citet{KoensLauga2018} for slender bodies in viscous flow. For axisymmetric particles, azimuthal variations in concentration, which are sub-leading order, have a leading order effect on the particle kinematics when the particle centreline is curved. As such, our theory is not only fully three-dimensional, but also algebraically-accurate to first-order in the filament slenderness. An interesting outcome of this analysis is that only centreline curvature, and not torsion, can have this leading order effect on the dynamics.}

%	{Following the method by which \citet{KoensLauga2018} derived a slender body theory for Stokes flow, we will derive our slender phoretic theory by expanding the boundary integral formulation of Laplace's equation.}

	{We will focus herein on the derivation of this theory, its validation, and the resultant analytical insights that arise, rather than the potential computational gains of our theory.} %While there are also clear computational advantages of the slender theory, these gains would be most apparent for multi-filament simulations, or dynamic fluid-structure interaction calculations for flexible filaments. We consider these extensions beyond the scope of the current work, and consider only rigid phoretic filaments.}	
	The paper is organised as follows. In section 2, we present the slender phoretic theory to obtain the surface distribution of chemical for a slender filament of arbitrary 3D centreline and surface chemical properties. This theory is then coupled to the recent slender body theory of \citet{KoensLauga2018} to obtain the filament's swimming velocities. Section 3 outlines the numerical solution of the resulting integral equation. { In Section 4, we validate our leading-order concentration field against analytical results for straight prolate spheroids, and in Section 5 we validate the first-order concentration calculation for curved planar filaments. In section 6, we examine further the change in dynamics from excluding azimuthal slip flows to curved planar filaments. Section 7 demonstrates the 3D capability of the theory with the simple test-case of an autophoretic helix, while section 8 concludes with a discussion.}

	\section{Slender Phoretic Theory}
	\subsection{Autophoretic propulsion}
	A phoretic swimmer achieves propulsion by catalysing a chemical reaction in the surrounding solute fuel, { denoted by its ``activity'' $\activity(\bv{x})$, for $\bv{x}$ a point on the swimmer surface. The activity represents concentration flux and may vary across the swimmer, with $\activity>0,\  \activity<0$ corresponding to release and or consumption of solute respectively. Local concentration gradients arise from spatial variation in the activity and confinement effects. These gradients result in} local pressure imbalances in a thin boundary layer at the swimmer surface, driving a surface slip flow \citep{Anderson1989, MichelinLauga2014}. {This slip flow is locally proportional to the concentration gradient, with the swimmer's ``mobility'' $\mobility(\bv{x})$ as the (spatially-varying) constant of proportionality. The mobility may be positive or negative, depending on whether the slip flow moves up or down the concentration gradient.}

    %Autophoretic motion
    We consider neutral solute self-diffusiophoresis, where electrokinetic effects are absent \citep{ebbens2014,brown2014}, and work in the limit of zero P\'eclet number \citep{GolestanianLiverpoolAjdari2007}
    where diffusion dominates and advection of the solute due to flow can be neglected. {This limit is appropriate provided that the particle is smaller than $r_c = D/U$, for $D$ solute diffusivity and $U$ the typical phoretic velocity~\citep{MichelinLauga2014}. For platinum-coated Janus particles in hydrogen peroxide solution, this critical radius corresponds to $r_c \approx 10-100\,\mu\mathrm{m}$ \citep{Howse2007}. For a detailed discussion of propulsion at finite P\'{e}clet number, see \citet{MichelinLauga2014}.
    
     At zero P\'{e}clet number, the solute dynamics decouples from the flow dynamics at any instant,} so that the concentration field $c(\x,t)$ is found by solving Laplace's equation 
	\begin{equation}
D\nabla^2c=0, \label{diffusion}
	\end{equation}
	in the region outside the swimmer, with surface $S$, subject to the boundary condition 
	\begin{equation}
		-D\no\cdot\boldsymbol{\nabla} c |_{S}=\activity(\bv{x}) |_{S}, \label{flux_activity}
	\end{equation}
	with $\no$ the outward normal to the swimmer's surface, pointing into the fluid, and $D$ the solute diffusivity. 
{ The solution of the diffusion problem is then used to calculate the surface slip velocity,}
\begin{equation}
	\vslip|_{S} = \mobility(\bv{x}) \left(\idmat - \no \no \right)\cdot\boldsymbol{\nabla} c,\label{vslip_mobility}
\end{equation}
which gives the body-frame boundary conditions for solving the Stokes flow problem
\begin{equation}
	\mu \nabla^2\mathbf{u} - \boldsymbol{\nabla}p = \mathbf{0}, \quad \boldsymbol{\nabla}\cdot \mathbf{u} = 0.
\end{equation}
This slip flow will in general propel the swimmer with translational velocity $\Uswim$, and angular velocity $\Omegaswim$, which are found by {enforcing the constraints that no net force or torque acts on the swimmer.}

\begin{figure}
	\begin{center}
	    \includegraphics{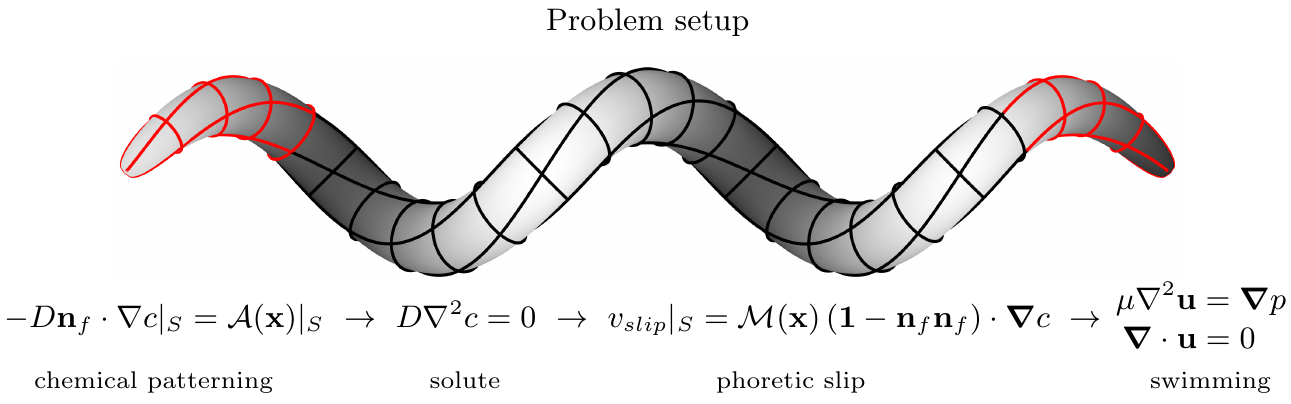}
	\end{center}
	\caption{A schematic of a slender phoretic filament in 3D, in the shape of a helix. Surface lines are used to emphasise curvature and torsion. Laplaces' equation is solved in the bulk, with flux boundary condition on the surface of the filament S determined by its activity $\mathcal{A}(\mathbf{x}|_S$. In later examples, we consider axisymmetric activity in the form of catalytic end caps (demarcated here by the red) with an inert midpiece, so that $\mathcal{A}(s) = -1,\ |s| > s_c,\ \mathcal{A}(s) = 0$ otherwise.}
	\label{activephoreticfilamentPlain}
\end{figure}

{ Thus, given an activity and mobility, the task of this paper is to find a convenient slender body approximation for the concentration field, and slip flow that  results from its surface gradient. This flow will drive the kinematics of the filament, which will {change depending on its centreline shape.} The setup of the problem and governing equations are shown in Fig.~\ref{activephoreticfilamentPlain}.}
%by developing a slender-body theory for active phoretic filaments. 

\subsection{Filament geometry}
We begin by describing the filament geometry, following the approach of \citet{KoensLauga2018}.
{The filament centreline $\rc(s)$ is parametrised by its arclength $s\in\left[-\length,\length\right]$, where $2l$ is the total contour length. The centreline tangent $\tanhat(s)$, normal $\norhat(s)$, and binormal $\binorhat(s)$ satisfy the Serret-Frenet equations,
\begin{subequations}\label{SerretFrenetEqns}
\begin{align}
 \dds{\tanhat}= \curvature \norhat(s), \quad %\label{SerretFrenetEqnst} \\
 \dds{\norhat}= -\curvature\tanhat(s) + \torsion \binorhat(s), \quad %\label{SerretFrenetEqnsn} \\
 \dds{\binorhat}=-\torsion \norhat(s),  %\label{SerretFrenetEqnsb}
\end{align}
\end{subequations}}

\subsubsection{Parametrising the surface}
The filament cross-sectional radius, which may vary along the filament, takes the value $\maxcrossradius \crossradius(s)$ at $s$, where $\maxcrossradius$ is the maximal radius and $\crossradius(s)\in\left[0,1\right]$. The surface of the filament is then parametrised by $s, \maxcrossradius \crossradius(s)$, and the azimuthal angle of the cross-section, $\theta\in\left[-\pi,\pi\right]$, by 
\begin{equation}
	\bv{S}(s,\theta) = \rc(s) + \maxcrossradius\crossradius(s) \erho  (s,\theta).
\end{equation}
The local radial unit vector perpendicular to the centreline tangent, $\erho(s,\theta)$, is given by
\begin{align}
	\erho(s,\theta)=\cos\dtheta\norhat(s) + \sin\dtheta\binorhat(s),
\end{align}
where $\dtheta=\theta-\thetators(s)$, 
with $\thetators(s)$, accounting for the torsion of the curve, chosen to satisfy
\begin{align}
	\dds{\thetators}&=\torsion(s), \label{theta_i_tors}
\end{align}	
as in \cite{KoensLauga2018}. With this choice, the derivative of $\erho$ with respect to $s$ simplifies to
\begin{align}
	\dds{\erho}&=-\curvature(s) \cos\dtheta\tanhat(s). \label{erhoderivs}
\end{align}

\subsubsection{Surface elements}
To obtain the surface element, used later in our integrals, we calculate the derivatives
\begin{subequations}
\begin{align}
 	\dds{\surf}
 	%= 	& \tanhat(s) + \maxcrossradius\crossradius(s)\left[-\curvature(s) \cos\dtheta\tanhat(s)\right]
 	%\nonumber\\&
% 	+ \maxcrossradius\crossradius'(s) \left[\cos\dtheta\norhat(s)+\sin\dtheta\binorhat(s)\right] \nonumber
 %	,\\
 	%
 	&= \tanhat(s)\left[1 - \maxcrossradius\crossradius(s)\curvature(s) \cos\dtheta\right]  \nonumber\\ 	&\quad
 	+ \maxcrossradius\ddrhods \left[\cos\dtheta\norhat(s)+\sin\dtheta\binorhat(s)\right],
\\
 	\ddth{\surf}&=  \maxcrossradius\crossradius(s)\left[-\sin\dtheta\norhat(s)+\cos\dtheta\binorhat(s)\right],
\end{align}
\end{subequations}
which gives the surface element as 
\begin{align}
 \dds{\surf}\times\ddth{\surf} %\nonumber\\ &
 %&= \maxcrossradius \crossradius(s)  \bigg\{ \maxcrossradius \crossradius'(s) \tanhat(s)
 %\\ 	&\qquad\quad
%	 -\left[1-\maxcrossradius\crossradius(s) \curvature(s) \cos\dtheta\right]
%	 \begin{pmatrix}\norhat(s)\cos\dtheta \\+ \binorhat(s) \sin\dtheta 
%	\end{pmatrix}      \bigg\}	 \label{surfelementvector} \\
	&=\maxcrossradius \crossradius(s)  \left( \maxcrossradius \ddrhods \tanhat(s)  
	 -\left[1-\maxcrossradius\crossradius(s) \curvature(s) \cos\dtheta\right]
	 \erho(s,\theta)    \right),\label{surfelementvector}
\end{align}
with magnitude
\begin{align}
\bigg|\dds{\surf}\times\ddth{\surf}\bigg| = \maxcrossradius \crossradius(s)  \sqrt{ \left[\maxcrossradius \ddrhods\right]^2 
+\left[1-\maxcrossradius\crossradius(s) \curvature(s) \cos\dtheta\right]^2
}.	 \label{surfelementscalar}
\end{align}

%\subsubsection{Projecting gradients onto the filament surface}
%In Eq.~\eqref{vslip_mobility}, with $\n=\erho$, we see that the operator $\idmat - \n \n$ projects $\nabla c$ on the surface of the filament, which is the plane spanned by $\tanhat, \etheta$, with $\etheta=\tanhat\times\erho$.
%Expressing this in the local cylindrical polars at the point $s$, the gradient operator projected onto the surface of the filament becomes 
%${\left( \idmat - \n \n \right)\cdot \nabla  =  \tanhat(s) \frac{\partial}{\partial s} + \frac{1}{\maxcrossradius\crossradius(s)}\etheta(s,\theta)\frac{\partial}{\partial \theta}}$, such that  Eq.~\eqref{vslip_mobility} simplifies to
%\begin{equation}
%\vslip(s,\theta) = \mobility(s,\theta) \left[ \tanhat(s) \frac{\partial c}{\partial s} + %\frac{1}{\maxcrossradius\crossradius(s)}\etheta(s,\theta)\frac{\partial c}{\partial \theta}  \right]. \label{vslip_project}
%\end{equation}
%We note the factor of $1/\maxcrossradius$ which indicates the importance of azimuthal concentration gradients that we will investigate in detail later on in the manuscript.  

\subsection{Boundary Integral Equation for the diffusion equation}
%{Following the method by which \citet{KoensLauga2018} derived a slender body theory for Stokes flow, we will derive our slender phoretic theory by expanding the boundary integral formulation of Laplace's equation.}
%
%to find the concentration field that will give the slip velocity on the surface of the body.
{ We begin with the well-known} Green's function for Laplace's equation in an unconfined, three-dimensional region, 
\begin{equation}
 \Greendiff (\x,\xo)= \frac{1}{4\pi |\x-\xo|},
\end{equation}
which solves Laplace's equation, forced by a point sink at $\xo$,
\begin{equation}
 \nabla^2\Greendiff(\x,\xo)=-\delta(\x-\xo).
\end{equation}
{Since in our unbounded domain we have translational invariance, we use the notation $\Greendiff(\x,\xo) = \Greendiff(\x-\xo) = 1/4\pi|\R|$, where $\R=\x-\xo$.}

Using Green's second identity for the functions $c(\xo)$ and $\Greendiff(\xo-\x)$, in the body of the fluid outside the filament $V$, bounded by filament surface $S$, with normal  $\no$ pointing out of the filament, and the notation $\partial/\partial n_f = \no \cdot\nabla_{\xo}$, we have%hence the minus sign
\begin{align}
&\int_{V}^{}\left( c(\xo)\nabla^2_{\xo} \Greendiff(\xo-\x) - \Greendiff(\xo-\x) \nabla^2_{\xo} c(\xo)\right) \dd V(\xo) \nonumber\\
&\quad
=
 -\int_{S}^{} \left( c(\xo) \no\cdot\frac{\partial \Greendiff(\xo-\x)}{\partial \xo} -  \Greendiff(\xo-\x) \no\cdot\frac{\partial c(\xo)}{\partial \xo} \right) \dd S(\xo),
\end{align}
which simplifies to { the classic boundary integral formulation}
% \begin{align}
% \lambdaf c(\x)&=
%  \int_{S}^{}\left[c(\xo)\left(\frac{\partial \Greendiff(\xo-\x)}{\partial n_0}\right) - \Greendiff(\xo-\x) \left(\frac{\partial c(\xo}{\partial n_0}\right)\right] \dd S(\xo) \nonumber\\
% % &=
% % \int_{S_{0}}^{}\left[c(\xo) \frac{\no \cdot\left(\x-\xo\right)}{4\pi |\x-\xo|^3} - \frac{1}{4\pi |\x-\xo|} \left(\frac{\partial c(\xo}{\partial n}\right)\right] \dd S_{\xo} \\
%  %&=\int_{S_{0}}^{}\left[c(\xo) \frac{\no \cdot\left(\x-\xo\right)}{4\pi |\x-\xo|^3} - \frac{1}{4\pi |\x-\xo|} \left(-\frac{\activity(\xo)}{D}\right)\right] \dd S_{\xo}\\
%  &=\frac{1}{4\pi}\int_{S}^{}\left[ \frac{c(\xo) \no \cdot\left(\x-\xo\right)}{ |\x-\xo|^3} + \frac{\activity(\xo)}{ D |\x-\xo|}\right] \dd S (\xo) ,\label{diffusionBIderived}
% \end{align}
\begin{equation}
\lambdaf c(\x)=\frac{1}{4\pi}\int_{S}^{}\left[ \frac{c(\xo) \no \cdot\left(\x-\xo\right)}{ |\x-\xo|^3} + \frac{\activity(\xo)}{ D |\x-\xo|}\right] \dd S (\xo) ,\label{diffusionBIderived}
\end{equation}
where $\lambda=1/2$ for $\x\in S$, $\lambda=1$ for $\x\in V$, and $\lambda=0$ for $\x\notin V$ \citep{Pozrikidis1992}. 

{We now substitute the filament geometry into Eq.~\ref{diffusionBIderived}. For $\x=\surf(s,\theta)$ and $\xo=\surf(\sdum,\thetadum)$ points on the filament surface, we set $\lambdaf=1/2$, and using the notation,} 
\begin{equation}
	\bvR \equiv\surf(s,\theta)-\surf(\sdum,\thetadum)  
	=\bvRo(s,\sdum) + \maxcrossradius\left[\crossradius(s)\erho(s,\theta) - \crossradius(\sdum)\erho(\sdum,\thetadum)\right], \label{bvRgeneral}
\end{equation}
with $\bvRo\equiv \bv{r}(s) - \bv{r}(\sdum)$, and $\no \dd S_{\xo} = - \ddsprime{\surf}\times\ddthprime{\surf}\dd \thetadum \dd \sdum$  (where the sign is due to definition of $\no$ pointing out of the filament),  we can write Eq.~\eqref{diffusionBIderived}
 as
\begin{align}
%c(s,\theta)
%&=\frac{1}{4\pi}\int_{S'}^{}\left[ \frac{c(\sdum,\thetadum) \no \cdot\bvR}{ |\bvR|^3} + \frac{\activity(\sdum,\thetadum)}{ D |\bvR|}\right] \dd S_{\xo}\\
%
%c^{(o)}(s,\theta)
2\pi c(s,\theta)=
&  \int\limits_{-l}^{l} \!\int\limits_{-\pi}^{\pi}
\left[
\frac{\activity(\sdum,\thetadum)}{ D |\bvR|} \bigg|\ddsprime{\surf}\times\ddthprime{\surf}\bigg| 
- \frac{c(\sdum,\thetadum)  \bvR}{ |\bvR|^3} \cdot\left(\ddsprime{\surf}\times\ddthprime{\surf}\right)
\right]\, 
\dd \thetadum \dd \sdum , \label{diffusionBIgeneral}
\end{align}
with the surface element and its magnitude given by Eqs.~\eqref{surfelementvector} and \eqref{surfelementscalar} respectively.

%\subsection{Scalings and non-dimensionalisation} %TDMJ taken out
{While the boundary integral~\eqref{diffusionBIgeneral} now includes the filament geometry, it does not yet use the approximation that the filament is slender. This approximation will allow us, after performing matched asymptotics (details in the appendices), to write the double integral equation~\eqref{diffusionBIgeneral} into a single integral formula for evaluating the concentration on the filament surface.}

We non-dimensionalise lengths by {$r_f$}, activity by a typical activity $[\curlyA]$, and concentration by a typical concentration $[c]$ taken as  $[c]\equiv[\curlyA]\maxcrossradius/D$. 
The last choice %for the relation $[c]\equiv[\curlyA]\maxcrossradius/D$
 comes from considering the boundary condition 
	$-D\no\cdot\nabla c |_{S}=\activity |_{S} $, 
	and noting that since $\no$ is mostly aligned with $\erho$, $\no\cdot\nabla c$ scales as $[c]/\maxcrossradius$. % to obtain  $[c]\approx [\activity]\maxcrossradius/D$.
%	multiplying both sides by $\dd S$ and substituting $\n \dd S$ by $\dds{\surf}\times\ddth{\surf} \dd s \dd \theta$,  to obtain $\activity \dd S\approx D\maxcrossradius\crossradius(s) \erho(s,\theta) \cdot \nabla c \dd s \dd \theta$.  Now $\dd S$ scales like $\maxcrossradius l$ and $\erho(s,\theta) \cdot \nabla c$ scales like $[c]/\maxcrossradius$.  All this gives $[c]\approx [\activity]\maxcrossradius/D$.
%
For a typical mobility scale $[\mobility]$, using that $[c]\sim [\activity] \maxcrossradius/D$, and scaling %(using Eq.~\eqref{vslip_project})
$\vslip \sim [\mobility][\activity] \maxcrossradius/ (D\length)$, (i.e. $[\partial c/\partial s]\sim [c]/\length)$.
 All quantities are henceforth non-dimensional, unless otherwise stated.

{
\subsection{Azimuthal vs. longitudinal slip flows in the slender limit}
Before deriving the slender body approximation of Eq.~\eqref{diffusionBIgeneral}, we sketch out the relevant terms in the theory, and the assumptions we will make. Defining $\epsslend=\maxcrossradius/l$, the slenderness parameter, we are interested in the leading-order swimming velocity, which is determined by the leading order slip velocity. 

For the majority of the filament, this slip velocity is equal to $\mobility \partial c/\partial s$ longitudinally, and $\mobility(\epsslend\crossradius)^{-1}\partial c/\partial \theta$ azimuthally. For nonaxisymmetric chemical patterning $\activity (s,\theta)$, we might expect variations of $c(s,\theta)$ in $s$ and $\theta$ to be of the same order, so that azimuthal slip flows will be $O(\epsslend^{-1})$ and dominate the dynamics. 

For axisymmetric chemical patterning $\activity (s)$, a straight rod will have no azimuthal concentration variation, by symmetry. However a curved rod will have small azimuthal variations in concentration arising from geometric confinement. We will show that these variations are in general $O(\epsslend)$, and as a consequence contribute to the slip flow at the same order as $\partial c/\partial s$. As such, a consistent leading order SPT expansion of the velocity requires $O(\epsslend)$ expansion of the concentration field, in contrast to other slender body theories~\citep{yariv2008slender, schnitzer2015osmotic, Ibrahim_2017, yariv2019self, Johnson1979,Gotz2000,KoensLauga2018}. 

For the derivation of the following theory, we will assume that the activity $\activity(s,\theta)$ and curvature $\kappa(s)$ of the filament are slowly varying with $s$. We will also assume that either the filament as prolate spheroidal cross-section $\rho(s) = \sqrt{(1-s^2)}$, or the activity is zero near the filament ends. From a practical calculation standpoint, our validation against boundary element simulation will demonstrate that these assumptions can often be relaxed. Finally, to allow a convenient decomposition of the azimuthal slip velocity into modes, for simple use in the Slender Body Theory of \citet{KoensLauga2018}, we will also assume axisymmetric mobility $\mobility(s)$, though this assumption does not come into the derivation of the Slender Phoretic Theory itself.

}

\subsection{Asymptotic expansion of the boundary integral kernels} 

The surface element from Eq.~\eqref{surfelementvector}, now in its non-dimensionalised form,  
%\begin{align}
%%	\bvR &= \bvRo(s,\sdum) + \epsslend \left[\crossradius(s)\erho(s,\theta) - \crossradius(\sdum)\erho(\sdum,\thetadum)\right]\\
%    %
% \ddsprime{\surf}\times\ddthprime{\surf} 
%&= \epsslend \crossradius  \bigg\{ \epsslend \ddrhodsprime \tanhat(\sdum)
%-\left[1-\epsslend\crossradius(\sdum) \curvature(\sdum) \cos\dthetaprime\right]
%\erho(\sdum,\thetadum)      \bigg\}\nonumber\\&
%=\epsslend\crossradius \left\{
%-\erho(\sdum,\thetadum) + \epsslend \left[ \ddrhodsprime\tanhat(\sdum) +\crossradius(\sdum) \curvature(\sdum) \cos\dthetaprime \erho (\sdum,\thetadum)\right]
%\right\}
%\end{align}
\begin{equation}
%	\bvR &= \bvRo(s,\sdum) + \epsslend \left[\crossradius(s)\erho(s,\theta) - \crossradius(\sdum)\erho(\sdum,\thetadum)\right]\\
    %
 \ddsprime{\surf}\times\ddthprime{\surf} =\epsslend\crossradius \left(
-\erho(\sdum,\thetadum) + \epsslend \left[ \ddrhodsprime\tanhat(\sdum) +\crossradius(\sdum) \curvature(\sdum) \cos\dthetaprime \erho (\sdum,\thetadum)\right]
\right)
\end{equation}
has magnitude
\begin{align}
\bigg|\ddsprime{\surf}\times\ddthprime{\surf} \bigg|
&= \epsslend \crossradius(\sdum)  \sqrt{ \epsslend^2 \left(\ddrhodsprime\right)^2 
+\left[1-\epsslend\crossradius(\sdum) \curvature(\sdum) \cos\dthetaprime\right]^2 }  \nonumber
\\
&= \epsslend \crossradius(\sdum)  \left[ 
1
-\epsslend\crossradius(\sdum)\curvature(\sdum)\cos\dthetaprime  + O(\epsslend^2)\right]. \label{surfelementscalar_nondim}
%
%&= \epsslend \crossradius(\sdum)  \left[ 
%1
%-\epsslend\crossradius(\sdum)\curvature(\sdum)\cos\dthetaprime  + \frac{1}{2}\epsslend^2 \left(\ddrhodsprime\right)^2  + O(\epsslend^3)\right]. \label{surfelementscalar_nondim}
\end{align}
The non-dimensionalised version of the boundary integral expression of  Eq.~\eqref{diffusionBIgeneral} is
\begin{align}
	2\pi c(s,\theta) = 
	\int\limits_{-1}^{1}\!\int\limits_{-\pi}^{\pi} 
	\left[\kerone(s,\theta,\sdum,\thetadum) + \kertwo(s,\theta,\sdum,\thetadum)\right] 
	\dd\thetadum\dd \sdum,  \label{diffusionBIgeneralnondim}
\end{align}
with the two kernels $\kerone(s,\sdum,\theta,\thetadum), \kertwo(s,\sdum,\theta,\thetadum)$ defined as
\begin{subequations}\label{kernels}
\begin{align}
\kerone&=	~\frac{\crossradius(\sdum)\activity(\sdum,\thetadum)}{ |\bvR|}    \quad\left[ 
1 - \epsslend\crossradius(\sdum)\curvature(\sdum)\cos\dthetaprime 
%+ \frac{1}{2}\epsslend^2 \left(\ddrhodsprime\right)^2 
+ O(\epsslend^2)\right], \label{keroneref} \\
\kertwo&=\epsslend\frac{\crossradius(\sdum)c(\sdum,\thetadum) }{ |\bvR|^3}\bvR\cdot   \bigg(
\erho(\sdum,\thetadum) 
%\nonumber\\&\qquad\qquad\qquad\qquad\qquad
- \epsslend \left[ \ddrhodsprime\tanhat(\sdum) +\crossradius(\sdum) \curvature(\sdum) \cos\dthetaprime \erho (\sdum,\thetadum)\right]
\bigg). \label{kertwodef}
\end{align}
\end{subequations}
%The fraction $1/|\bvR|$ in each of $\kerone$ and $\kertwo$ integrands diverges as the integration variable $\sdum$ approaches the arclength parameter of the concentration evaluation $s$, however this singularity will be integrated over. 

In the slender limit where $\epsslend$ is small, for each evaluation parameter $s$ two regions of the integration variable $\sdum$ arise, according to how $\sdum-s$ compares to $\epsslend$. In the outer region, $\sdum-s = O(1)$, the integration variable is far away from the evaluation arclength parameter. In the inner region, $\sdum-s=O(\epsslend)$, the integration variable is within $\epsslend$ arclength distance from the evaluation arclength. We proceed with a matched asymptotic expansion of the integral kernels prior to their integration. 

%The full derivations of 
%
%\subsection{Asymptotic expansion of the boundary integral kernels}

\subsubsection{Outer region}
In the outer region, $s-\sdum=O(1)$, hence  $\bvRo(s,\sdum)=O(1)$, so we can approximate $\bvR \approx\bvRo$. 
%In the outer region, $s-\sdum=O(1)$, hence  $\bvRo(s,\sdum)=O(1)$, so we %can approximate $\bvR \approx\bvRo$.
%have 
To first order,
\begin{align}
\bvR=\bvRo(s,\sdum) + \epsslend \Deltaerhorho, \label{bvR_outer}
\end{align}
where 
%\begin{align}
%   \Delta^{(s,\theta)}_{(\sdum,\thetadum)} f= f(s,\theta) - f(\sdum,\thetadum), 
%\end{align}
% such that 
 \begin{align}
    \Deltaerhorho\equiv \crossradius(s)\erho(s,\theta)- \crossradius(\sdum)\erho(\sdum,\thetadum).   \label{Deltaerhorho_outer}
 \end{align} 
 Expanding and collecting the orders of $\epsslend$ (see App.~\ref{OuterRegion_App} for details), gives the outer expansions $\outerexp{\kerone}, \outerexp{\kertwo}$ of the kernels $\kerone,\kertwo$, defined in Eqs.~\eqref{keroneref}-\eqref{kertwodef}, as  
\begin{align}
&  \outerexp{\kerone}   =	~\frac{\crossradius(\sdum)\activity(\sdum,\thetadum)}{ |\bvRo(s,\sdum)|}    
\left(1-\epsslend \left[\crossradius(\sdum)\curvature(\sdum)\cos\dthetaprime+\frac{\bvRo}{|\bvRo|^2}\cdot\Deltaerhorho \right]  +  O(\epsslend^2)\right)
\label{K1_outer},\\
&\outerexp{\kertwo}  =\epsslend\frac{\crossradius(\sdum)c(\sdum,\thetadum)}{ |\bvRo|^3}
\Bigg(
\bvRo\cdot\erho(\sdum,\thetadum) + \epsslend \left(\Deltaerhorho\right)\cdot\erho(\sdum,\thetadum)\nonumber \\
&\qquad\qquad\qquad\qquad\qquad
- \epsslend \bvRo\cdot\left[ \ddrhodsprime\tanhat(\sdum) +\crossradius(\sdum) \curvature(\sdum) \cos\dthetaprime \erho (\sdum,\thetadum)\right]  \nonumber \\
&\qquad\qquad\qquad\qquad\qquad\qquad\qquad
- \epsslend (\bvRo\cdot\erho(\sdum,\thetadum)) \frac{3\bvRo\cdot\Deltaerhorho}{|\bvRo|^2} 
+O(\epsslend^2) 
\Bigg).
    \label{K2_outer}
\end{align}

\normalsize

\subsubsection{Inner region}
In the inner region, $s-\sdum=O(\epsslend)$,  %, i.e. $\sdum$ is of order $\epsslend$ of $s$,
and we let $\sdum=s+\epsslend\chi$, where $\chi$ is O(1). 
The expansion and calculation, which we will now summarise, are given in detail in App.~\ref{InnerRegion_App}. Taylor-expanding functions of $\sdum$ around $s$ (for example, $\crossradius(\sdum)=\crossradius(s) + \epsslend\chi \dd \crossradius(s)/\dd s  + O(\epsslend^2)$) gives the inner approximation for  Eq.~\eqref{bvRgeneral} as
\begin{align}
\bvR&=\epsslend\left[\bvRione + \epsslend \bvRitwo\right] + O(\epsslend^3),
\end{align}
where
\begin{subequations}
\begin{align}
\bvRione
&=-\chi\tanhat(s) +\crossradius(s)\left[\erho(s,\theta)- \erho(s,\thetadum)\right],  \label{bvRione}
\\
\bvRitwo&=-\left[ \frac{1}{2} \chi^2 \curvature \norhat(s) + \chi\ddrhods \erho(s,\thetadum) - \chi\crossradius(s)  \curvature(s) \cos\theta_m(s,\thetadum)\tanhat(s)\right],
\label{bvRitwo} \\
|\bvRione + \epsslend \bvRitwo|
%&=\sqrt{|\bvRione|^2 + 2\epsslend \bvRione\cdot\bvRitwo + O(\epsslend^2)} 
%\nonumber\\
%&=|\bvRione|\sqrt{1 + 2\epsslend \frac{\bvRione\cdot\bvRitwo}{|\bvRione|^2} + O(\epsslend^2)} \nonumber\\
&=|\bvRione|\left[1 + \epsslend \bvRione\cdot\bvRitwo/|\bvRione|^2 + O(\epsslend^2)\right].
\end{align}
\end{subequations}
Noting that
\begin{align}
|\bvRione|&=\sqrt{\chi^2+\gamma^2},
\end{align}
with 
\begin{equation}
\gamma^2=2\crossradius^2(s)\left[1-\cos(\theta-\thetadum)\right],  
\end{equation}
the expansion of $1/\bvR$ will give rise to a factor $1/\sqrt{\chi^2+\gamma^2}$. As a result, powers of $1/\sqrt{\chi^2+\gamma^2}$ with different exponents appear in different terms of the expression of $\innerexp{\kerone}, \innerexp{\kertwo}$, given in App.~\ref{Inner_Simpl_App}. 

Performing the integration in $\sdum$, we treat the integrals of the form
$I^{i}_{j}=\int_{-1}^{1}\frac{\chi^i}{\epsslend\sqrt{\chi^2+\gamma^2}^j}\dd \sdum$, where $i,j$ positive constants, as in \cite{KoensLauga2018}, see App.~\ref{Inner_Eval_App}. Some of these integrals give rise to logarithmic terms, and we arrive at
\begin{align}
\int_{-1}^{1}\innerexp{\kerone} \dd \sdum
&=
\crossradius(s)\activity(s,\thetadum) \log\left(\frac{2(1-s^2)}{\epsslend^2 \crossradius^2(s)\left[1-\cos(\theta-\thetadum)\right]}\right)\ 
-2s
\partial_s\left[\crossradius(s)\activity(s,\thetadum) \right] 
\nonumber\\
&\quad
+\epsslend\Bigg\{-\activity(s,\thetadum)\crossradius^2(s)\curvature(s)\cos\theta_m(s,\thetadum) \log\left(\frac{2(1-s^2)}{\epsslend^2 \crossradius^2(s)\left[1-\cos(\theta-\thetadum)\right]}\right)  
\nonumber\\
&\qquad\qquad
+\crossradius^2(s)\activity(s,\thetadum)
\frac{1}{2}\curvature(s)  \left[\cos\theta_m(s,\thetadum) + \cos\theta_m(s,\theta)\right]\times\nonumber\\&
\qquad\qquad\qquad\qquad\qquad\left[\log\left(\frac{2(1-s^2)}{\epsslend^2 \crossradius^2(s)\left[1-\cos(\theta-\thetadum)\right]}\right)-2 \right]\Bigg\}
\nonumber\\
&\quad
+ \frac{2s\epsslend^2}{s^2-1}\crossradius^2(s)\activity(s,\thetadum)\ddrhods [\cos(\theta-\thetadum)-1]
%\nonumber\\
%&\quad
+ O(\epsslend^2),
\\
\int_{-1}^{1}\innerexp{\kertwo}\dd \sdum
&=
-c(s,\thetadum)  \nonumber\\
&\quad
+\frac{1}{2}\epsslend c(s,\thetadum)
\left[\log\left(\frac{2(1-s^2)}{\epsslend^2 \crossradius^2(s)\left[1-\cos(\theta-\thetadum)\right]}\right)\right]
\curvature(s)\crossradius(s)
\cos\theta_m(s,\thetadum) \nonumber\\
&\quad
-\epsslend c(s,\thetadum) 
\frac{1}{2}\crossradius(s)\curvature(s) \left[\cos\theta_m(s,\thetadum) + \cos\theta_m(s,\theta)\right] \nonumber\\
& \quad
+
\frac{2s\epsslend^2}{s^2-1} \crossradius(s) \left[\cos(\theta-\thetadum)-1\right]\partial_s\left[\crossradius(s) c(s,\thetadum)\right]  + O(\epsilon^2 c).
\end{align}
While terms incorporating the fraction $2\epsslend^2\frac{s\crossradius^2(s)}{s^2-1}\ddrhods(s)$ are $O(\epsilon^2)$, we have written them explicitly above as they generally diverge as $s\to\pm 1$, and can thus become leading order {in a very small region $\epsslend^2$ from the ends. This divergence can be circumvented by assuming a prolate spheroidal shape filament $\crossradius(s) \sim \sqrt{1-s^2}$, or ensuring the activity $\activity$ decays to zero at either end. For a more detailed scaling argument, see App.~\ref{app:errors}.}

%We note that it is permissible to ignore the terms with fractions   $2\epsslend^2\frac{s\crossradius^2(s)}{s^2-1}\ddrhods(s)$ for   prolate spheroidal ends, as the $\crossradius^2(s)$   on the numerator cancels the singulatity in the denominator at $s=\pm1$). 

%
\normalsize

\subsubsection{Matching: common part}
We follow the Van Dyke matching method, and use $\epsslend\chi=\sdum-s$ expanding the inner region in terms of the outer region variable $\sdum$, and the outer region in terms of the inner variable $\chi$, and finding the common part, expected to be the same, by expanding in $\epsslend$.
We use the superscripts  $(i)\in(o)$ for the expansion of the inner region kernel in terms of the outer variable, and
$(o)\in(i)$ for the expansion of the outer region kernel in terms of the inner variable.

In order to obtain $\expouterkernelininner{\kerone}$ and $\expouterkernelininner{\kerone}$, we substitute $\sdum=s+\epsslend\chi$ in the expressions for $\outerexp{\kerone}$ and $\outerexp{\kertwo}$ and expand. The resulting expressions for  $\expouterkernelininner{\kerone}$ and $\expouterkernelininner{\kerone}$ are given in App.~\ref{MatchingRegion_App} (we note that these are the same as for $\expinnerkernelinouter{\kerone}, \expinnerkernelinouter{\kertwo}$, as expected).  
%The expressions $\expinnerkernelinouter{\kerone}, \expinnerkernelinouter{\kertwo}$ are the same as those for $\expouterkernelininner{\kerone}, \expouterkernelininner{\kertwo}$, as expected.
Following integration,
\begin{align}
&\int\limits_{-1}^{1}\!\int\limits_{-\pi}^{\pi}\expouterkernelininner{\kerone}\dd \sdum\dd\thetadum
=
\int_{-1}^{1}\frac{\crossradius(s)}{|\sdum-s|}\int_{-\pi}^{\pi}\activity(s,\thetadum)\dd\thetadum \dd \sdum 
-2s\partial_s\int_{-\pi}^{\pi}\left[\crossradius(s)\activity(s,\thetadum)\right] \dd\thetadum
\nonumber\\
& 
-\frac{1}{2}\epsslend\crossradius^2(s)\curvature(s)\int_{-1}^{1}\frac{1}{|\sdum-s|}\int_{-\pi}^{\pi}\activity(s,\thetadum)\left[
\cos\theta_m(s,\thetadum)-\cos\theta_m(s,\theta)
\right]\dd\thetadum \dd \sdum
 + O(\epsslend^2),
\\
&\int\limits_{-1}^{1}\int\limits_{-\pi}^{\pi}\expouterkernelininner{\kertwo}\dd \sdum\dd\thetadum=O(\epsslend^2).
\end{align}

\subsubsection{Composite Solution}
{The boundary integral~\eqref{diffusionBIgeneral} is} approximated by adding the outer and inner expansions and subtracting from each the common part,
\begin{align}
	2\pi c(s,\theta) \approx 
	\int\limits_{-1}^{1}\!\int\limits_{-\pi}^{\pi} 
	\bigg( \outerexp{\kerone} + \innerexp{\kerone} - \expinnerkernelinouter{\kerone}  + \outerexp{\kertwo} + \innerexp{\kertwo} - \expinnerkernelinouter{\kertwo} \bigg) 
	\dd\thetadum\dd \sdum.
\end{align}
The full expansion for the concentration field is given in  App.~\ref{full_sum_appendix}.  
We now use the notation $\zerothorder{c}$ and $\firstorder{c}$ for the leading and first order algebraic corrections of $c$, 
\begin{align}
c=\zerothorder{c} + \epsslend\firstorder{c} + O(\epsslend^2),
\end{align}
and consider each in turn. The leading order includes all terms of $O(1)$ and $O(\log\epsslend)$. The correction to the leading order,  $\epsslend\firstorder{c}$, vanishes as $\epsslend\to 0$, and includes both $O(\epsslend)$ and $O(\epsslend \log\epsslend)$. The details of simplifications are given in Appendices \ref{LeadingOrder_app} and \ref{NextOrder_App}. We now present the resulting expressions.

\subsection{Leading order concentration field}
Writing $\langle f(s)\rangle\equiv \int_{-\pi}^{\pi}f(s,\thetadum)\dd \thetadum$, the leading order expression for the concentration field (originating from  Eq.~\eqref{c0_App} in App.~\ref{LeadingOrder_app}) is given by
\begin{align}
	2\pi \zerothorder{c}(s,\theta)&+
    \langle\zerothorder{c}(s)\rangle
=  
+	\int_{-1}^{1} \left[
\frac{\crossradius(\sdum)\langle\activity(\sdum)\rangle}{ |\bvRo(s,\sdum)|} 
-\frac{\crossradius(s)\langle\activity(s)\rangle}{|\sdum-s|} \right]
 \dd \sdum \nonumber \\
&+
\crossradius(s)\langle\activity(s)\rangle\log\left(\frac{2(1-s^2)}{\epsslend^2 \crossradius^2(s)}\right)
-\crossradius(s)\int_{-\pi}^{\pi}\activity(s,\thetadum)\log\left[1-\cos(\theta-\thetadum)\right]\dd \thetadum.\label{c_zerothorder_1}
\end{align}
Integrating this again over $\theta$ allows us to evaluate $\langle\zerothorder{c}(s)\rangle$ 
{
and then subtract it from Eq.~\eqref{c_zerothorder_1}, the details of which are given in App.~\ref{LeadingOrder_app}. We thus arrive at the leading order slender boundary integral equation  
%\begin{tcolorbox}[colback=red!5!white,colframe=red!75!black,title=Leading Order Slender BI equation]
\begin{align}
	4\pi \zerothorder{c}(s,\theta)
= & 
	\int_{-1}^{1} \left[
\frac{\crossradius(\sdum)\langle\activity(\sdum)\rangle}{ |\bvRo(s,\sdum)|} 
-\frac{\crossradius(s)\langle\activity(s)\rangle}{|\sdum-s|} \right]
 \dd \sdum 
+
\crossradius(s)\langle\activity(s)\rangle\log\left(\frac{(1-s^2)}{\epsslend^2 \crossradius^2(s)}\right)
\nonumber\\&-2\crossradius(s)\int_{-\pi}^{\pi}\activity(s,\thetadum)\log\left[1-\cos(\theta-\thetadum)\right]\dd \thetadum.\label{c_zerothorder_nonaxi}
\end{align}
%\end{tcolorbox}
We thus see that for nonaxisymmetric filaments, there is a leading-order contribution to the concentration that varies with $\theta$. As such, we expect the slip velocity, and resulting dynamics, of such filaments, to be dominated by the final term in Eq.~\eqref{c_zerothorder_nonaxi}.}

For filaments with axisymmetric activity, $\activity(s,\theta)\equiv\activity(s)$, and  $\langle\activity(s,\theta) \rangle = 2\pi \activity(s)$, and  
using $\int_{-\pi}^{\pi}\log\left[1-\cos(\theta-\thetadum)\right]\dd \thetadum=-2\pi\log(2)$, we arrive at 
%\begin{tcolorbox}[colback=red!5!white,colframe=red!75!black,title=Axisymmetric activity - Leading Order Slender BI equation]
\begin{align}
	2 \zerothorder{c}(s,\theta)
= & 
+	\int_{-1}^{1} \left[
\frac{\crossradius(\sdum)\activity(\sdum)}{ |\bvRo(s,\sdum)|} 
-\frac{\crossradius(s)\activity(s)}{|\sdum-s|} \right]
 \dd \sdum 
+
\crossradius(s)\activity(s)\log\left(\frac{4(1-s^2)}{\epsslend^2 \crossradius^2(s)}\right).\label{c_zerothorder}
\end{align}
%\end{tcolorbox}
The two terms inside the integrand of Eq.~\eqref{c_zerothorder} both include non-local effects, {while the logarithmic term, and the $\thetadum$ integral in the nonaxisymmetric case~\eqref{c_zerothorder_nonaxi}, are local}. {Physically, Eq.~\eqref{c_zerothorder} represents} a line distribution of point sources located on the filament centreline, weighted by the filament activity and radius, with a local correction arising because we are evaluating the concentration on the filament. 

%Additionally, we can see that if we increase the curvature of the filament close to some value of the parameter $s$ while keeping its total arclength fixed, $\bvRo(s,\sdum)$ in the denominator of the first term decreases. Assuming, without loss of generality, that the activity is positive, this increases the value of the concentration near that arclength parameter $s$ due to the non-local effects and confinement, as expected intuitively.

Note that the terms of the integrand inside the square brackets both diverge when the integration variable $\sdum$ passes through the evaluation arclength parameter $s$, however the two singularities cancel each other, such that the integrand is regular. 
%result of the integral does not diverge, as the two singularities cancel each other.
%This can be seen for example by expanding the first term around $s$. 
As explained in detail in \citet{KoensLauga2018}, the fraction inside the logarithmic term of Eq.~\eqref{c_zerothorder} means that close to the endpoints, the cross-sectional radius $\crossradius(s)$ must decrease as $\sqrt{(1-s^2)}$ for the logarithmic term not to diverge, i.e. the analysis is valid for prolate spheroidal ends.

%on the centreline (outer solution) + singular part associated with the local contribution? 
%FOR_SEB: 

%DISCUSSION OF THE RESULT: NON LOCAL EFFECTS, LOCAL CURVATURE, ENDPOINTS, CANCELLING SINGULARITY

\subsection{Next Order Correction}
Eq.~\eqref{c1_full} in  App.~\ref{NextOrder_App} gives the full expression for $ \firstorder{c}(s,\theta) $ for a general activity $\activity(s,\theta)$. Simplifying for the axisymmetric case $\activity(s,\theta)\equiv\activity(s)$, after some algebra (see App.~\ref{NextOrder_App}) we finally arrive at 
\begin{align}
	2\pi \firstorder{c}(s,\theta) +\langle\firstorder{c}(s)\rangle =
&\quad
+\pi\crossradius^2(s)\curvature(s)\activity(s)\cos\theta_m(s,\theta)
\left[\log\left(\frac{4(1-s^2)}{\epsslend^2 \crossradius^2(s)}\right)-3\right]
\nonumber\\
&
-2\pi\crossradius(s)\int_{-1}^{1}\crossradius(\sdum)\activity(\sdum)\frac{\bvRo}{|\bvRo|^3}\dd \sdum\cdot \erho(s,\theta) 
\nonumber\\
&
-\pi\crossradius^2(s)\curvature(s)\int_{-1}^{1}\frac{1}{|\sdum-s|}\activity(s)\cos\theta_m(s,\theta)
\dd \sdum.
\end{align}
%
%Taking $\int_{-\pi}^{\pi} \dd \theta$ of the above
Integrating over $\theta$ we see that $\langle\firstorder{c}(s)\rangle=0$, and the first order correction to the slender boundary integral expression in the case of axisymmetric activity becomes
%\begin{tcolorbox}[colback=red!5!white,colframe=red!75!black,title=Axisymmetric activity - First Order Correction to the Slender BI equation]
\begin{align}
	2\firstorder{c}(s,\theta)=
&\left[\log\left(\frac{4(1-s^2)}{\epsslend^2 \crossradius^2(s)}\right)-3\right]\crossradius^2(s)\curvature(s)\activity(s)\cos\left[\theta-\thetators(s)\right]    \nonumber\\
&
-\crossradius(s)\int_{-1}^{1}
\left[\frac{2\crossradius(\sdum)\activity(\sdum)\bvRo\cdot \erho(s,\theta)}{|\bvRo|^3}+\frac{\crossradius(s)\curvature(s)\activity(s)\cos\left[\theta-\thetators(s)\right]}{|\sdum-s|}\right]
\dd \sdum . \label{c_firstorder_Aaxisym}
\end{align}
%\end{tcolorbox}
Importantly (though perhaps not obviously), the integral converges. This can be seen by substituting
\begin{equation}
{\bvRo=-(\sdum-s)\left[\tanhat(s) + \frac{1}{2}(\sdum-s)\curvature(s)\norhat(s) + O((\sdum-s)^2)\right]},
\end{equation}
{in the first fraction of the integrand, {whereupon the leading order term of the expansion of the first fraction of the integrand cancels the second term of the integrand, $\crossradius(s)\curvature(s)\activity(s)\cos\left[\theta-\thetators(s)\right]/|\sdum-s|$, regularising the singularity. Physically, Eq.~\eqref{c_firstorder_Aaxisym} represents a line distribution of source dipoles located on the filament centreline, weighted by the filament activity, radius, and curvature, with a local correction.}
%
%Summarising, 
%in the box below we provide the expressions for the concentration at first order.

Thus, for axisymmetric activity the full surface concentration is given (up to $O(\epsslend^2)$ corrections) by
%\begin{tcolorbox}[colback=white!5!white,colframe=white!75!black,title=Axisymmetric activity - Leading Order Slender BI equation]
\begin{align}
%	2 \zerothorder{c}(s,\theta)
%=&
%+	\int_{-1}^{1} \left[
%\frac{\crossradius(\sdum)\activity(\sdum)}{ |\bvRo(s,\sdum)|} 
%-\frac{\crossradius(s)\activity(s)}{|\sdum-s|} \right]
% \dd \sdum 
%+
%\crossradius(s)\activity(s)\log\left(\frac{4(1-s^2)}{\epsslend^2 \crossradius^2(s)}\right)  \label{c_0_axisymm}
%\\
%
%
%	2\firstorder{c}(s,\theta)=&
%\left[\log\left(\frac{4(1-s^2)}{\epsslend^2 \crossradius^2(s)}\right)-3\right]\crossradius^2(s)\curvature(s)\activity(s)\cos\left[\theta-\thetators(s)\right]\nonumber\\
%&
%-\crossradius(s)\int_{-1}^{1}
%\left[\frac{2\crossradius(\sdum)\activity(\sdum)\bvRo\cdot \erho(s,\theta)}{|\bvRo|^3}+\frac{\crossradius(s)\curvature(s)\activity(s)\cos\left[\theta-\thetators(s)\right]}{|\sdum-s|}\right]
%\dd \sdum       \label{c_1_axisymm}
%\\
%
%
2c(s,\theta)=&	\int_{-1}^{1} \left[
\frac{\crossradius(\sdum)\activity(\sdum)}{ |\bvRo(s,\sdum)|} 
-\frac{\crossradius(s)\activity(s)}{|\sdum-s|} \right]
 \dd \sdum 
+
\crossradius(s)\activity(s)\log\left(\frac{4(1-s^2)}{\epsslend^2 \crossradius^2(s)}\right)\nonumber\\
&+\epsslend \bigg[
\left[\log\left(\frac{4(1-s^2)}{\epsslend^2 \crossradius^2(s)}\right)-3\right]\crossradius^2(s)\curvature(s)\activity(s)\cos\left[\theta-\thetators(s)\right]
\nonumber\\
&\quad
-\crossradius(s)\int_{-1}^{1}
\left[\frac{2\crossradius(\sdum)\activity(\sdum)\bvRo\cdot \erho(s,\theta)}{|\bvRo|^3}+\frac{\crossradius(s)\curvature(s)\activity(s)\cos\left[\theta-\thetators(s)\right]}{|\sdum-s|}\right]\bigg] \dd \sdum
\nonumber\\
&
+ O(\epsslend^2). \label{c_tot_0and1_axisymm}
\end{align}
%\end{tcolorbox}

%DISCUSSION OF THE RESULT: NON LOCAL EFFECTS, LOCAL CURVATURE, ENDPOINTS, CANCELLING SINGULARITY
\normalsize
In contrast to slender body equations for viscous flows, Eq.~\eqref{c_tot_0and1_axisymm} is explicit; given the filament activity and geometry, one can directly calculate the concentration field by simply evaluating a line integral, rather than having to solve an integral equation (cf Eq.~\eqref{kertwodef}, that has the concentration in the kernel $\kertwo$). %Given the activity and filament geometry, we can directly calculate the concentration field.
Note further that, since we have non-dimensionalised with respect to the filament radius for asymptotic convenience, in order to make comparisons with existing results (non-dimensionalised with respect to the filament semiaxis), the entire solution must be premultiplied by $\epsslend$.

% {
% As with the leading orderThe terms of the integrand inside the square brackets in the $O(\epsslend)$ contribution both diverge when the integration variable $\sdum$ passes through the evaluation arclength parameter $s$, however the result of the integral does not diverge, as the two singularities cancel each other. This can be seen by expanding the first term around $s$, similarly to the leading order contribution.  Again we can see  non-local effects also arise at first order as seen in the terms of the last integral. 

%\subsubsection{Azimuthal effects for curved rods with axisymmetric activity} \label{AzimEffTheorSec}
{We can now clearly see that there is azimuthal variation in the concentration field for curved axisymmetric filaments that is $O(\epsslend)$. As such, these azimuthal variations will in general have a leading-order effect on the filament swimming velocity.} For a straight filament, all the $O(\epsslend)$ terms vanish in Eq.~\eqref{c_tot_0and1_axisymm}, which means that the leading order expression for the concentration of a straight filament with axisymmetric activity is correct to $O(\epsslend^2)$. 

%Let's contrast these first order terms of the concentration field to the leading order concentration field  for axisymmetric activity given by  Eq.~\eqref{c_zerothorder}. The expression for $\zerothorder{c}$ in  Eq.~\eqref{c_zerothorder} includes non-local effects through the integrals with terms that include $\bvR_0(s,\sdum)$ and $|s-\sdum|$, and end-point effects through the log terms, but there is no azimuthal dependence. This means that the fact that a filament is curved will affect the leading order concentration field through the terms that include $\bvR_0(s,\sdum)$, but no azimuthal dependence will arise at leading order. 
%
%In contrast, in addition to non-local effects and end-point effects, the first order concentration field in Eq.~\eqref{c_firstorder_Aaxisym} has a clear azimuthal dependence which involves the curvature of the filament, through the terms that involve the local curvature $\kappa(s)$. 

This azimuthal dependence has a natural modal form in terms of  $\sin\left[\theta-\thetators(s)\right]$ and $\cos\left[\theta-\thetators(s)\right]$, which will carry through as a modal expression for the phoretic slip velocity, and will naturally lead to a Fourier modes approach for the kinematics which we now describe. 

%In our results, we will highlight the importance of azimuthal effects  in  curved planar filaments of circular arc and S-shapes: demonstrating their leading contribution to  the kinematics by switching the azimuthal terms on and off, and showing the large discrepancies that would arise should one naively calculated the kinematics from the leading concentration only.
% }

%The two terms inside the  integrand of the $O(\epsslend)$ term of  Eq.~\eqref{c_tot_0and1_axisymm} both include non-local effects, which mean that the concentration at $\sdum$ depends on the activity of the entire filament.

%In addition, 

%As explained in detail in \citet{KoensLauga2018}, the fraction inside the logarithmic term of Eq.~\eqref{c_zerothorder} means that close to the endpoints, the cross-sectional radius $\crossradius(s)$ must decrease as $(1-s^2)$ for the logarithmic term not to diverge, i.e. the analysis is valid for ellipsoidal ends.

%\section{Validation for an ellipsoid}

%c(s,theta)
	%can validate against Seb's solution
\subsection{Slip velocity - azimuthal modes}
\normalsize
%\section{Slip velocity calculation}
%\subsubsection{Projecting gradients onto the filament surface}
{ We now proceed to calculate the slip velocity for slender curved filaments with axisymmetric activity. In experimental systems, activity arises from deposition of catalyst at the surface, and so axisymmetric activity implies no azimithal variation in surface chemistry. As such, we may also assume axisymmetric mobility $\mobility(s)$. 

As in the above, the following analysis is not valid in a very small region $O(\epsslend^2)$ from the filament ends (typically 0.01\% of the total filament length for the examples we consider). The contribution to the dynamics from this region is discussed, and shown to be negligible, in App.~\ref{app:errors}.}
The leading order slip velocity (Eq.~\eqref{vslip_mobility}) for axisymmetric activity is given by
\begin{equation}
\vslip(s,\theta) = \mobility(s) 
\left[  \tanhat(s) \frac{\partial \zerothorder{c}}{\partial s} + \etheta(s,\theta)\frac{1}{\crossradius(s)}\frac{\partial \firstorder{c}}{\partial \theta}  \right]. \label{vslip_project_nondim_leading}
\end{equation}
%
%Eq.~\eqref{vslip_project_nondim_leading} shows the importance of azimuthal effects; although they appear  %While these azimuthal effects appear at first order in the concentration field, they have  a leading order contribution to the kinematics,  due to the slenderness of the filament and the  $1/\rm{radius}$ prefactor in the azimuthal component of the gradient. This is in the $\partial \firstorder{c}/\partial \theta$ term and we will now proceed to express it in terms of the activity. 
Taking the $\theta$-derivative of  Eq.~\eqref{c_tot_0and1_axisymm}, we see that 

%PK_comm: I 'm not sure if this expression for dc0/ds is 100% correct, and we don't use it, so best to remove Eq 2.50. and the part of the sentence above: Using a similar matched asymptotics approach for $\partial \zerothorder{c}/\partial s$ and
%
%{\color{red} HERE}
%PK3: Must check how I implement the code for dc0/ds. If I don't use the following equation then it might be better to remove it. 
\begin{align}
% 2\frac{\partial \zerothorder{c}}{\partial s} =&
% -\int_{-1}^{1} \left[
% \frac{\crossradius(\sdum)\activity(\sdum)\tanhat(s) \cdot \bvRo}{ |\bvRo|^3} 
% +\frac{\crossradius(s)\activity(s) \sign{\sdum-s}}{(\sdum-s)^2} \right]
%  \dd \sdum \nonumber\\
%  &
% +
% \partial_s\left[\crossradius(s)\activity(s)\right]\log\left(\frac{4(1-s^2)}{\epsslend^2 \crossradius^2(s)}\right),
% \\
%
%
%
\frac{2}{\crossradius(s)}\frac{\partial \firstorder{c}}{\partial \theta} =& 
-\bigg\{\int_{-1}^{1} \left[
\frac{2\crossradius(\sdum)\activity(\sdum)\bvRo\cdot\etheta(s,\theta) }{ |\bvRo|^3} 
-\frac{\crossradius(s)\curvature(s)\activity(s) \sin\left[\theta-\thetators(s)\right]}{|\sdum-s|} \right]
 \dd \sdum \nonumber\\
 &
\qquad\qquad
+
\left[\log\left(\frac{4(1-s^2)}{\epsslend^2 \crossradius^2(s)}\right) -3\right]\crossradius(s)\curvature(s)\activity(s)\sin\left[\theta-\thetators(s)\right] \bigg\},  \label{dc_1dtheta_axisymm}
\end{align}
where $\etheta(s,\theta)=-\norhat(s)\sin\left[\theta-\thetators(s)\right]  +  \binorhat(s)\cos\left[\theta-\thetators(s)\right]$.
Using the notation $\thetamode\equiv\theta-\thetators(s)$,  
%this becomes  
%\begin{align}
%    \frac{2}{\crossradius(s)}\frac{\partial \firstorder{c}}{\partial \theta} =& 
%\int_{-1}^{1} 
%\bigg[
%-\frac{2\crossradius(\sdum)\activity(\sdum)
%\left[- \bvRo\cdot\norhat(s)\sin\thetamode +  \bvRo\cdot\binorhat(s)\cos\thetamode \right] }{ |\bvRo|^3}   \nonumber \\
%&\qquad\quad
%+\frac{\crossradius(s)\curvature(s)\activity(s) \sin\thetamode}{|\sdum-s|} 
%\bigg]
% \dd \sdum \nonumber\\
% &
%-
%\left[\log\left(\frac{4(1-s^2)}{\epsslend^2 \crossradius^2(s)}\right) -3\right]\crossradius(s)\curvature(s)\activity(s)\sin\thetamode.
%\end{align}
%Thus,
%\begin{align}
 %   \frac{1}{\crossradius(s)}\frac{\partial \firstorder{c}}{\partial \theta} =& 
%+\sin\thetamode\bigg\{\int_{-1}^{1} \bigg[ \frac{\crossradius(\sdum)\activity(\sdum)\left[ \bvRo\cdot\norhat(s)\right]}{ |\bvRo|^3}   +\frac{\crossradius(s)\curvature(s)\activity(s) }{2|\sdum-s|} \bigg] \dd \sdum  \nonumber\\&\qquad\qquad\qquad\quad  -
%\frac{1}{2}\left[\log\left(\frac{4(1-s^2)}{\epsslend^2 \crossradius^2(s)}\right) -3\right]\crossradius(s)\curvature(s)\activity(s) 
%\bigg\}
%\nonumber \\ 
%
%
%&
%-\cos\thetamode\int_{-1}^{1} \frac{
%\crossradius(\sdum)\activity(\sdum)
%\bvRo\cdot\binorhat(s)}{ |\bvRo|^3}  
% \dd \sdum.
%\end{align}
we can rearrange Eq.~\eqref{dc_1dtheta_axisymm} as
\begin{align}
    \frac{1}{\crossradius(s)}\frac{\partial \firstorder{c}}{\partial \theta} =& 
+\coeffsinmode(s)\sin\thetamode
+\coeffcosmode(s)\cos\thetamode , \label{dc1dthetamodeexpression}
\end{align}
where 
\begin{align}
    \coeffsinmode(s)=& 
\int_{-1}^{1} \bigg[
\frac{\crossradius(\sdum)\activity(\sdum)
\left[ \bvRo\cdot\norhat(s)\right]}{ |\bvRo|^3}   +\frac{\crossradius(s)\curvature(s)\activity(s) }{2|\sdum-s|} \bigg] \dd \sdum  \nonumber\\&\qquad\qquad\qquad\quad  -
\frac{1}{2}\left[\log\left(\frac{4(1-s^2)}{\epsslend^2 \crossradius^2(s)}\right) -3\right]\crossradius(s)\curvature(s)\activity(s) ,  \label{coeffsinemode}  \\
    \coeffcosmode(s)=&-\int_{-1}^{1} \frac{
\crossradius(\sdum)\activity(\sdum)
\bvRo\cdot\binorhat(s)
}{ |\bvRo|^3}  
 \dd \sdum  .  \label{coeffcosmode}
\end{align}
Substituting Eq.~\eqref{dc1dthetamodeexpression} into Eq.~\eqref{vslip_project_nondim_leading} gives
\begin{align}
\vslip(s,\theta) = 
\mobility(s)\bigg(  \tanhat(s) \frac{\partial \zerothorder{c}}{\partial s} 
 + &\left[\coeffsinmode(s)\sin\thetamode
+\coeffcosmode(s)\cos\thetamode\right]\times\nonumber\\
&\left[-\norhat(s)\sin\thetamode  +  \binorhat(s)\cos\thetamode\right] \bigg),
\end{align}
and using the double angle trigonometrical formulae we find that the slip velocity has only zero and second modes,
\begin{align}
\vslip(s,\theta) =
\frac{1}{2}\mobility(s)  
\bigg\{&
\left[
2\tanhat(s) \frac{\partial \zerothorder{c}}{\partial s}
-\norhat(s)\coeffsinmode(s)  +  \binorhat(s)\coeffcosmode(s)\right] 
\nonumber\\
&
+\cos2\thetamode\left[ +\norhat(s)\coeffsinmode(s)  + \binorhat(s) \coeffcosmode(s) \right] 
\nonumber\\
&
+ \sin2\thetamode\left[ -\norhat(s)\coeffcosmode(s)  + \binorhat(s) \coeffsinmode(s) \right]  \bigg\} \label{vslip_fourier}.
\end{align}
In the case of planar filaments, $\bvRo(s,\sdum)\cdot\binorhat(s)\equiv 0$ and hence $\coeffcosmode=0$. This reduces Eq.~\eqref{vslip_fourier} to 
\begin{align}
\vslip(s,\theta) =
\frac{1}{2}\mobility(s)  
\bigg( 
2\tanhat(s) \frac{\partial \zerothorder{c}}{\partial s}  +\coeffsinmode(s)\left[\norhat(s)\left(\cos2\thetamode -1\right)
+\binorhat(s)  \sin2\thetamode \right]\bigg) \label{vslip_planar_fourier}.
\end{align}
Given the activity, mobility and geometry of the filament, we can find the concentration, the concentration gradients and then the phoretic slip velocity field  on the surface of the filament  according to Eq.~\eqref{vslip_fourier}.

\subsection{Phoretic swimming kinematics}
We now turn to the problem of finding the leading order swimming (rigid body) dynamics resulting from the slip velocity forcing defined in Eq.~\eqref{vslip_fourier}, namely  the translational velocity $\Uswim$ and rotational velocity $\Omegaswim$. 

\subsubsection{Fourier modes of surface velocities and tractions}
Eqs.~\eqref{vslip_fourier},\eqref{vslip_planar_fourier} already have the correct form in order to use the Fourier mode description of surface velocity described by \citet{KoensLauga2018}, who decomposed the surface velocity and traction in Fourier modes
\begin{subequations}
\begin{align}
    2\pi \bv{U}(s,\theta) &= \bv{U}_0(s) + \sum_{n=1}^{\infty} 
    \left[ \Ucoeffcosmode{n}(s)\cos n\thetamode 
    +  \Ucoeffsinmode{n}(s)\sin n\thetamode \right], \label{U_fourier_lyndon}
    \\
    2\pi \crossradius(s) \bv{f}(s,\theta) &= \bvfo(s) + \sum_{n=1}^{\infty} \left[\fcoeffcosmode{n}(s)\cos n\thetamode +  \fcoeffsinmode{n}(s)\sin n\thetamode\right]  \label{f_fourier_lyndon}.
\end{align}
\end{subequations}
We write the rigid body motion as
\begin{align}
    \Uswim + \Omegaswim\times\surf(s,\theta) 
    =& \Uswim + \Omegaswim\times\rc(s) + \epsslend\crossradius(s) \Omegaswim\times\erho(s,\theta) \nonumber \\
    =& \Uswim + \Omegaswim\times\rc(s) \nonumber \\
    &+ \epsslend\crossradius(s) \Omegaswim\times\norhat(s)\cos\thetamode \nonumber \\
    &+ \epsslend\crossradius(s) \Omegaswim\times\binorhat(s)\sin\thetamode.  \label{SwimKin}
\end{align}
The contributions of the first sine and cosine modes of the surface velocity due to rigid body motion  are  $O(\epsslend)$, hence the leading order Fourier mode decomposition of the total surface velocity (i.e. the sum of the rigid body motion and phoretic slip velocity) is
\begin{subequations}
\begin{align}
\bvUo(s) 
&=2\pi \left[\Uswim+\Omegaswim\times\rc(s)\right]  
%\nonumber \\&\qquad
+ \pi\mobility(s) 
\left[
2\tanhat(s) \frac{\partial \zerothorder{c}}{\partial s}
-\norhat(s)\coeffsinmode(s)  +  \binorhat(s)\coeffcosmode(s)\right],
\label{U0mode_coeffs_expr}
\\
\Ucoeffcosmode{2}(s)
&=\pi\mobility(s) \left[ +\norhat(s)\coeffsinmode(s)  + \binorhat(s) \coeffcosmode(s) \right],  
\label{Ucosmode_coeffs_expr}  %\cos2\thetamode
\\
\Ucoeffsinmode{2}(s)
&= \pi\mobility(s) \left[ -\norhat(s)\coeffcosmode(s)  + \binorhat(s) \coeffsinmode(s) \right], 
\label{Usinmode_coeffs_expr}  %\sin2\thetamode  
\end{align}
\end{subequations}
where the rigid body kinematics $(\Uswim,\Omegaswim)$ are to be found by imposing the force and torque balances on the filament. The total force and torque on the filament are
\begin{subequations}
\begin{align}
  \bv{F}_{tot}&=\int\limits_{-1}^{1}\!\int\limits_{-\pi}^{\pi} \bv{f}(\sdum,\thetadum) \bigg|\ddsprime{\surf}\times\ddthprime{\surf} \bigg| \dd \thetadum \dd \sdum,    \\
  \bv{T}_{tot}&=\int\limits_{-1}^{1}\!\int\limits_{-\pi}^{\pi} \surf(\sdum,\thetadum)\times\bv{f}(\sdum,\thetadum)    \bigg|\ddsprime{\surf}\times\ddthprime{\surf} \bigg| \dd \thetadum \dd \sdum,  
\end{align}
\end{subequations}
and substituting the surface element form given by Eq.~\eqref{surfelementscalar_nondim} yields 
\begin{subequations}
\begin{align}
  %\bv{F}_{tot}
  %&=\int_{-1}^{1}\int_{0}^{2\pi} \bv{f}(\sdum,\thetadum) \epsslend \crossradius(\sdum)  \left[ 1-\epsslend\crossradius(\sdum)\curvature(\sdum)\cos\dthetaprime  + \frac{1}{2}\epsslend^2 \left(\crossradius'(\sdum)\right)^2  + O(\epsslend^3)\right] \dd \thetadum \dd \sdum .
  \bv{F}_{tot}&=\int_{-1}^{1}\int_{0}^{2\pi} \bv{f}(\sdum,\thetadum) \epsslend \crossradius(\sdum)  \left[ 1+ O(\epsslend)\right] \dd \thetadum \dd \sdum ,\label{Ftot_expanded}
  \\
  \bv{T}_{tot}&=\int_{-1}^{1}\int_{0}^{2\pi}
  	\left[ \rc(\sdum) + \epsslend\crossradius(\sdum) \erho  (\sdum,\thetadum)\right]\times
   \bv{f}(\sdum,\thetadum) \epsslend \crossradius(\sdum)  \left[ 1+ O(\epsslend)\right] \dd \thetadum \dd \sdum .\label{Ttot_expanded}
\end{align}
\end{subequations}
With $\bv{f}(s,\theta)$ given by Eq.~\eqref{f_fourier_lyndon}, due to the periodicity of the cosine and sine modes having vanishing contributions, we see that the force and torque balances to leading order are
\begin{subequations}
\begin{align}
    \int_{-1}^{1} \bvfo(\sdum) \dd \sdum &=\bv{0}, \label{forcebalanceleading} \\
        \int_{-1}^{1} \rc(\sdum)\times\bvfo(\sdum) \dd \sdum &=\bv{0}. \label{torquebalanceleading}
\end{align}
\end{subequations}
%Combining Eq.~\eqref{Ufcoeffs_0mode_rel},\ref{U0mode_coeffs_expr} and \ref{forcebalanceleading},

\subsubsection{Phoretic swimming kinematics from slender body theory of viscous propulsion}
After expanding the boundary integral equation for Stokes flow for a 3D filament of arbitrary geometry, \citet{KoensLauga2018} derived the following relation between the surface traction modes on the filament and the modes of the surface velocities,
\begin{subequations}
\begin{align}
4 \bvUo(s)=&\int_{-1}^{1} \dd \sdum 
\left(\frac{\bv{1} + \hat{\bvR}_0 \hat{\bvR}_0}{|\bvRo|}\cdot \bvfo(\sdum) - \frac{\bv{1} + \tanhat \tanhat}{|\sdum-s|}\cdot \bvfo(s) \right)   \nonumber \\
&+\left[\log\left(\frac{4\crossradius^2(s) }{\epsslend^2(1-s^2)}\right)\left(\bv{1} +\tanhat \tanhat \right) + \bv{1} - 3 \tanhat \tanhat \right]\cdot\bvfo(s) + O(\epsslend),    \label{Ufcoeffs_0mode_rel}\\
2n\Ucoeffcosmode{n}(s)=& (\bv{1} + \tanhat \tanhat)\cdot\fcoeffcosmode{n}(s) + O(\epsslend), \qquad (n>0) \label{Ufcoeffs_cosmode_rel}  \\
2n\Ucoeffsinmode{n}(s)=& (\bv{1} + \tanhat \tanhat)\cdot\fcoeffsinmode{n}(s) + O(\epsslend), \qquad (n>0). \label{Ufcoeffs_sinmode_rel}
\end{align}
\end{subequations}
Thus in order to find the leading order rigid body kinematics $(\Uswim,\Omegaswim)$ of the filament, one must {solve} the following system for the unknown force per unit length $\bvfo(s)$,
%\begin{align}
%8\pi &\left[\Uswim+\Omegaswim\times\rc(s)\right] + 4\pi\mobility(s) % 
%\left[2\tanhat(s) \frac{\partial \zerothorder{c}}{\partial s}-\norhat(s)\coeffsinmode(s)  +  \binorhat(s)\coeffcosmode(s)\right] \nonumber\\
%&\qquad=\int_{-1}^{1} \dd \sdum 
%\left(\frac{\bv{1} + \hat{\bvR}_0 \hat{\bvR}_0}{|\bvRo|}\cdot \bvfo(\sdum) - \frac{\bv{1} + \tanhat \tanhat}{|\sdum-s|}\cdot \bvfo(s) \right) \nonumber \\
%&\qquad\qquad\qquad
%+\left[\log\left(\frac{4\crossradius^2(s) }{\epsslend^2(1-s^2)}\right)\left(\bv{1} +\tanhat \tanhat \right) + \bv{1} - 3 \tanhat \tanhat \right]\cdot\bvfo(s), \label{Uswimforceintegralequation}
%\end{align}
\begin{align}
8\pi &\left[\Uswim+\Omegaswim\times\rc(s) + \uophor(s)\right] \nonumber\\
&\qquad=
\int_{-1}^{1}   
\left[\Gtensor(s,\sdum)\cdot \bvfo(\sdum) - \Jtensor(s,\sdum)\cdot \bvfo(s) \right]\dd \sdum 
+ \Ltensor(s)\cdot\bvfo(s), \label{Uswimforceintegralequation_concise}
\end{align}
where the unknowns $(\Uswim,\Omegaswim)$ are such that zero net force~\eqref{forcebalanceleading} and torque~\eqref{torquebalanceleading} act on the swimmer, and where 
\begin{align}
    \uophor(s) = \frac{1}{2} \mobility(s)  
\left[
2\tanhat(s) \frac{\partial \zerothorder{c}}{\partial s}
-\norhat(s)\coeffsinmode(s)  +  \binorhat(s)\coeffcosmode(s)\right],  \label{u0phor}
\end{align}
is the average slip velocity (as in  Eq.~\eqref{vslip_fourier}) over $\theta$, i.e. $\uophor(s)~=~\frac{1}{2\pi} \int_{-\pi}^{\pi} \vslip(s,\theta) \dd \theta$, that is the zeroth mode of the phoretic slip velocity according to the definition for modes, Eq.~\eqref{U0mode_coeffs_expr} divided by $2\pi$. We also have defined the tensors 
%For a more concise notation we define the tensors
\begin{subequations}
\begin{align}
    \Gtensor (s,\sdum)&= \frac{\bv{1} + \hat{\bvR}_0 (s, \sdum) \hat{\bvR}_0(s, \sdum)}{|\bvRo(s, \sdum)|},\\
    \Jtensor (s,\sdum)&=
    \frac{\bv{1} + \tanhat(s) \tanhat(s)}{|\sdum-s|}, \\
    \Ltensor (s)&=\log\left(\frac{4\crossradius^2(s) }{\epsslend^2(1-s^2)}\right)\left(\bv{1} +\tanhat(s) \tanhat(s) \right) + \bv{1} - 3 \tanhat(s) \tanhat(s),
\end{align}
\end{subequations}
for a more compact notation. Note that $\Gtensor(s,\sdum)=\Gtensor(\sdum,s)$ and $\Jtensor(s,\sdum)=\Jtensor (\sdum,s)$.   
%Eq.~\eqref{Uswimforceintegralequation} becomes
%\begin{align}
%8\pi &\left[\Uswim+\Omegaswim\times\rc(s) + \uophor(s)\right] \nonumber\\
%&\qquad=
%\int_{-1}^{1}   
%\left[\Gtensor(s,\sdum)\cdot \bvfo(\sdum) - \Jtensor(s,\sdum)\cdot \bvfo(s) \right]\dd \sdum 
%+ \Ltensor(s)\cdot\bvfo(s), \label{Uswimforceintegralequation_concise}
%\end{align}
%where 
%\begin{align}
%    \uophor(s) = \frac{1}{2} \mobility(s)  
%\left[
%2\tanhat(s) \frac{\partial \zerothorder{c}}{\partial s}
%-\norhat(s)\coeffsinmode(s)  +  \binorhat(s)\coeffcosmode(s)\right].  \label{u0phor}
%\end{align}

{ A numerical implementation of this system of equations is given in section \S\ref{NumericalImplementationSPT}. Following a validation of our theory in sections \S\ref{JanusProlateSpheroidValidationSec},\ref{planar_valid_sec}, we present some simple results in sections \S\ref{planar_results_sec},\ref{JanusHelix_results_sec}.}

%Eqs. can be used in Eqs. to solve for the fourier modes of the surface tractions $\bvfo$, $\fcoeffcosmode$, $\fcoeffsinmode$ from which the surface traction $\bv{f}(s,\theta)$ can be reconstructed and integrated over the entire filament to impose force balance and find the unknown rigid body velocity.

%slip velocity
%Eqs.~\eqref{Ufcoeffs_0mode_rel}-\ref{Ufcoeffs_sinmode_rel} can be used to solve for 
%
%Eqs.~\eqref{vslip_fourier},\ref{vslip_planar_fourier} only have zero and second Fourier modes.  

%%%%%%%%%%%%

%{\color{red} HERE}
%PK3: We need to say how we calculate the concentration  field, using a sum over integral of elements (done with quadrature), and we calculate derivatives using splines. And then explain that we couple the SPT with SBT according to Lyndon.

\section{Implementation of Slender Phoretic Theory} \label{NumericalImplementationSPT}
%PK_comms:
%SEB: Just a side note but which could be important to argue what is the main contribution here: when reading this sentence and referring the cited equation, one could believe that "slender phoretic theory" is only the hydrodynamic part (i.e. slender body theory). I think it is important to indicate that the main novelty of this work is the chemical part -- and yes, the motion of hte particle is a by-product in some way.
%PK: He is right, the current section is rather numerical implementation of SBT according to Lyndon.  We need to add details of how we implement numerically the phoretic  details: The elements, the midpoints of elements and quadrature points and splines  etc. Right?  Please edit/expand the red text below.

%Slender phoretic theory (SPT) provides the surface concentration field as well as the phoretic  surface slip velocity.  The latter give rise to surface tractions which set the filament into motion, with the resulting swimming kinematics to be found from the force and torque balances, Eqs.~\eqref{forcebalanceleading}-\eqref{torquebalanceleading}, and Eqs.~\eqref{Uswimforceintegralequation_concise}-\eqref{u0phor}. 

In this section, we describe how to numerically implement SPT to find the surface concentration field and phoretic slip velocity given the activity on a filament of arbitrary shape, and then couple it with slender body theory of \citet{KoensLauga2018} to find the swimming kinematics. { Since the aim of this work is the theoretical development of the SPT, the following implementation is somewhat rudimentary, but will be shown to be sufficiently accurate and fast for our purposes.} We note that calculating the concentration $c(s,\theta)$ with SPT amounts to merely evaluating a line integral, rather than solving an integral equation, as required for boundary element approaches. % and subsequently calculation of the swimming velocity with slender body theory.

%In order to numerically implement slender phoretic theory, we solve the system of Eqs.~\eqref{forcebalanceleading}-\eqref{Uswimforceintegralequation} to find the swimming velocity for various filament geometries. 

\subsection{Numerical implementation}

The filament is first partitioned into $\Nelts$ segments of equal length $l_{tot}/\Nelts$. The $n^{\rm{th}}$ segment, denoted by $\Eelt{n}$ has contour length $\lelt{n}$ and parameterised by its contour length parameter $s$ taking values in $(\seltmidpt{n}-\lelt{n}/2, \seltmidpt{n}+\lelt{n}/2)$ where $\seltmidpt{n}$ is the midpoint of $\Eelt{n}$ (according to contour length), which we will refer to as the $n^{\rm{th}}$ collocation point.

%and approximate the force per unit length $\bvfo(s)$ as piecewise constant over the filament. 

We simply evaluate the concentration field on the filament surface  according to Eq.~\eqref{c_tot_0and1_axisymm}, using Gaussian quadrature over each segment. { When the evaluation point lies in the element over which the integration takes place, it is sufficient to simply use a high-order even quadrature rule, so that no quadrature point lies at $\sdum = s$. Although the integrand is nonsingular at this point, this regularity arises from two singularities cancelling, and so for numerical purposes the point should be avoided. } 

Using a bicubic spline interpolant of the concentration field $c(s,\theta)$ with  periodic BCs in $\theta$ and clamped BCs for $s$, we can use functional derivatives to find the partial derivatives of the concentration along $s, \theta$, and hence evaluate the phoretic slip velocity field on the surface of the filament from Eq.~\eqref{vslip_project_nondim_leading}.

For the kinematics, we will only need the zeroth mode of the phoretic slip velocity, $\uophor$, at the collocation points, which we can calculate according to Eq.~\eqref{u0phor}.  For the evaluation of $\partial  \zerothorder{c}/\partial s$, we use the functional derivative of a cubic spline data interpolation for $\zerothorder{c}$ based on its values at the collocation points. The coefficients $\coeffsinmode, \coeffcosmode$ are evaluated from Eqs.\ref{coeffsinemode}-\ref{coeffcosmode} using Gaussian quadrature for the definite integrals.

With the zeroth mode of the phoretic slip velocity at hand, we can now proceed to find the swimming kinematics by numerically implementing slender body theory according to Ref.~\cite{KoensLauga2018}. The force per unit length $\bvfo(s)$ along $\Eelt{n}$ is approximated as being constant on each segment, taking the value $\bvfoelt{n}$. 
Then Eqs.~\eqref{forcebalanceleading}-\eqref{torquebalanceleading} become  
\begin{subequations}
\begin{align}
    &\sum_{n=1}^{\Nelts} \lelt{n} \bvfoelt{n}  =\bv{0}, \label{forcebalanceleading_elts} \\
    &\sum_{n=1}^{\Nelts} \left(\int_{\Eelt{n}}^{} \rc(\sdum)\dd \sdum\right)\times\bvfoelt{n}  =\bv{0}. \label{torquebalanceleading_elts}
\end{align}
\end{subequations}
For $\lelt{n}$ small enough, we can approximate $\rc(\sdum)$ with $\sdum \in \Eelt{n}$ as 
\begin{align}
    \rc(\sdum)= \rc(\seltmidpt{n}) + (\sdum-\seltmidpt{n})\tanhat(\seltmidpt{n}) + \frac{1}{2} (\sdum-\seltmidpt{n})^2 \curvature(\seltmidpt{n}) \norhat(\seltmidpt{n}) + O(\lelt{n}^3)
\end{align} and use 
\begin{align}
    \int_{\seltmidpt{n}-\lelt{n}/2}^{\seltmidpt{n}+\lelt{n}/2} (\sdum-\seltmidpt{n})^m \dd \sdum = \frac{\lelt{n}^{m+1}}{2^{m+1}(m+1)} \left[1 - (-1)^{m+1} \right]
\end{align}
to obtain
\begin{align}
\int_{\Eelt{n}}^{} \rc(\sdum)\dd \sdum&= \rc(\seltmidpt{n})\lelt{n} + \frac{1}{24} \curvature(\seltmidpt{n}) \norhat(\seltmidpt{n}) \lelt{n}^3 + O(\lelt{n}^5)
\end{align}
We assume that all segments have equal contour length $\lelt{n}=\leltconst ~ \forall n$. Eqns.~\eqref{forcebalanceleading_elts}-\eqref{torquebalanceleading_elts} become
 \begin{subequations}
\begin{align}
    &\sum_{n=1}^{\Nelts}  \bvfoelt{n}  =\bv{0}, \label{forcebalance_elts} \\
    &\sum_{n=1}^{\Nelts} \left[ \rc(\seltmidpt{n}) + \frac{\leltconst^2}{24}\curvature(\seltmidpt{n}) \norhat(\seltmidpt{n}) + O(\leltconst^4)\right]\times\bvfoelt{n}  =\bv{0}, \label{torquebalance_elts}
\end{align}
\end{subequations}
%PK_comm: I didn't include the curvature term for the torque calculation in the code. I leave it to you whether to keep it or not.
noting that for reasonable curvatures and small segments, the term $\leltconst^2\curvature(\seltmidpt{n})/24$ is also negligibly small. Evaluating Eq.~\eqref{Uswimforceintegralequation_concise} at the $i^{\rm{th}}$ collocation point, $\seltmidpt{i}$,  
\begin{align}
8\pi &\left[\Uswim+\Omegaswim\times\rc(\seltmidpt{i}) + \uophor(\seltmidpt{i})\right] \nonumber\\
&\qquad=
\int_{-1}^{1}   
\left[\Gtensor(\seltmidpt{i},\sdum)\cdot \bvfo(\sdum) - \Jtensor(\seltmidpt{i},\sdum)\cdot \bvfo(\seltmidpt{i}) \right]\dd \sdum 
+ \Ltensor(\seltmidpt{i})\cdot\bvfo(\seltmidpt{i}), \label{Uswimforceintegralequation_concise_elt}
\end{align}
For our element-based approach, we break the integral over the entire filament into a sum of integrals over the segments, in each of which $\bvfo(s)$ takes a constant value.  
\begin{align}
8\pi &\left[\Uswim+\Omegaswim\times\rc(\seltmidpt{i}) + \uophor(\seltmidpt{i})\right] \nonumber\\
&\qquad=
\sum_{j=1}^{\Nelts} \int_{\Eelt{j}}^{}    
\left[\Gtensor(\seltmidpt{i},\sdum)\cdot \bvfoelt{j} - \Jtensor(\seltmidpt{i},\sdum)\cdot \bvfoelt{i}  \right]\dd \sdum 
+ \Ltensor(\seltmidpt{i})\cdot\bvfoelt{i} , \label{Uswimforceintegralequation_concise_elt}
\end{align}
We aim to obtain a system of equations of the form 
\begin{align}
8\pi \left[\Uswim+\Omegaswim\times\rc(\seltmidpt{i}) + \uophor(\seltmidpt{i})\right]=\sum_{j=1}^{\Nelts} \Mmatrix_{ij}\cdot\bvfoelt{j}. \label{integraleqn_matrix_elt}
\end{align}
and solve Eqs.~\eqref{integraleqn_matrix_elt}, \ref{forcebalance_elts} and \ref{torquebalance_elts} for the unknowns $\{\bvfoelt{i},~{i=1,\dots,\Nelts}\}, \Uswim,\Omegaswim$. 
%
%Care needs to be taken in the expression to be used in numerics, as both $\Gtensor(s,\sdum)$ and $\Jtensor(s,\sdum)$ diverge when $s=\sdum$ and in our integral expressions these two singularities cancel each other. This cancelling needs to happen in the numerical calculation too: 
%We must not calculate the integrals of $\Gtensor(\seltmidpt{i},\sdum)$ and $\Jtensor(\seltmidpt{i},\sdum)$ over the segment $\Eelt{i}$ where the singularity occurs  separately, but instead have their difference, which is regular, integrated over $\Eelt{i}$.  
Each submatrix $\Mmatrix_{ij}$ is given by
\begin{align}
    \Mmatrix_{ij}= 
    \begin{cases}
    \int_{\Eelt{j}}^{} \Gtensor(\seltmidpt{i},\sdum)\dd \sdum\quad \text{if}\quad i\neq j\\
    \int_{\Eelt{i}}^{} \left[\Gtensor(\seltmidpt{i},\sdum)-\Jtensor(\seltmidpt{i},\sdum)\right]\dd \sdum - \int_{\Omega\backslash\Eelt{i}}^{} \Jtensor(\seltmidpt{i},\sdum)\dd \sdum + \Ltensor(\seltmidpt{i}) \quad\text{if}\quad i= j\\
    \end{cases}
\end{align}
Note for the diagonal submatrices, $\Mmatrix_{ij}$ with $i=j$, we avoid any singularities by having the difference $\left[\Gtensor(\seltmidpt{i},\sdum)-\Jtensor(\seltmidpt{i},\sdum)\right]$ (which is regular over $\Eelt{i}$ as the singularities have cancelled each other) integrated over $\Eelt{i}$, and $\Jtensor(\seltmidpt{i})$ (which is singular in $\Eelt{i}$) integrated  over the entire filament except $\Eelt{i}$, denoted by $\Omega\backslash\Eelt{i}$. We formulate the entire system as:
\begin{align}
    \begin{pmatrix}
    \Mmatrix_{11}& \cdots& \Mmatrix_{1\Nelts} & -\bv{1}_{3\times3}& \boldsymbol{\epsilon} \cdot\rc(\seltmidpt{1})\\
    \vdots&\ddots&\vdots&\vdots&\vdots\\
    \Mmatrix_{\Nelts1}& \cdots& \Mmatrix_{\Nelts\Nelts} & -\bv{1}_{3\times3}& \boldsymbol{\epsilon} \cdot\rc(\seltmidpt{\Nelts})\\
    %\hline
    \bv{1}_{3\times3}& \cdots& \bv{1}_{3\times3}&\bv{0}_{3\times3}&\bv{0}_{3\times3}\\
    %\hline
    \boldsymbol{\epsilon} \cdot\rc(\seltmidpt{1})& \cdots&\boldsymbol{\epsilon} \cdot\rc(\seltmidpt{\Nelts})&\bv{0}_{3\times3}&\bv{0}_{3\times3}
    \end{pmatrix}
    \begin{pmatrix}
    \bvfoelt{1}\\\vdots \\ \bvfoelt{\Nelts}\\ \Uswim\\\Omegaswim
    \end{pmatrix}
    =
    \begin{pmatrix}
    \uophor(\seltmidpt{1}) \\ \vdots \\ \uophor(\seltmidpt{\Nelts})\\ \bv{0}_{3\times1}\\ \bv{0}_{3\times1}
    \end{pmatrix}  \label{full_matrix_equation}
\end{align}
where $\boldsymbol{\epsilon}$ is the Levi-Civita tensor such that 
%$\boldsymbol{\epsilon} \cdot\rc$ is the matrix $[0,-z,y; z,0,-x;-y,x,0]$  and 
$\left(\boldsymbol{\epsilon} \cdot\rc\right)\cdot\Omegaswim =   \rc\times\Omegaswim$. We use the notation $\bv{0}_{3\times1}$ for the $3\times1$ zero vector, the notation $\bv{0}_{3\times3}$ for the $3\times3$ zero matrix, and similarly $\bv{1}_{3\times3}$ for the  $3\times 3$ identity matrix.
The last two block-lines %(6 lines in component form) 
of the matrix on the LHS of Eq.~\eqref{full_matrix_equation} enforce force and torque balances. {Note that it is also possible to calculate the swimming velocity using a version of the Reciprocal Theorem that is appropriate for filaments, which may be used to give more detailed insight into the total contributions of azimuthal slip flows.}

{\subsection{Computational cost}
Before we proceed with the method validation and results, it is worth briefly discussing the computational efficiency gains in employing SPT over the Boundary Element Method, and providing some benchmarks. In doing so, we will focus on the gains in calculating the chemical solute concentration, which represents our novel contribution.

The computational gains from employing SPT over the BEM arise in two parts of the code, a) generating the filament geometry, in other words computational mesh and quadrature points, and b) solving for the surface solute concentration. Once $c(s,\theta)$ at a cloud of points the surface is available, the same techniques may be used to evaluate the concentration gradient for the slip velocity, and so this process is comparable (and takes a negligible time). Similarly, once this slip velocity is available, either Slender Body Theory or the BEM may be used to solve for the hydrodynamics (with SBT significantly faster). The benchmark results are given in table~\ref{tab:cost}.

\begin{table*}
\centering
\begin{tabular}{l r r}
\toprule   
{} & SPT & BEM\\
 \midrule 
 Discretisation in $s$, $N_s$ & 100 & 100 \\
 Discretisation in $\theta$, $N_{\theta}$ & 20 & 20 \\
 Evaluation points $N_{\text{eval}}$ & 2000 & 2000 \\
 Quad points/element (singular) & 20 & 190 \\
 Quad points/element (nonsingular) & 20 & 28 \\
 \midrule
 Geometry runtime (secs) & 0.014 & 0.26 \\
 Concentration runtime (secs) & 0.044 & 12.9 \\
\bottomrule
\end{tabular}
\caption{An indication of the computational savings that can be made by employing SPT over the boundary element method. Simulations run on a Macbook Pro 2.9 GHz Intel Core i7 (2017) with 16 GB 2133 MHz LPDDR3 RAM.} 
\label{tab:cost}
\end{table*}

Simulations were performed on a 2017 Macbook Pro using Matlab. It should be noted that the code is efficient, but not precompiled. The BEM code uses Fekete quadrature over quadrilateral triangles, with a high-order rule (190 points) when integrating a triangle where the evaluation point lies on a vertex, and a lower-order rule (28 points) for other triangles. Note that while these values seem much higher than the SPT quadrature, the fact that these points are spread over two dimensions makes the high-order rule comparable, and the low-order rule significantly coarser. Calculating the slip velocity via spline interpolation took $O(0.01)$ seconds for both methods.

A key reason for this speed up is not only the reduction in the dimension of the integral equation from 2 (BEM) to 1 (SPT), but also the fact that SPT gives the surface concentration by evaluating an integral, whereas BEM requires one to solve an integral equation (which entails setting up and solving a matrix system). As a crude estimate, evaluation requires $O(N_s^2)$, whereas the solution for BEM requires $O(N_{\text{eval}}^3)$ steps ($N_{\text{eval}} \gg N_s$) to solve for a direct solver, though this could be reduced using iterative solvers such as GMRES.  
}

\subsection{Layout of validation and results}
In the following sections, we validate and apply SPT to various filament geometries, cross-sections, and activity patterns of increasing complexity, as summarised in table~\ref{Geom_Activ_summary}. Throughout, we set the mobility $\mobility = -1$, so that slip flows go from high surface solute concentrations to low. Firstly, straight rods are  used to validate our  SPT results against analytical formulae for prolate spheroids. Secondly, planar curved rods (with uniform cross-section) are validated against Boundary Element computations using the authors' previously published regularised singularity code~\citep{montenegro2015regularised,varma2018clustering}, paying particular attention to the important azimuthal variation. Next, planar curved rods are analysed with SPT - a circular arc and sinusoidal S-shaped centreline. The impact of azimuthal phoretic effects on the resulting kinematics is  calculated. Finally, we consider 3D helical phoretic filaments, focusing on the Janus helix, an autophoretic swimmer with the ability to explore space on a helical trajectory, relevant to sensing and enhanced 3D mixing applications.

\begin{table}
\includegraphics[width=\textwidth]{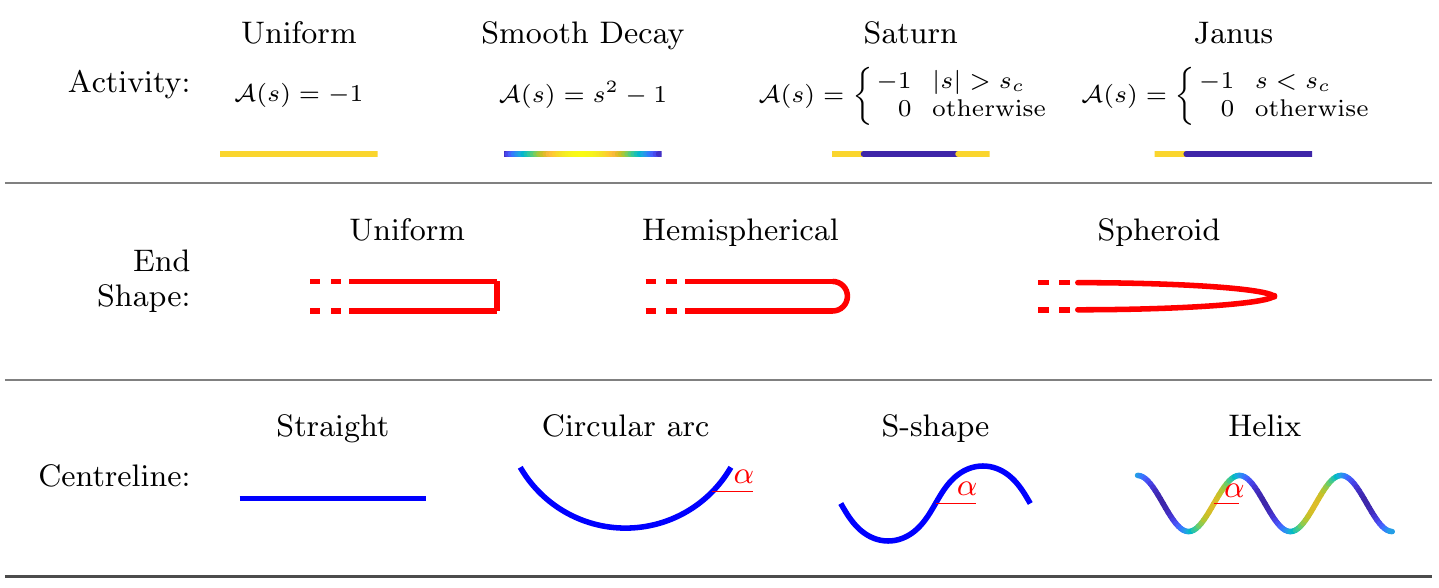}
	\caption{Summary of the different activity patterns (uniform $\activity(s) = -1$, smooth decay $\activity(s) = s^2-1$, Saturn  - 2 symmetric active caps, Janus - 1 active cap ), cross-sectional radius profiles (uniform, hemispherical cap used for BEM, prolate spheroid) and centreline geometries (straight, circular arc, S-shape, helix), used throughout the results section.}  
	\label{Geom_Activ_summary}
\end{table}

	\section{Validation against analytics for autophoretic prolate spheroids}\label{JanusProlateSpheroidValidationSec}
%	\subsection{Straight rods: Validation against analytical formulae  for spheroids 
	%for straight rods (spheroids)	 and computational simulations with BEM}

We first validate our results from the numerical implementation of SPT against the analytical solution of \cite{michelin2017geometric} for straight, spheroidal particles.  \cite{michelin2017geometric} used spheroidal polar coordinates and decomposed the concentration field into the associated Legendre polynomials to solve Laplace's equation for Janus prolate and oblate spheroids. Given the concentration field, the Reciprocal Theorem was then employed to find the translational kinematics, using the classical solution of Oberbeck for the stress field of  translating spheroids \citep{Lamb1932Hydrodyn,HappelBrenner1965LowReHydro}.  Fig.\ref{michelin_validation}a shows the root mean square percentage error in the surface concentration of a uniformly active prolate spheroid as calculated by SPT compared to   \cite{michelin2017geometric}, as a function of the slenderness parameter $\epsslend$. We used 20-point Gaussian quadrature over 100 curved segments for the SPT calculation, and 801 modes of the  series of the analytical solution by \cite{michelin2017geometric}. The error decays like $\epsslend^2$, as predicted by SPT: Eq.~\eqref{c_tot_0and1_axisymm} shows that the leading order expression for the concentration field is correct with $O(\epsslend^2)$ error. 
%%PK: Tom, if needed, please add the 
%Also, I don't think we need to show their formulae, do we? TDMJ - No

Fig.~\ref{michelin_validation}b shows the surface concentration along the filament, parameterised by its arclength,  $-1<s<1$, for a set of Janus prolate spheroids with $\epsslend= 0.01$ and varying catalytic coverage. Our SPT predictions show good agreement with the solution of \cite{michelin2017geometric} (dashed black line). Note that this close agreement is in some ways slightly surprising at the edge of the Janus cap; our formulation requires that $\mathcal{A}(s)$ varies slowly across the filament surface. In practice, our result here shows that the discontinuity may be effectively handled by ensuring that the discontinuity lies at the end point between the segments of either side, hence the use of 100 segments.
%PK: perhaps we need to add that we use the midpoint of elts in the implementation of SPT

In Fig.~\ref{michelin_validation}c we plot the swimming velocity of a slender ($\epsslend=0.01$) Janus prolate spheroid as a function of the catalytic cap coverage $s_c$ (-1 all inert, 1 all active) as calculated by SPT versus the solution of \cite{michelin2017geometric}, showing further good agreement. { The spheroid swims with the inert side at the front, ie to the right in table~\ref{Geom_Activ_summary}.} 
%%PK: Tom, I was thinking why is the optimum at coverage=1. Is this because for coverage=1 the slip velocity at s=1  will be all horizontal and thus contributing more to the swimming speed? Does this mean that if we have a straight rod with its radius non-uniform such that $\rho(s)$ is   maximised at some s not equal to 1 then we could shift this maximum coverage? T
%DMJ - Not sure this is super interesting, let's leave out for now for brevity.

%to validate our numerical method in Fig\ref{michelin_validation}. 

%Fig.~\ref{michelin_validation} shows the error between ... 

\begin{figure}
	\includegraphics{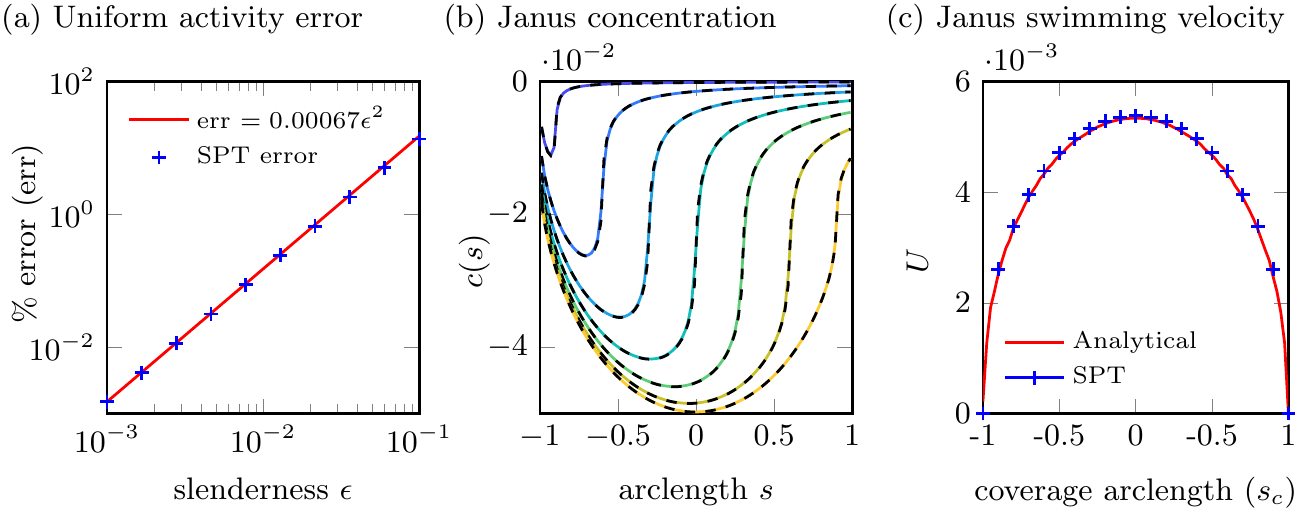}
	\caption{Validation of SPT vs the series solution of %Michelin and Lauga 
	\citet{michelin2017geometric}. a) The root mean square percentage error in the surface concentration of a uniformly active prolate spheroid as calculated by SPT, as a function of the slenderness parameter $\epsilon$.  b) The surface concentration as a function of arclength $s$ of a set of ``Janus'' prolate spheroids with $\epsilon = 0.01$, with active caps such that $s_c < -0.9,-0.6,\ldots 0.9$ from dark to light respectively. The solution of Michelin and Lauga is given in each case by the dashed black line. c) Comparison of the swimming velocities of Janus prolate spheroids with $\epsilon = 0.01$ as a function of $s_c$ (-1 all inert, 1 all active).}  
	\label{michelin_validation}
\end{figure}

%\section{Planar rods with uniform end-shape: validation against BEM}
\section{Validation against BEM: planar filaments with uniform end-shape}	\label{planar_valid_sec} %\label{IUS_azim_subsubsection}
	
%	The framework of SPT can then be used to analyse the phoretic dynamics of curved planar filaments. As emphasized in section \ref{AzimEffTheorSec}, the finite curvature of the filament introduces an azimuthal dependence in the surface concentration, driving an azimuthal surface slip flow. We wish to examine the contribution of this azimuthal slip, which arises from curvature, on two archetypal planar curves: circular arc and S-shaped filaments. Before we investigate the azimuthal variation in detail (in \S\ref{planar_results_sec}), we first validate SPT using BEM.

{We now validate the $O(\epsslend)$ calculation of SPT by comparing against  boundary element method simulations.} For slender body theory~\citep{KoensLauga2018}, the natural choice of cross-sectional radius profile in SPT is that of a prolate spheroid. {However, slender spheroidal ends are difficult to handle with regularised boundary element methods that include the double-layer term. This is because the support of the regularised singularity in the domain depends on local curvature~\citep{varma2018clustering}, and local curvature diverges near the tips of the filaments in the slender limit.}

Thus, in order to validate SPT for curved rods against Boundary Element simulations, we consider a uniform radius profile $\rho(s)=1$ for SPT, with hemispherical end-caps for the BEM (see table~\ref{Geom_Activ_summary}). However, for the uniform cross-section, the logarithmic term in SPT diverges at either end. To avoid this divergence, we thus validate against the quadratic activity profile $\mathcal{A}(s) = s^2-1$.

The centreline of a planar filament is uniquely defined by specifying the local angle $\psi(s)$ between the tangent $\tanhat(s)$ to the centreline  and a fixed direction (here  the $x$-axis). Then the shape of the curve can be constructed using the function $\psi(s)$ through integration,  $ x(s) = \int \cos(\psi(\sdum)) \dd \sdum$, and $y(s) = \int \sin(\psi(\sdum)) \dd \sdum$. Following the non-dimensionalisation in the previous sections, we parametrise the centreline by the contour length parameter $s$, such that $-1<s<1$. Circular arcs of varying curvature are parameterised such that $\psi(s) = s \angleampl$, where $\angleampl$ is equal to the curvature $\kappa$. S-shaped filaments are parameterised a sinusoidal tangent angle $\psi(s)=\angleampl\cos(\pi s)$, which gives the smooth curvature $\kappa(s)=-\angleampl\pi\sin(\pi s)$. For both curves, the limiting case of $\angleampl\to 0$ corresponds to a straight filament.

%\subsection{Validation} \label{conc_field_subsection}

{Since the concentration for a slender filament scales with the slenderness, azimuthal variations in the filament concentration are in fact $O(\epsslend^2)$. As such, we require a very refined Boundary Element Method in order to capture this variation accurately. For accurate implementations of SPT and BEM, we can therefore expect a difference between our BEM/SPT calculations of at least $O(\epsslend^3)$, arising from the truncation of SPT. In practice, at this very small order for slender filaments, other numerical errors begin to affect the BEM solution.

With this caveat in mind,} we plot the azimuthal variation in the concentration as a function of arclength for the circular arc and S-shapes in Fig.~\ref{first_order_validation}. For the case of a) a circular arc with $\angleampl=\pi/2$ (ie a semicircle) and b) an S-shaped centreline with $\angleampl=\pi/3$, we find good agreement between the SPT and Boundary Element calculations. We show the variation of $\epsslend^2c_1$ with $s$  at the azimuthal positions $\theta=0$ and $\theta=\pi$, as these bound the values of $\epsslend^2c_1$ for the rest azimuthal positions by symmetry.

%PK: Tom, I thought $\theta=0$ was facing the centre of the semicircle in which case it should be the lower line on Fig3a. Do we have our theta definitions off by pi?? Please check, fill in the gaps above and let me know. TDMJ - yes, good spot

\begin{figure}
	\begin{center}
        \includegraphics{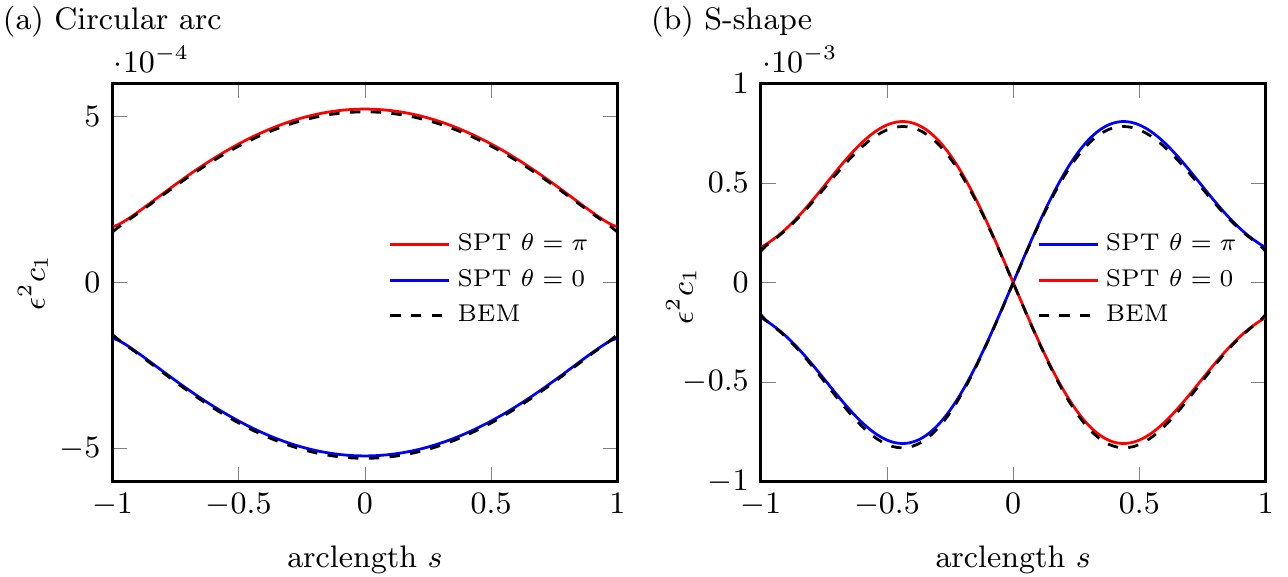}
	\end{center}
	\caption{Validation of the first order concentration correction $c_1$ of SPT vs Boundary Element Simulation
	%you had straight rod,  
	%PK: straight??  do you mean uniform radius?? Please correct
	using a regularised singularity approach~\citep{montenegro2015regularised} with $\mathcal{A}(s) = s^2-1$ for a) a U-shaped filament with $\alpha=\pi/2$ and b) an S-shaped filament with $\alpha=\pi/3$, showing good agreement.}  
	\label{first_order_validation}
\end{figure}

%Given the concentration field, 
We now validate the gradients of this concentration field (Fig.~\ref{slip_vel_validation}), which provide the phoretic slip velocity according to Eq.~\eqref{vslip_project_nondim_leading}. Fig.~\ref{slip_vel_validation}a shows $\partial c/\partial s$ for the case of a circular arc filament of uniform activity, showing the expected divergence in the gradient when calculated with SPT at the ends of the filament due to the divergence of the logarithmic term in the concentration field for uniform cross-sections. However, there is otherwise generally good agreement with the BEM, except in this small region around $s=\pm 1$. This discrepancy is removed (as expected) by taking the quadratic activity profile $\activity(s) = s^2-1$. Furthermore, Fig.~\ref{slip_vel_validation}c shows that our SPT captures the azimuthal gradients in the concentration profile with good accuracy, which is crucial for accurate kinematics calculations.

\begin{figure}
	\begin{center}
		\includegraphics{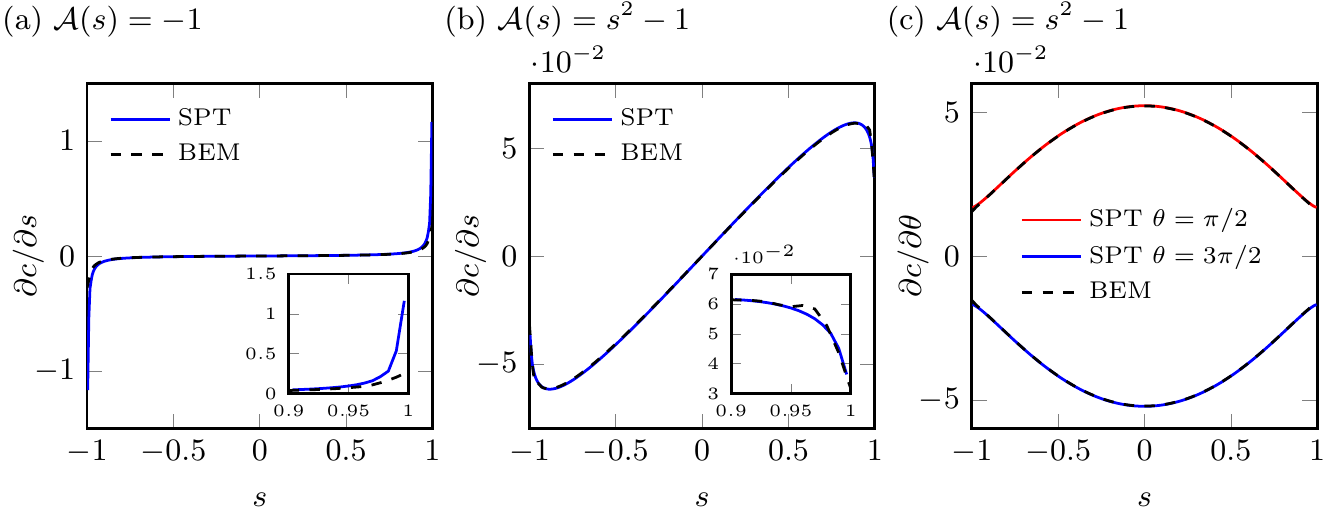}
	\end{center}
	\caption{Validating concentration gradients against Boundary Element simulation, for a semicircular rod of constant radius (required for BEM). a) For uniform activity, the slip velocity along $\theta = \pi/2$ diverges at the ends when calculated via SPT, due to the logarithmic term. b) This divergence may be regularised by ensuring the activity smoothly decays to zero at either end. c) Azimuthal slip flows show good agreement between SPT and BEM.}
	\label{slip_vel_validation}
\end{figure}

Finally, we compare the translational and rotational velocities of circular arc and S-shape filaments as a function of the angle amplitude $\angleampl$ (a proxy for curvature). { The geometry is as shown in table~\ref{Geom_Activ_summary}, where a positive translational velocity for the circular arc means swimming upwards, and a positive angular velocity for the S-shape means anticlockwise rotation about the centroid.}

Since we have non-vanishing slip velocity at $s=\pm 1$, the Slender Body Theory solution also diverges at either end. However, since this is over a very small region, the swimming velocity (which can be thought of as arising from an integral of the surface slip) is not greatly impacted by this divergence. Figure~\ref{kinematics_validation} shows that the SPT kinematics calculations agree well with the full Boundary Element simulations. In particular, we note that the kinematics are changed at leading order when azimuthal slip flows are neglected {(i.e. $A_s = A_c = 0$ in Eq.~\eqref{u0phor})}, and that SPT captures this effect. Having validated our SPT against analytical and Boundary Element calculations, we now proceed with using SPT to study curved filaments with activity profiles that are more relevant to fabrication.

\begin{figure}
	\begin{center}
        \includegraphics{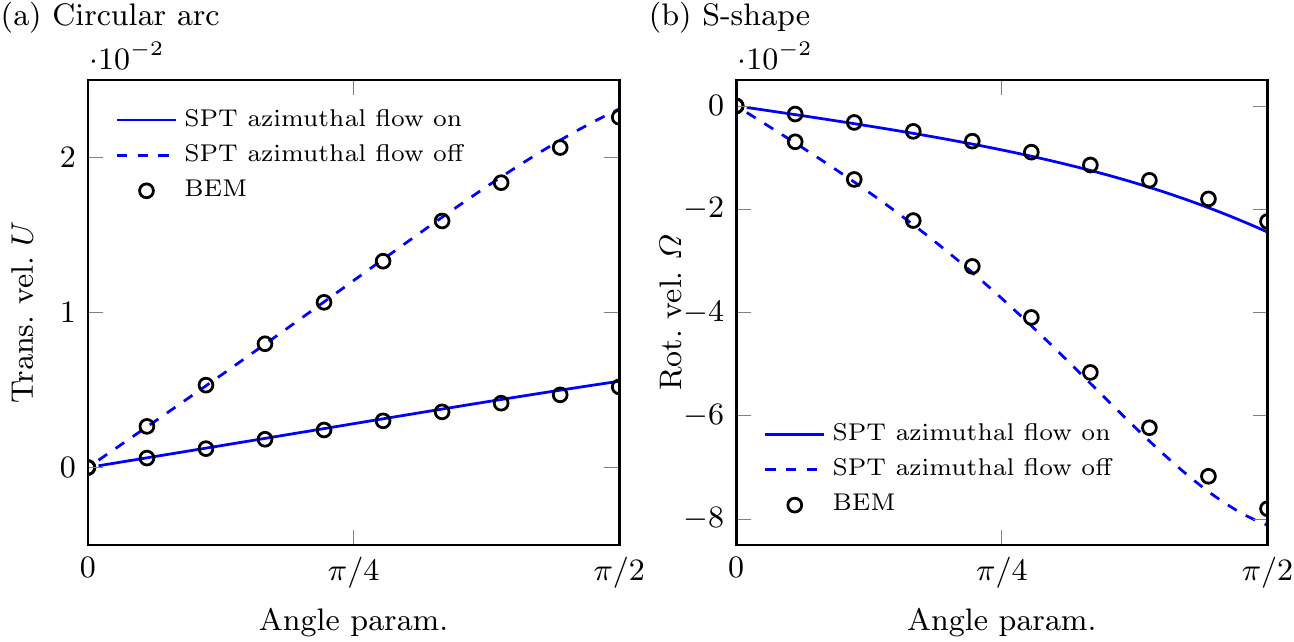}
	\end{center}
	\caption{Validation of the swimming kinematics of a series of circular arcs and S-shapes as a function of the angle parameter $\angleampl$, with $\mathcal{A}(s) = s^2 - 1$ showing good agreement between SPT and BEM. { Results obtained using the full slip flows (azimuthal and longitudinal, solid) are compared to those obtained when neglecting azimuthal slip flow ($A_s = A_c = 0$ in Eq.~\eqref{u0phor}, dashed).}}  
	\label{kinematics_validation}
\end{figure}

\section{Curved planar filaments and azimuthal effects} \label{planar_results_sec}
\subsection{Azimuthal variation of the concentration field}
We now use SPT to examine the physics of slender autophoretic filaments of curved centreline. Throughout this section we consider a spheroidal end shape to ensure the regularity of the SPT formulation. Fig.~\ref{conc_s_theta} shows colour maps of the concentration field (lines of same colour indicate the contours of the concentration field) at different arclength values ($s$) and azimuthal ($\theta$) positions for the three shapes (straight a, circular arc b and S-shaped c) in the case of a uniform catalytic coating, $\activity(s) = -1$, that depletes its surrounding solute.   

For the straight filament (Fig.~\ref{conc_s_theta}a), we see axisymmetry (ie no variation of colour with different $\theta$ at a given $s$) in the concentration field, as expected. We note that for the straight filament the first order concentration field vanishes, hence Fig.~\ref{conc_s_theta}a shows the zeroth order concentration field which is axisymmetric, but non-uniform in $s$ due to the non-local and end-point effects already present in Eq.~\eqref{c_zerothorder}. For example, the closer to the endpoints, the more space there is available for diffusion, hence the solute depletion is less and the solute reactant is at a higher concentration. 

For circular arc filaments, (Fig.~\ref{conc_s_theta}b), we see a minimum of the concentration field occurring at $s=0$ and $\theta=0 (2\pi)$, i.e. the part of the surface that is facing the centre of the semicircular centreline. This is a confinement effect arising from the curvature of the centreline, as this inner part of the surface experiences more depletion of the reactant from the local surface and the adjacent active parts of the surface. Close to the outer part of the surface, there is more space (due to curvature) for diffusion of solute from the bulk to the filament surface thus the solute that has been depleted due to the reaction can be replenished more easily. 
%
%In all planar filaments, there is a symmetry $c(s,\theta)=c(s,2\pi-\theta)$ (or equivalently $c(s,\pi + \theta)=c(s,\pi-\theta)$) as expected; 
%%for centrelines lying on the xy plane, both the filament's surface and its activity (here assumed a function of $s$ only,  $\activity=\activity(s)$) are symmetric in projections $z\to-z$.  
Straight and circular arc filament centreline shapes share the symmetry $c(s,\theta)=c(-s,\theta)$ for the concentration field, arising naturally from the centreline symmetry in $s\to -s$. This symmetry is broken by S-shaped filaments, for which $c(s,\theta) = c(-s,-\theta)$, as in Fig.~\ref{conc_s_theta}c. At $s=0$, the concentration is minimised by the azimuthal positions $\theta=0,\pi$ with $c(0, 0) =c(0, \pi)$, as these lie on the $x,y$ plane and therefore are closer to the nearby active surfaces (by the same amount by symmetry), and $c(0,\pi/2)=c(0,3\pi/2)$.

\begin{figure}
	\begin{center}
		\includegraphics{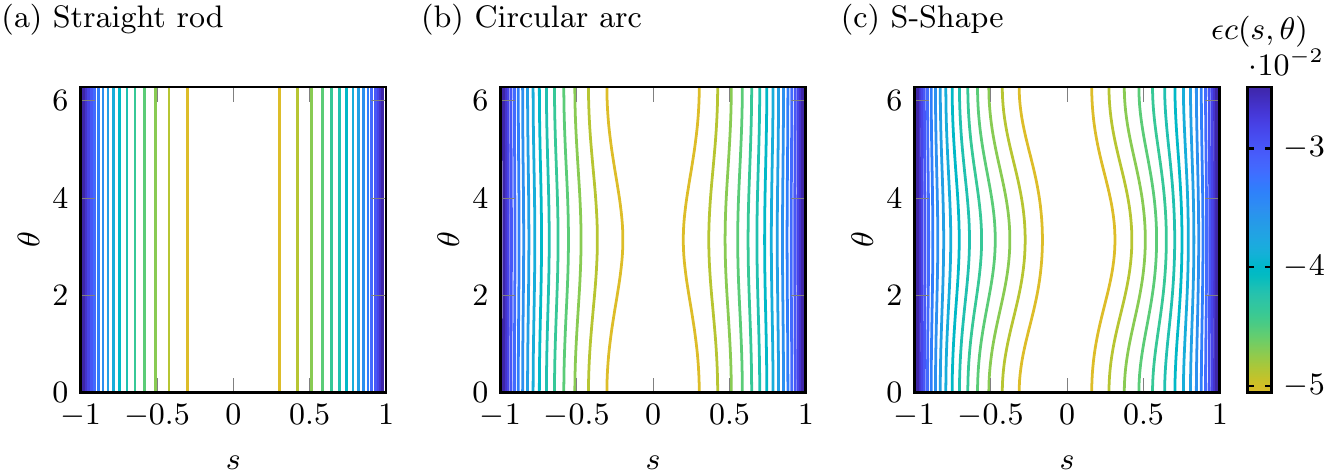}
	\end{center}
	\caption{The surface concentration of uniformly active phoretic filaments with a prolate spheroidal end-shape and planar centrelines with a) straight {which is stationary but pumps fluid}, b) circular arc {which translates}, and c) S-shape {which rotates}. Small azimuthal variation due to non-local interactions and confinement effects from curvature are visible in the circular arc and S-shaped examples.} \label{conc_s_theta}
\end{figure}

%Turning now to the kinematics of these filaments, we note that azimuthal effects arise from the $\partial \firstorder{c}/\partial \theta$ term in  Eq. \ref{vslip_project_nondim_leading}, and although they appear at the next order in the concentration field, they have a leading order contribution to the kinematics. This is due to the $1/\rm{radius}$ prefactor in the azimuthal component of the gradient. The importance of curvature-induced azimuthal effects is demonstrated in Fig.~\ref{U_S_ellipsoid_onoff_28Jun_unif_2cat_caps} by plotting the translational speed and rotation rate for circular arc and S -shaped centrelines respectively, as a function of the angle parameter $\angleampl$, and comparing the resulting kinematics when the azimuthal terns are switched on (solid lines) and off (dashed lines). 

\subsection{Kinematics:  Leading order contribution of azimuthal variations of the concentration field} 
Turning now to the kinematics of these filaments, Fig.~\ref{U_S_ellipsoid_onoff_28Jun_unif_2cat_caps} shows the translational and rotational speeds for circular arc and S-shaped filaments when the azimuthal terms are included (`on' - solid lines) vs neglected (`off' - dashed), for two different activity profiles. The azimuthal effect contribution to the kinematics vanishes as  $\angleampl$ decreases to $0$ in all the plots of Fig.~\ref{U_S_ellipsoid_onoff_28Jun_unif_2cat_caps}, as expected for straight filaments that have zero curvature ($\angleampl=0$) and an axisymmetric concentration field. As $\angleampl$ increases, so does the curvature, and hence the azimuthal effects too.

%Importantly, we note that missing the azimuthal terms (and the $c1$ correction) completely fails to predict the swimming velocity properly as both effects (contributions of the longitudinal and azimuthal concentration gradients to the swimming kinematics) are of the same order, and azimuthal effects can in fact determine the swimming directions. 

The importance of the azimuthal effects is quite profound in the case of uniform activity, where neglecting their contribution results in an incorrect prediction in the direction of motion, as shown by the dashed lines in  Fig.~\ref{U_S_ellipsoid_onoff_28Jun_unif_2cat_caps}a,b.  In order to understand this, let us consider the uniformly active, circular arc filament (see also the qualitative schematic Fig.~\ref{schematic_azim_on_off}b). If we discount azimuthal effects, the higher solute concentration at the ends drives a {longitudinal} surface slip flow to the middle, giving rise to translation in the opposite direction, hence the positive velocity shown by the dashed blue line in Fig. \ref{U_S_ellipsoid_onoff_28Jun_unif_2cat_caps}a. 

{Since the filament consumes solute, confinement effects} mean that there is a lower surface concentration in the inner filament surface that faces the centre of the circular arc centreline. This drives an azimuthal tangential surface slip flow {from the outer to the inner side of the curve. This means that, locally, the filament is acting as a 2D squirmer~\citep{blake1971self}, with symmetry about $\theta = 0, \pi$. This outer to inner flow locally exerts a flow forcing towards the outer side of the cylinder, propelling the filament in that direction. This forcing is in the opposite direction to that arising from longitudinal concentration gradient}, and this contribution is large enough that the overall translational velocity has a negative sign, as shown by the solid blue line in Fig. \ref{U_S_ellipsoid_onoff_28Jun_unif_2cat_caps}a. 
%PK: Tom, please add the schematic tikz files and fill in their names in the subcaption boxes below and uncomment/edit the commented text in the paragraph above.

\begin{figure}
	\begin{center}
        \includegraphics{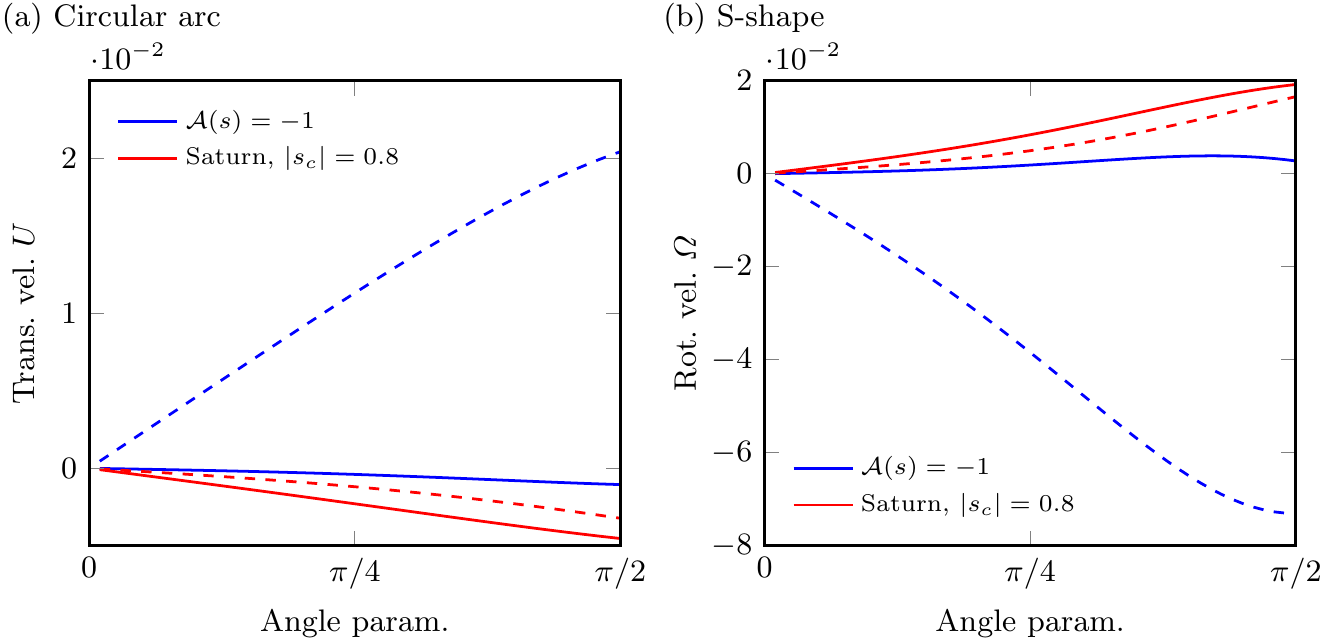}
	\end{center}
	\caption{Leading order contribution of azimuthal effects to kinematics arising from curvature and confinement.   
	Kinematics of (a) circular arc and (b) S-shaped filaments vs the angle parameter based on SPT, for uniform and Saturn activity patterns, {where the Saturn particle has two active end caps and an inert midpiece (see table~\ref{Geom_Activ_summary}. Here,} $\epsslend = 0.01$. Solid lines include azimuthal slip flows, dashed lines have this contribution removed, {i.e. $A_s = A_c = 0$ in Eq.~\eqref{u0phor}}.}  
	\label{U_S_ellipsoid_onoff_28Jun_unif_2cat_caps}
\end{figure}

For Saturn particles, {which have two active caps and an inert midpiece (see table~\ref{Geom_Activ_summary}),} the active caps ensure that the tangential concentration gradients dominate (due to the fast velocity at the change in activity) and hence the kinematics with the azimuthal terms on/off have the same directionality. 
In more detail (see also the qualitative schematic Fig.~\ref{schematic_azim_on_off}c), 
let us consider with the case of two catalytic caps of length $\Delta s=0.2$. The catalytic end-caps in the regions $|s|>0.8$ give rise to a lower concentration of solute at the end region, compared to the region $|s|< 0.8$, driving a slip flow across the interface at $s=0.8$ directed from the high to the low solute concentration region (ie towards the ends). There is of course a small region of higher solute concentration at the very ends due to the end confinement effect (magnified by the presence of the prolate spheroidal ends - see the qualitative schematic Fig.~\ref{schematic_azim_on_off}a), driving a small slip flow at the very ends directed to the middle, but overall the slip flow at the interface at $|s|=0.8$ dominates, leading to motion of the filament in the negative direction, as shown by the red, dashed line in Fig.~\ref{U_S_ellipsoid_onoff_28Jun_unif_2cat_caps}a.  %and explained qualitatively in the schematic of Fig.~\ref{}? 

Now consider the azimuthal effect, focusing on the middle region around $s=0$, which by the confinement effect, has higher solute concentration close to the `outer' surface that looks away from the centre of the semicircle. This drives an azimuthal flow around the filament towards the centre of the semicircle, which contributes to a negative direction, ie in the same direction as the kinematics with the azimuthal effects switched `off', hence the overall speed is increased when the azimuthal effects are included, as shown by the red, solid  line in Fig.~\ref{U_S_ellipsoid_onoff_28Jun_unif_2cat_caps}a. 

Similar explanations hold for the kinematics of S-shaped filaments shown in Fig.~\ref{U_S_ellipsoid_onoff_28Jun_unif_2cat_caps}b. 

%and explained qualitatively in the schematic of Fig.~\ref{}?
%PK: DTom, do we need a second schematic for the 2 catalytic caps explanation?

\begin{figure}
   \begin{center}
		\includegraphics{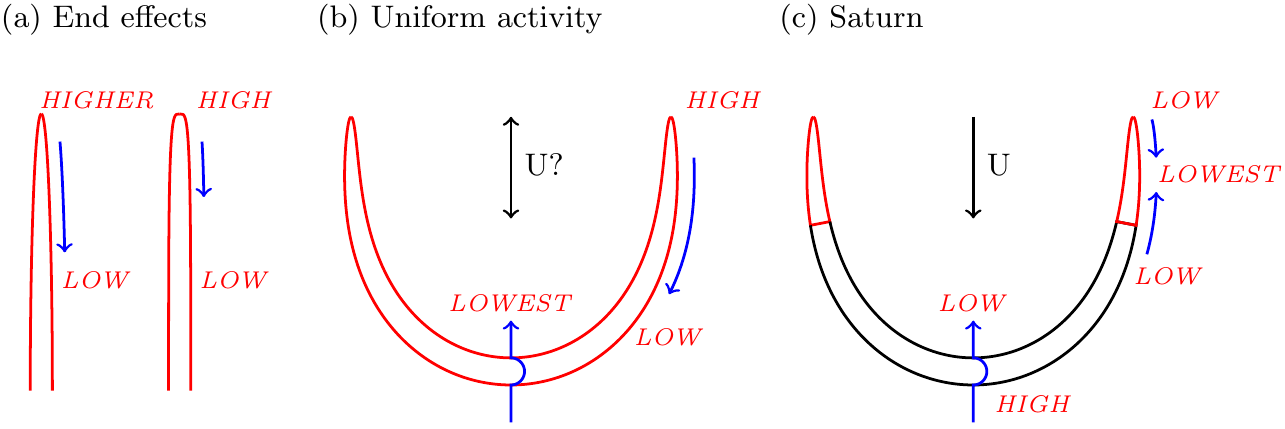}
    \caption{Qualitative schematic: understanding end shape and azimuthal effects. Red text indicates the relative concentration of solute in each figure, with slip flow (blue) moving from higher to lower concentrations, assuming constant, negative mobility. a) The shape of the filaments affects the geometric confinement at the ends, resulting in different strength of slip flows. b) A uniformly active prolate spheroid. For our parameter choices, azimuthal flows work in opposition to axial flows and hence the swimming direction is sensitive to which effect is stronger, which in turn depends on geometry and end shape. c) In a capped filament azimuthal flows and axial flows work together, however there is an opposing end-effect flow (as with the Janus prolate spheroid) that slows the swimmer. } \label{schematic_azim_on_off}
    \end{center}
\end{figure}

\section{Janus helix}  \label{JanusHelix_results_sec}
Finally, we apply SPT to fully 3D geometries, focusing specifically on helical centrelines. As we will see in \S\ref{ExplorationofSpace}, Janus helical filaments give rise to helical trajectories, which offer novel exploration of space capabilities which are not currently accessible to phoretic microswimmer designs of Janus spheres, rods and prolate spheroids. 

\begin{figure}
	\begin{center}
	    \includegraphics{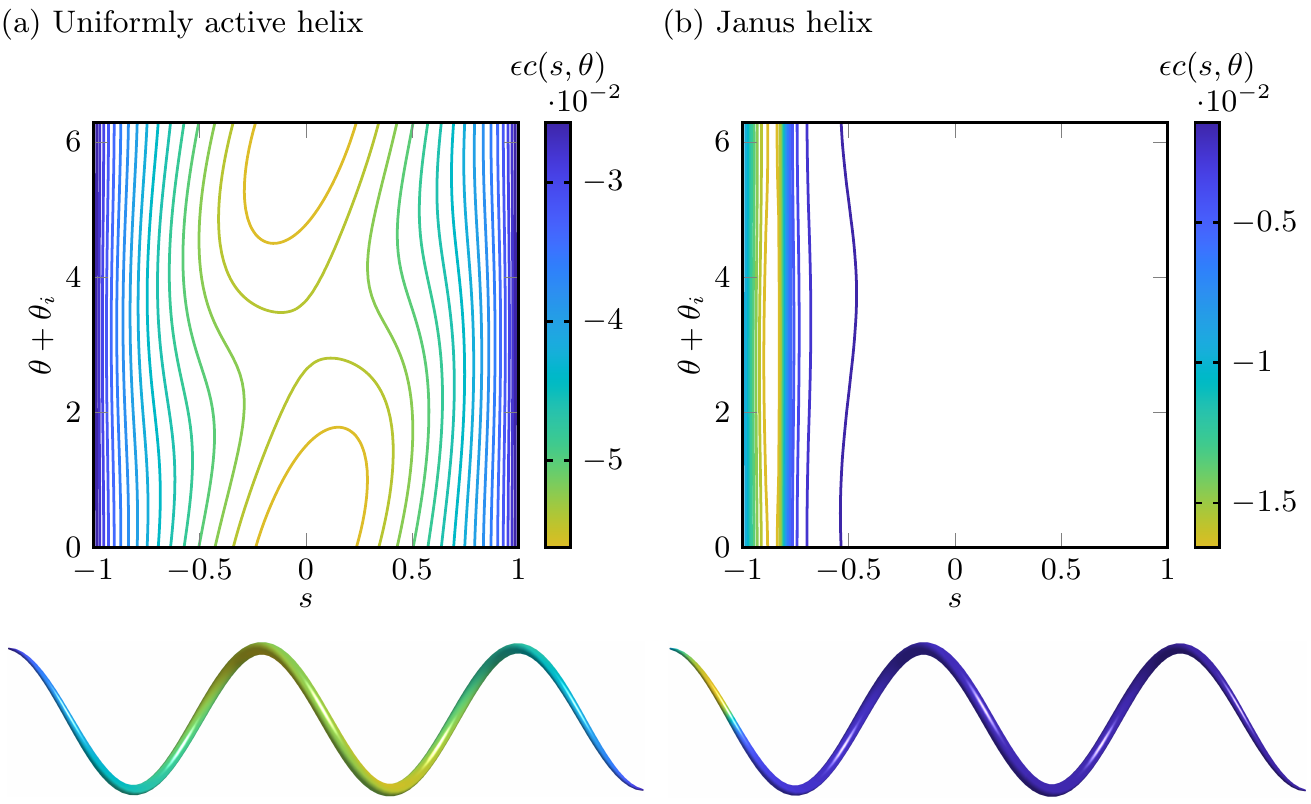}
	\end{center}
 	\vspace{-0.0cm}
	\caption{The surface concentration of a) a uniformly active helix and b) a Janus helix, as calculated with SPT, with $\epsslend = 0.01,\alpha=\pi/3, s_c = -0.8,$ and 2.5 turns.} \label{HelixConcentrationPlots}
\end{figure}

\subsection{Helix geometry, concentration and kinematics}
% \subsubsection*{Geometry}
We consider a helical centreline with chirality $h$ (taking the values $\pm1$ for right-handed and left-handed helices respectively), helical angle $\angleampl$, and $n$ turns,
\begin{equation}
   %\Rhelix= \frac{\sin\angleampl}{n\pi}, \quad
(x,y,z) =  \left(\Rhelix\cos\frac{(s+1)\sin \angleampl}{\Rhelix},  h\Rhelix\sin\frac{(s+1)\sin \angleampl}{\Rhelix},   s\cos \angleampl \right), 
\end{equation}
where the helical radius $\Rhelix$, curvature $\curvature$ and torsion $\torsion$ are given by
\begin{equation}
    \Rhelix= \frac{\sin\angleampl}{n\pi}, \quad \curvature =  \frac{\sin^2\angleampl}{\Rhelix}, \quad
        \torsion   =  h\frac{\sin\angleampl \cos\angleampl}{\Rhelix},
\end{equation}
and its pitch by $(2\cos\angleampl)/n$. %Its curvature $\curvature$ and torsion $\torsion$ are given by
%\begin{equation}
 %      \curvature =  \frac{\sin^2\angleampl}{\Rhelix}, \quad
  %      \torsion   =  h\frac{\sin\angleampl \cos\angleampl}{\Rhelix} \cdot
%\end{equation} 
%
We begin by validating a constant cross-section rod against BEM for a helix with 2.5 turns and $\angleampl = \pi/3$, with regularised activity $\mathcal{A}(s) = -(s+1)(1-\tanh(2s))$ and mobility $\mathcal{M}= -(s+1)(1-\tanh(4s))$. This regularisation is somewhat similar to a single capped end, but with a smoothly decaying velocity. We achieve approximately 1\% error in the concentration field, and 2\% and 7\% errors in translational and angular velocities respectively, { where angular velocity is calculated about the helix centroid. We note that this larger relative error in the swimming velocity arises from the fact that this regularised case only swims relatively slowly.}

Continuing with SPT for a filament with prolate spheroidal ends, the surface concentration of a uniformly active helix is shown in Fig.~\ref{HelixConcentrationPlots}a for varying arclengths ($s$) and azimuthal positions, using $\theta+\theta_i(s)$ in order to ensure that $\theta$ lies in $[0,2\pi)$ (recall that by definition, $\thetators(s)$ follows the torsion of the filament - Eq.~\eqref{theta_i_tors}). The apparent shear about $\theta(s) = \pi$ is a result of this choice for $\thetators(s)$. In Fig.~\ref{HelixConcentrationPlots}b, we show the surface concentration of a Janus helix capped for $s \leq -0.8$, with $\epsslend = 0.01$, showing strong local gradients around the capped end.

\subsection{Exploration of space}\label{ExplorationofSpace}
	Many swimming microswimmers make use of helical trajectories for exploring their environment and sensing, and thus also perform chemotaxis towards food resources. Helical trajectories could offer enhanced sensing and 3D mixing to artificial phoretic microswimmer applications. 
	
	As shown {schematically in Fig.~\ref{fig:helix_traj_schematic}}, Janus helices can explore space on a helical trajectory, with radius and pitch that vary with the geometry of the helical filament. Fig.~\ref{JanusHelix}a shows the trajectories of Janus helices of the same arclength ($2$), helical angle ($\pi/3$) and length of catalytic cap ($s_c = -0.9$), but different number of turns. All simulations run over the same time interval ($t=30$ in dimensionless units). In real space, the axis of the trajectories is aligned with the $\boldsymbol{\Omega}$ vector. In Fig.\ref{JanusHelix}a the trajectories have been rotated to be aligned vertically for the purpose of comparison. In order to quantify the ability of the different Janus helices to explore space, in Fig.~\ref{JanusHelix}b we plot the {explorative area of the helical trajectory, which is equal to $\pi R_T^2$ for $R_T$ the helix radius as in Fig.~\ref{fig:helix_traj_schematic}.} Fig.~\ref{JanusHelix}c shows the velocity along the axis of each helical trajectory vs the number of turns.
	
	\begin{figure}
	\begin{center}
	    \includegraphics{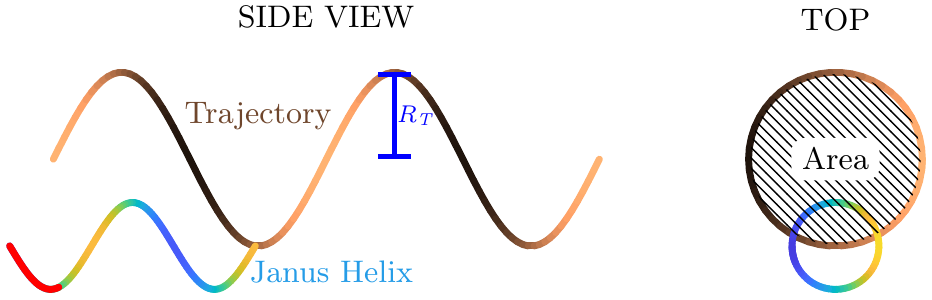}
	\end{center}
 	\vspace{-0.0cm}
	\caption{Schematic showing a Janus Helix, on a helical trajectory of radius $R_T$, from a side and top view, demonstrating the ``exploratory area'' of the swimmer $\pi R_T^2$} \label{fig:helix_traj_schematic}
\end{figure}
	
	These plots show interesting nonlinear behaviour, with a number of transition points, the most dramatic of which  is close to when the number of turns $n$ increases above a single full turn.  There are two limiting behaviours. Firstly, as $n \to 0$, the helix tends to a straight rod, and the axial velocity in turn tends to that of a straight rod. However, the explorative area does not seem to approach zero in this limit - rather the nearly straight helix may explore a non-zero or even infinite area, but take an infinite time to do it as $n\to 0$.  
	%It may be possible to derive this limiting behaviour analytically for nearly straight helices.
	On the other hand, as $n\to\infty$, the helix again approaches a straight rod, { and we would expect the axial velocity to once more tend to the straight case. However, we cannot analyse this limit within our framework, as it entails divergent curvature $\kappa$; our analysis remains valid provided $\kappa = n\pi\sin\alpha \ll 1/\epsslend$.}
	
	We note that between these limits, we expect a decrease in axial velocity (since propulsion force does not lie solely along the axial direction), and hence we expect a minimum in axial velocity - obtained somewhere in $n\in(0.75,1)$ for the particular values of parameters we have chosen here. This behaviour is not monotonic, but displays periodic fluctuations that change depending on whether the helix has an odd or even number of turns, {that are not clearly visible in figure~\ref{JanusHelix}a} . Finally, we note that smart, stimulus-responsive materials, such as thermoresponsive hydrogel composites, could assist in exploiting both features by changing the number of turns of a Janus helix, as can occur naturally in the polymorphism of bacterial flagella~\citep{spagnolie2011comparative}, so that it can have a slow, explorative mode with less than one turn, and a faster, close to straight swimming mode at higher number of turns. 
	
% 	For $n>n_*$ the explorative area and axial velocity decrease monotonically as 
% 	 the number of turns increases. Thus from the Janus helix trajectories plotted in Fig.~\ref{JanusHelix}, the helices with $n<n_*$ are best suited for exploration of space purposes, but are slower compared to the helices with more turns  $n>2.5$, indicating that enhanced  exploration of space might come at a the price of reduced speed. 

	%PK2: Tom, it would be nice if we could zoom in around n=0. For n=0 the explorative area should be 0. TDMJ - no the explorative area in this case is infinite - but it takes infinitely far to explore!
	
    %PK2: Tom, do we want to add anything else for the Janus helix?

	%If made from a  thermoresponsive material, the geometry could be vary to alter the explorative radius of the  	trajectories - TDMJ I like this!

\begin{figure}
	\begin{center}
	    \includegraphics{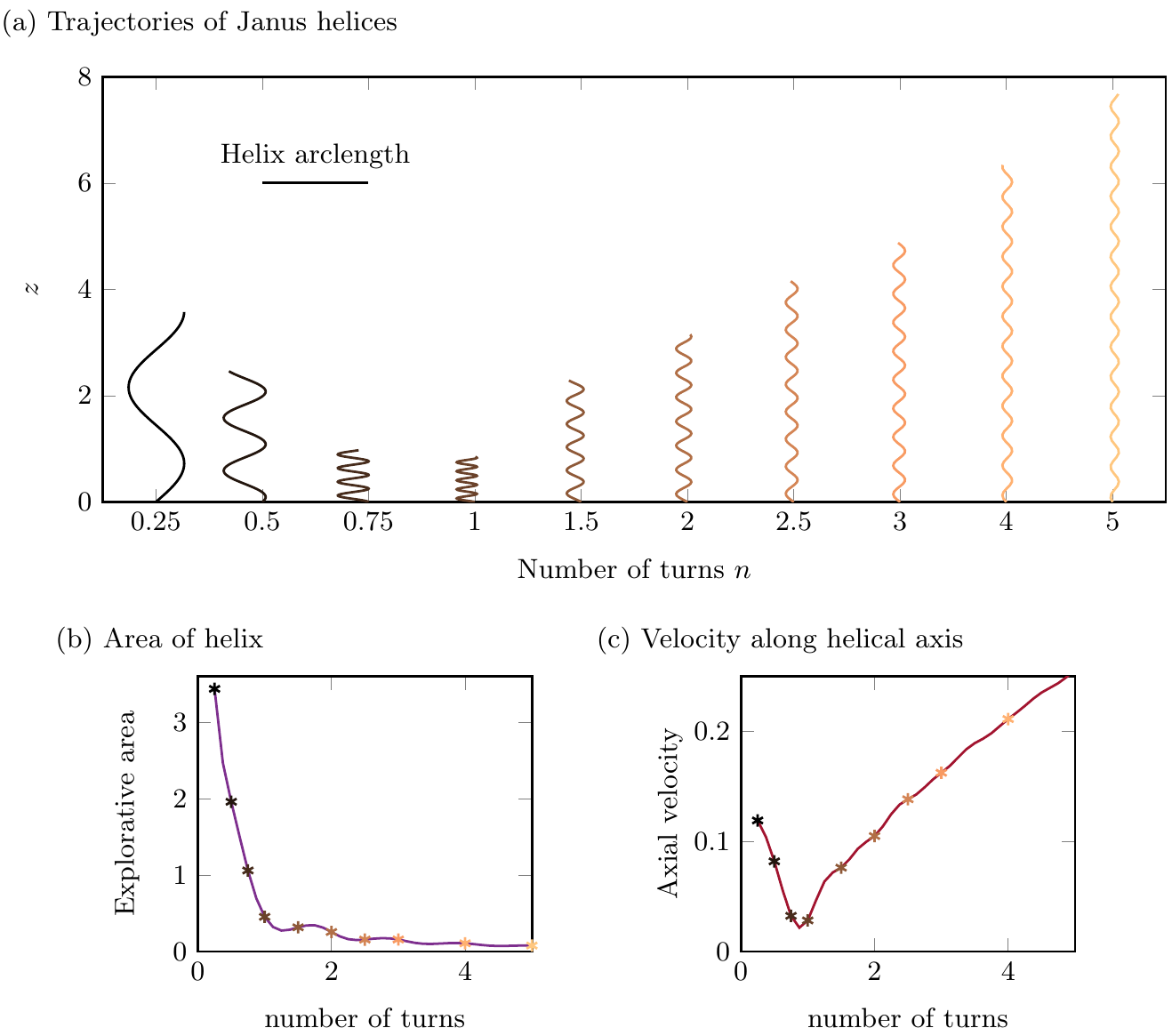}
	\caption{Trajectories of helical filaments with one catalytic end of length $0.1$, i.e. $s_c = -0.9$, helical angle $\pi/3$, and varying number of turns. 
(a) The different helical trajectories arising from varying the number of turns. All simulations run over the same time interval ($t=30$ in dimensionless units). The axis of the trajectories is aligned with the $\Omega$ vector in real space. The trajectories have been rotated to be aligned vertically for the purposes of this plot. 
(b) {Area of helical trajectories, i.e. $\pi R_T^2$ for helical radius $R_T$, vs  number of turns of the Janus helix.}
(c) Velocity along the axis of each helical trajectory vs number of turns of the Janus helix.
} \label{JanusHelix}
	\end{center}
\end{figure}

\section{Discussion}
This paper provides a novel and complete framework to analyse the dynamics of slender (auto)-phoretic filaments with centrelines that are not limited to straight geometries, but are fully 3D. These (auto)-phoretic filaments self-propel by catalysing a reaction of a solvent in their surroundings. Catalytic patterning on the filament surface induces differential surface reaction, which generates solute concentration gradients that drive a propulsive surface slip flow.  
% In self-diffusiophoresis(autophoresis), surface patterning of a particle with a catalyst gives rise to differentialsurface  reaction,  allowing  the  particle  to  self-generate  solute  concentration  gradientswhich  drive  a  propulsive  slip  flow
 %These modes of motion offer enhanced control compared to typical Janus spheres, especially when shape-transformation is applied. 

Previous analytical theories have considered straight slender phoretic filaments~\citep{yariv2008slender,schnitzer2015osmotic,yariv2019self}. %Numerically, solving the underlying partial differential equations for the motion of slender filaments in 3D would render dynamic calculations prohibitively slow, owing to the need to resolve both the dramatically different length scales of width and length. This is before considering more complex problems, such as interacting filaments and coupled fluid-structure interactions. 
We developed a Slender Phoretic Theory which addresses slender phoretic filaments of arbitrary shape.  We used matched asymptotics to expand the boundary integral representation of the solution of the diffusion equation in the slender limit, similarly to the approach of  \citet{KoensLauga2018} for filaments moving in viscous fluids. In contrast to many other slender body theories, which finish at leading order, the form of the azimuthal slip flow,  necessitates that SPT be derived to first order in the slenderness of the filament. These first order azimuthal variations to the concentration arise from confinement effects from curvature, drive azimuthal slip flows  and have a leading order contribution to the kinematics. The latter can be as profound as reversing the direction of motion.  
%For axisymmetric activity profiles, the structure of the slip flow implies azimuthal concentration gradients arise due the presence of curvature and have a have a leading order contribution to the kinematics (eg the swimming speed of a U-shaped filament with 2 catalytic caps). This required us to find the next order correction ($O(\epsslend$) to the concentration field.  
%SPT, to address an axisymmetric activity profile

% the structure of the slip flow implies that azimuthal effects
%	Crucially, unlike other slender body theories that finish at leading order, the structure of the slip flow implies azimuthal effects have a leading order contribution to the swimming velocity, and arise whenever curvature is present by a confinement effect of the diffusion. 
%	This is in agreement to numerical evidence in \citet{MontenegroJohnson2018Microtransformers}, where reversal reversal of azimuthal flow led to reduction to  two fifths of the swimming velocity.
%	Here we show analytically that the azimuthal flow has a leading order contribution to kinematics whenever the filament is curved. 

{ SPT is more computationally efficient than BEM. However, although the speed-up in calculating the surface concentration distribution is considerable ($\approx 300\times$), for the examples we considered herein, the boundary element calculation is not prohibitively slow at $O(10)$ seconds. As with classical slender body theory for Stokes flows, which has proven to be efficient and popular in the study of many biological flows, such as ciliary flows, the full capability of SPT lies in two possible extensions to the theory, that we believe will make fruitful avenues of future research. The first is in efficiently evaluating the solute dynamics of multiple filaments. This would likely use a representation of neighbouring curved filaments by line distributions of sources and dipoles, weighted by activity and curvature, but will require significant work to rigorously derive. The second use is in dynamic fluid-structure interaction problems for flexible chemically active filaments. There have been a number of recent advances in efficient simulation elastohydrodynamics combining local drag theory or slender body theory for Stokes flow with elastic beam theory \citep{moreau2018asymptotic, hall2019efficient, walker2019filament, schoeller2019methods}. Taking \citet{hall2019efficient} as an example, a typical simulation took 0.0007 seconds per timestep, with a single beat of the active filament model (the minimum for interesting dynamics) taking around 20 seconds to resolve, ie around 30000 timesteps. To consider similar chemoelastohydrodynamic simulations, coupling the concentration solution via SPT would increase the total run time to around 22 minutes, whereas coupling via the boundary element method would increase the total run time to 100 hours. It is also worth noting that our SPT code has as yet not been optimised for speed, and in fact more efficient implementations will certainly be able to reduce this further.
}

Furthermore, the form of the SPT integral equation reveals the underlying structure of the various contributions to the concentration field (and swimming velocity), providing opportunities for new insights. In particular, we note that the torsion is absent from the SPT equation, as it appears only at $O(\epsslend^2)$. The natural decomposition of the azimuthal slip flow into Fourier modes furthermore makes the theory ideal for coupling with the viscous flow slender body theory of \citet{KoensLauga2018}, providing further gains in efficiency. We hope that this new theory will provide the { basis} for researchers to begin to look at complex, interacting, flexible phoretic filament flows { in 3D}.

%PK_comms: 
\subsection*{Acknowledgements}
TDM-J and PK gratefully acknowledge funding from EPSRC Bright Ideas grant no. EP/R041555/1. SM acknowledges funding from the European Research Council (ERC) under the European Union's Horizon 2020 research and innovation programme (grant agreement 714027 to SM).  The authors would like to thank Eric Lauga for helpful discussions on autophoretic theory, and Lyndon Koens for helpful discussions on Slender Body Theory. \\

\noindent Declaration of Interests. The authors report no conflict of interest.

\appendix
\section{Expansion of the kernels}
\subsection{Outer region} \label{OuterRegion_App}
Expanding in the outer region, $s-\sdum=O(1)$,  (ie $\bvRo(s,\sdum)=O(1)$ and we can approximate $\bvR \approx\bvRo$), and using Eqs.~\ref{bvR_outer},\ref{Deltaerhorho_outer}, 
%In the outer region, $s-\sdum=O(1)$, hence  $\bvRo(s,\sdum)=O(1)$, so we %can approximate $\bvR \approx\bvRo$.
%have 
%To first order,
Expanding $1/|\bvR|$, 
\begin{align}
\frac{1}{|\bvR|} 
%&=\frac{1}{|\bvRo(s,\sdum) + \epsslend \Deltaerhorho|}\nonumber\\
&=\left[|\bvRo|^2 + 2\epsslend \bvRo\cdot\Deltaerhorho  + O(\epsslend^2)\right]^{-1/2} 
\nonumber\\
%&=\frac{1}{|\bvRo|}\left[1 + 2\epsslend \frac{\bvRo}{|\bvRo|^2}\cdot\Deltaerhorho  + O(\epsslend^2)\right]^{-1/2} 
%\nonumber\\
&=\frac{1}{|\bvRo|}\left[1-\epsslend \frac{\bvRo}{|\bvRo|^2}\cdot\Deltaerhorho + O(\epsslend^2)\right], \end{align}
%
%\begin{align}
%4\pi c(s,\theta)
%=&-\epsslend\int\int \frac{\crossradius(\sdum)c(\sdum,\thetadum)  \left[\bvRo(s,\sdum)\cdot\erho(\sdum,\thetadum)\right]}{ |\bvRo(s,\sdum)|^3} \dd \sdum \dd \thetadum \nonumber\\
%&+ \epsslend\int\int\frac{\tilde{\activity}(\sdum,\thetadum)\crossradius(\sdum)}{ |\bvRo(s,\sdum)|} \dd \sdum \dd \thetadum + O(\epsslend^2)
%\end{align}
%\begin{align}
%\outerexp{\kerone}(s,\sdum,\theta,\thetadum)&=  \frac{\crossradius(\sdum)\activity(\sdum,\thetadum)}{|\bvRo|}\left[1-\epsslend \frac{\bvRo}{|\bvRo|^2}\cdot\Deltaerhorho + O(\epsslend^2)\right]\left[ 1 - \epsslend\crossradius(\sdum)\curvature(\sdum)\cos\dthetaprime  + \frac{1}{2}\epsslend^2 \left(\crossradius'(\sdum)\right)^2  + O(\epsslend^3)\right] , \label{K1_outer}
%\end{align}
%
%
%\begin{align}
%4\pi c(s,\theta)
%=&-\epsslend\int\int \frac{\crossradius(\sdum)c(\sdum,\thetadum)  \left[\bvRo(s,\sdum)\cdot\erho(\sdum,\thetadum)\right]}{ |\bvRo(s,\sdum)|^3} \dd \sdum \dd \thetadum \nonumber\\
%&+ \epsslend\int\int\frac{\tilde{\activity}(\sdum,\thetadum)\crossradius(\sdum)}{ |\bvRo(s,\sdum)|} \dd \sdum \dd \thetadum + O(\epsslend^2)
%\end{align}
%\scriptsize
which gives the outer expansions $\outerexp{\kerone}, \outerexp{\kertwo}$ of the kernels $\kerone,\kertwo$, defined in Eqs.~\eqref{keroneref}-\eqref{kertwodef}, as  
\begin{align}
  \outerexp{\kerone}  
%\nonumber\\
%
&=	~\frac{\crossradius(\sdum)\activity(\sdum,\thetadum)}{ |\bvRo(s,\sdum)|}    \left[1-\epsslend \frac{\bvRo\cdot\Deltaerhorho}{|\bvRo|^2} + O(\epsslend^2)\right]\left[ 
1 - \epsslend\crossradius(\sdum)\curvature(\sdum)\cos\dthetaprime  +  O(\epsslend^2)\right] \nonumber \\
&=	~\frac{\crossradius(\sdum)\activity(\sdum,\thetadum)}{ |\bvRo(s,\sdum)|}    
\left[1-\epsslend \left[\crossradius(\sdum)\curvature(\sdum)\cos\dthetaprime+\frac{\bvRo\cdot\Deltaerhorho}{|\bvRo|^2} \right]  +  O(\epsslend^2)\right]
\label{K1_outer_App},\\
\outerexp{\kertwo} %\nonumber\\
&=\frac{\epsslend\crossradius(\sdum)c(\sdum,\thetadum)}{ |\bvRo|^3}
\begin{bmatrix}
1 - \frac{3\epsslend\bvRo\cdot\Deltaerhorho}{|\bvRo|^2}\\ 
%\\ + O(\epsslend^2) \qquad\qquad\qquad
%
\end{bmatrix}\times \nonumber\\
&\hspace{-2em}%\qquad
\begin{bmatrix}\bvRo + \epsslend \Deltaerhorho
\end{bmatrix}
 \cdot 
\begin{bmatrix}
\erho(\sdum,\thetadum)  - \epsslend 
\ddrhodsprime\tanhat(\sdum) 
- \epsslend \crossradius(\sdum) \curvature(\sdum) \cos\dthetaprime \erho (\sdum,\thetadum)
\end{bmatrix}
%\nonumber\\ &\qquad
+ O(\epsslend^3) \nonumber\\
&=\epsslend\frac{\crossradius(\sdum)c(\sdum,\thetadum)}{ |\bvRo|^3}
\Bigg\{
\bvRo\cdot\erho(\sdum,\thetadum) + \epsslend \left(\Deltaerhorho\right)\cdot\erho(\sdum,\thetadum)\nonumber \\
&\qquad\qquad\qquad\qquad
- \epsslend \bvRo\cdot\left[ \ddrhodsprime\tanhat(\sdum) +\crossradius(\sdum) \curvature(\sdum) \cos\dthetaprime \erho (\sdum,\thetadum)\right]  \nonumber \\
&\qquad\qquad\qquad\qquad\qquad\qquad\qquad\quad
- \epsslend (\bvRo\cdot\erho(\sdum,\thetadum)) \frac{3\bvRo\cdot\Deltaerhorho}{|\bvRo|^2} 
+O(\epsslend^2) 
\Bigg\}
    \label{K2_outer_App}
\end{align}
In the above expressions we keep the first two orders to ensure all relevant contributions are accounted for, in particular in anticipation of the matching to follow (A.3). 
\normalsize

\subsection{Inner region} \label{InnerRegion_App}
\subsubsection{Inner region expansion} \label{Inner_Exp_App}
In the inner region, $s-\sdum=O(\epsslend)$  %, i.e. $\sdum$ is of order $\epsslend$ of $s$,
and we let $\sdum=s+\epsslend\chi$, where $\chi$ is O(1). We proceed by Taylor-expanding functions of $\sdum$ around $s$, using 
the Serret-Frenet equation $\dds{\tanhat}= \curvature \norhat(s)$, and from Eq.~\eqref{erhoderivs},  
$\dds{\erho}=-\curvature(s) \cos\dtheta\tanhat(s)$, %the Taylor expansion for $\erho(\sdum,\thetadum)$,   
\begin{subequations}
\begin{align}
\crossradius(\sdum)&=\crossradius(s) + \epsslend\chi \ddrhods  + O(\epsslend^2),\\
%\end{align}
%For the centreline position vector at $\sdum$ we have 
%\begin{align}
%\rc(\sdum)
%&=\rc(s) + (\epsslend\chi)\rc'(s)+ \frac{1}{2} (\epsslend\chi)^2 \rc''(s) + O(\epsslend^3) \nonumber\\
%\end{align}
%and using the Serret-Frenet equation $\dds{\tanhat}= \curvature \norhat(s)$,
%\begin{align} \dds{\tanhat}&= \curvature \norhat(s), \qquad  \dds{\norhat}= -\curvature\tanhat(s) + \torsion \binorhat(s),\qquad  \dds{\binorhat}=-\torsion \norhat(s), \end{align}
%we obtain
%\begin{align}
%\rc(\sdum) 
\rc(\sdum)&=\rc(s) + \epsslend\chi\tanhat(s) + \frac{1}{2} (\epsslend\chi)^2\curvature(s) \norhat(s) + O(\epsslend^3) \\
\erho(\sdum,\thetadum)&=\erho(s,\thetadum)-\epsslend \chi  \curvature(s) \cos\theta_m(s,\thetadum) \tanhat(s),\\
%\end{align}
%\begin{align}
%\erho(\sdum,\thetadum)&=\erho(s,\thetadum)-\epsslend \chi  \curvature(s) \cos\theta_m(s,\thetadum) \tanhat(s), 
%\end{align}
%allows us to expand 
%\begin{align}
%\surf(\sdum,\thetadum)
%&=\rc(\sdum) + \epsslend \crossradius(\sdum)\erho(\sdum,\thetadum) \nonumber\\
%&=\rc(s) + \epsslend\chi\tanhat(s) + \frac{1}{2} (\epsslend\chi)^2\curvature(s) \norhat(s)  
%\nonumber\\
%&\quad
%+ \epsslend \left[ \crossradius(s) + \epsslend\chi \crossradius'(s)  + O(\epsslend^2) \right]\left[\erho(s,\thetadum)-\epsslend \chi  \curvature(s) \cos\theta_m(s,\thetadum) \tanhat(s)  \right]+ O(\epsslend^3)\nonumber\\
%
\surf(\sdum,\thetadum)&=\rc(s) + \epsslend\left[\chi\tanhat(s) +\crossradius(s)\erho(s,\thetadum) \right]
\nonumber\\
&\quad
+\epsslend^2 \bigg[ \frac{1}{2} \chi^2 \curvature \norhat(s) + \chi\ddrhods \erho(s,\thetadum) 
%\nonumber\\&\qquad\qquad\quad
- \crossradius(s) \chi \curvature(s) \cos\theta_m(s,\thetadum)\tanhat(s)\bigg] + O(\epsslend^3).
\end{align}
\end{subequations}
%
%Substituting this in Eq.~\eqref{bvRgeneral},
%\begin{align}
%-\bvR
%&=\left[\rc(\sdum) + \epsslend \crossradius(\sdum)\erho(\sdum,\thetadum)\right] - \left[\rc(s) + \epsslend \crossradius(s)\erho(s,\theta)\right]  \nonumber\\
%&=
% \epsslend\left[\chi\tanhat(s) +\crossradius(s)\left[\erho(s,\thetadum)- \erho(s,\theta)\right]\right]
% \nonumber\\
%&\quad
%+\epsslend^2 \left[ \frac{1}{2} \chi^2 \curvature \norhat(s) + \chi\crossradius'(s) \erho(s,\thetadum) - \chi\crossradius(s)  \curvature(s) \cos\theta_m(s,\thetadum)\tanhat(s)\right] + O(\epsslend^3).
%\end{align}
%
%We write this as
Eq.~\eqref{bvRgeneral} gives
\begin{align}
\bvR&=\epsslend\left[\bvRione + \epsslend \bvRitwo\right] + O(\epsslend^3),
\end{align}
where
\begin{subequations}
\begin{align}
\bvRione
&=-\chi\tanhat(s) +\crossradius(s)\left[\erho(s,\theta)- \erho(s,\thetadum)\right],  \label{bvRione_App}
\\
\bvRitwo&=-\left[ \frac{1}{2} \chi^2 \curvature \norhat(s) + \chi\ddrhods \erho(s,\thetadum) - \chi\crossradius(s)  \curvature(s) \cos\theta_m(s,\thetadum)\tanhat(s)\right], \label{bvRitwo_App} \\
%\end{align}
%and 
%\begin{align}
|\bvRione + \epsslend \bvRitwo|
%&=\sqrt{|\bvRione|^2 + 2\epsslend \bvRione\cdot\bvRitwo + O(\epsslend^2)} 
%\nonumber\\
%&=|\bvRione|\sqrt{1 + 2\epsslend \frac{\bvRione\cdot\bvRitwo}{|\bvRione|^2} + O(\epsslend^2)} \nonumber\\
&=|\bvRione|\left[1 + \epsslend \bvRione\cdot\bvRitwo/|\bvRione|^2 + O(\epsslend^2)\right].
\end{align}
\end{subequations}
%Calculations given in Appendix~\ref{InnerRegion_Appendix} give rise to 
%obtain $\kerone$ and $\kertwo$,
Using these, the inner expansions of $\kerone, \kertwo$ become
\begin{align}
\epsslend |\bvRione| \innerexp{\kerone}&=
\crossradius(s)\activity(s,\thetadum)
%\nonumber\\ &\quad 
+
\epsslend 
\Bigg(
+ \chi \left[\crossradius(s)\partial_s\activity(s,\thetadum) +\ddrhods\activity(s,\thetadum)\right] 
\nonumber\\
&\quad %\qquad\qquad \quad
-\crossradius(s)\activity(s,\thetadum)\left[\crossradius(s)\curvature(s)\cos\theta_m(s,\thetadum) + \bvRione\cdot\bvRitwo/|\bvRione|^2  \right] 
\Bigg)
%\nonumber\\ &\quad 
+ O(\epsslend^2),
%\\
%
\nonumber\\
\epsslend|\bvRione|^3 \innerexp{\kertwo}
&=
\crossradius(s)c(s,\thetadum) \bvRione\cdot
\erho(s,\thetadum) 
%\nonumber \\&\quad
+
\epsslend\chi \partial_s\left[\crossradius(s) c(s,\thetadum)\right] \bvRione\cdot 
\erho(s,\thetadum) \nonumber\\
&\quad
+
\epsslend \crossradius(s)c(s,\thetadum)
\Bigg[
\bvRitwo\cdot\erho(s,\thetadum)- \left(\chi  \curvature(s) \cos\theta_m(s,\thetadum) +   \ddrhods\right)\bvRione\cdot\tanhat(s)
\nonumber\\
&\quad %\qquad%\qquad%\qquad\quad \quad
- \left(\crossradius(s) \curvature(s) \cos\theta_m(s,\thetadum) 
+\frac{3\bvRione\cdot\bvRitwo}{|\bvRione|^2} \right)\bvRione\cdot\erho(s,\thetadum)
\Bigg]
%\nonumber\\ &\quad
+ O(\epsslend^2 c).
\end{align}

\normalsize

\subsubsection{Simplifying the inner integrals} \label{Inner_Simpl_App}
From Eq.~\eqref{bvRione_App} we have that 
\begin{align}
|\bvRione|&=\sqrt{\chi^2+\gamma^2},
\end{align}
where
\begin{equation}
\gamma^2=2\crossradius^2(s)\left[1-\cos(\theta-\thetadum)\right].    
\end{equation}
and 
\begin{subequations}
\begin{align}
\bvRione\cdot\tanhat(s) &= - \chi,\\
\bvRione\cdot\erho(s,\thetadum) &=\crossradius(s) \left[\cos(\theta-\thetadum)-1\right],\\
\bvRitwo\cdot\erho(s,\thetadum)  &=-\left[\frac{1}{2}\chi^2\curvature(s)\cos\theta_m(s,\thetadum) + \chi\ddrhods\right],
\end{align}
\end{subequations}
where we used that $\erho(s,\theta)\cdot\erho(s,\thetadum)=\cos(\theta-\thetadum)$. 

\noindent Rewriting Eqs.~\eqref{bvRione_App}, \eqref{bvRitwo_App} in the $\{\tanhat,\norhat,\binorhat\}$ frame,
\begin{align}
    \bvRione 
   &=-\chi\tanhat(s) +\crossradius(s)\left[\erho(s,\theta)- \erho(s,\thetadum)\right] 
\\    &\hspace{-2em} =
-\chi\tanhat{s}+ \crossradius(s)\left\{\left[
\cos\theta_m(s,\theta) - \cos\theta_m(s,\thetadum)\right]\norhat(s)
%\nonumber\\&\qquad\qquad
+  \left[
\sin\theta_m(s,\theta) - \sin\theta_m(s,\thetadum)\right]\binorhat(s)\right\}
,
\\
\bvRitwo 
&=-\left[ \frac{1}{2} \chi^2 \curvature \norhat(s) + \chi\ddrhods \erho(s,\thetadum) - \chi\crossradius(s)  \curvature(s) \cos\theta_m(s,\thetadum)\tanhat(s)\right]\\
&=
\chi\crossradius(s)\curvature(s)\cos\theta_m(s,\thetadum)\tanhat(s)
-\left[\frac{1}{2}\chi^2\curvature(s) + \chi\ddrhods\cos\theta_m(s,\thetadum) \right]\norhat(s)\nonumber\\
&\qquad\qquad\qquad\qquad\qquad\qquad\quad
-\chi\ddrhods\sin\theta_m(s,\thetadum)\binorhat(s) ,
\end{align}
we obtain 
%\scriptsize
\begin{align}
\bvRione\cdot\bvRitwo 
%&=\begin{pmatrix}
%-\chi\tanhat{s}\\ \crossradius(s)\left[
%\cos\theta_m(s,\theta) - \cos\theta_m(s,\thetadum)\right]\norhat(s)\\ \crossradius(s)\left[ \sin\theta_m(s,\theta) - \sin\theta_m(s,\thetadum)\right]\binorhat(s)\end{pmatrix}
%\cdot
%\begin{pmatrix}\chi\crossradius(s)\curvature(s)\cos\theta_m(s,\thetadum)\tanhat(s)\\ -\left[\frac{1}{2}\chi^2\curvature(s) + \chi\crossradius'(s)\cos\theta_m(s,\thetadum) \right]\norhat(s)\\-\chi\crossradius'(s)\sin\theta_m(s,\thetadum)\binorhat(s) \end{pmatrix} \nonumber\\
%
&=-\chi\crossradius(s) \bigg[\frac{1}{2}\curvature(s) \chi \left(\cos\theta_m(s,\thetadum) + \cos\theta_m(s,\theta)\right) %\nonumber\\
%&\qquad\qquad\qquad\qquad\qquad\quad 
+ \ddrhods (\cos(\theta-\thetadum)-1)\bigg].
\end{align}

% for $\innerexp{\kerone}$,
%\begin{align}
%\innerexp{\kerone}&=\frac{\crossradius(s)\activity(s,\thetadum)}{ \epsslend\sqrt{\chi^2+\gamma^2}}\quad +\quad \epsslend \frac{1}{\epsslend\sqrt{\chi^2+\gamma^2}} \begin{pmatrix}&-\crossradius(s)\activity(s,\thetadum)\left[\crossradius(s)\curvature(s)\cos\theta_m(s,\thetadum) + \frac{\bvRione\cdot\bvRitwo}{\chi^2+\gamma^2}  \right] \\& \qquad \quad + \chi \partial_s\left[\crossradius(s)\activity(s,\thetadum) \right] \end{pmatrix}+O(\epsslend^2)\end{align}

%\begin{align}
%\innerexp{\kerone}&=
%\frac{\crossradius(s)\activity(s,\thetadum)}{ \epsslend\sqrt{\chi^2+\gamma^2}}
%\nonumber\\
%&\quad
%-\epsslend\activity(s,\thetadum) \frac{1}{\epsslend\sqrt{\chi^2+\gamma^2}}
%\crossradius^2(s)\curvature(s)\cos\theta_m(s,\thetadum)
%\nonumber\\
%
%&\quad
%-\epsslend\crossradius(s)\activity(s,\thetadum) \frac{1}{\epsslend\sqrt{\chi^2+\gamma^2}^3}
%\left[\bvRione\cdot\bvRitwo  \right]
%\nonumber\\
%
%&\quad
%+\epsslend \frac{\chi }{\epsslend\sqrt{\chi^2+\gamma^2}}
%\partial_s\left[\crossradius(s)\activity(s,\thetadum) \right] 
%\nonumber\\
%&\quad+ O(\epsslend^2)
%\end{align}

\noindent This allows us after some rearranging, to express the inner kernels $\innerexp{\kerone}, \innerexp{\kertwo}$ as
\begin{align}
\innerexp{\kerone}&=
\frac{\crossradius(s)\activity(s,\thetadum)}{ \epsslend\sqrt{\chi^2+\gamma^2}}
-\epsslend\activity(s,\thetadum) \frac{1}{\epsslend\sqrt{\chi^2+\gamma^2}}
\crossradius^2(s)\curvature(s)\cos\theta_m(s,\thetadum)
\nonumber\\
&\quad
+\epsslend\crossradius^2(s)\activity(s,\thetadum)
\bigg[\frac{1}{2} \frac{\curvature(s)\chi^2}{\epsslend\sqrt{\chi^2+\gamma^2}^3} \left[\cos\theta_m(s,\thetadum) + \cos\theta_m(s,\theta)\right] \nonumber\\
&\qquad\qquad\qquad\qquad\qquad\qquad\qquad
+ \frac{\chi}{\epsslend\sqrt{\chi^2+\gamma^2}^3}\ddrhods [\cos(\theta-\thetadum)-1]\bigg]
\nonumber\\
&\quad
+\epsslend \frac{\chi }{\epsslend\sqrt{\chi^2+\gamma^2}}
\partial_s\left[\crossradius(s)\activity(s,\thetadum) \right] 
+ O(\epsslend^2),
\end{align}

%\begin{align}
%\innerexp{\kertwo}&=
%\frac{\crossradius(s)c(s,\thetadum)  }{ \epsslend\sqrt{\chi^2+\gamma^2}^3} \crossradius(s) \left[\cos(\theta-\thetadum)-1\right] \nonumber\\
%&
%\quad
%-\quad
%\epsslend\frac{\crossradius(s)c(s,\thetadum) }{ \epsslend\sqrt{\chi^2+\gamma^2}^3}
%\begin{pmatrix}
%\frac{1}{2}\chi^2\curvature(s)\cos\theta_m(s,\thetadum) + \chi\crossradius'(s)
%-\chi^2  \curvature(s) \cos\theta_m(s,\thetadum) 
%\\
%-3\frac{\chi\left[\frac{1}{2}\curvature(s) \chi \left(\cos\theta_m(s,\thetadum) + \cos\theta_m(s,\theta)\right) + \crossradius'(s) (\cos(\theta-\thetadum)-1)\right]}{\chi^2+\gamma^2} \crossradius^2(s) \left[\cos(\theta-\thetadum)-1\right]
%\\
% -\chi\crossradius'(s)+\crossradius(s) \curvature(s) \cos\theta_m(s,\thetadum) \crossradius(s) \left[\cos(\theta-\thetadum)-1\right]
%\end{pmatrix}\nonumber\\
%\quad &\quad 
%+\quad
%\epsslend\frac{\chi}{ \epsslend\sqrt{\chi^2+\gamma^2}^3}  \crossradius(s) \left[\cos(\theta-\thetadum)-1\right]\partial_s\left[\crossradius(s) c(s,\thetadum)\right]
%\end{align}

\begin{align}
\innerexp{\kertwo}
&=
\frac{\crossradius(s)c(s,\thetadum)  }{ \epsslend\sqrt{\chi^2+\gamma^2}^3} \crossradius(s) \left[\cos(\theta-\thetadum)-1\right] \nonumber\\
&+
\epsslend\frac{\chi}{ \epsslend\sqrt{\chi^2+\gamma^2}^3}  \crossradius(s) \left[\cos(\theta-\thetadum)-1\right]\partial_s\left[\crossradius(s) c(s,\thetadum)\right]\nonumber\\
&
-\epsslend\frac{\crossradius(s)c(s,\thetadum)\curvature(s)\cos\theta_m(s,\thetadum)}{\epsslend\sqrt{\chi^2+\gamma^2}^3} 
\left[-\frac{\chi^2}{2}
+\crossradius^2(s)  \left[\cos(\theta-\thetadum)-1\right]\right]
\nonumber\\
&+3\epsslend\frac{\crossradius^3(s)c(s,\thetadum)}{\epsslend\sqrt{\chi^2+\gamma^2}^5}
\left[\cos(\theta-\thetadum)-1\right]
\bigg\{\chi\ddrhods \left[\cos(\theta-\thetadum)-1\right]\nonumber\\
&\qquad\qquad\qquad\qquad\qquad\qquad\qquad\qquad
+\frac{\chi^2 }{2}\curvature(s) \left[\cos\theta_m(s,\thetadum) + \cos\theta_m(s,\theta)\right] 
\bigg\} \nonumber\\
&+O\left( \epsslend c/\sqrt{\chi^2+\gamma^2}^3\right). 
\end{align}
Note we have deliberately not simplified $\epsslend$ in the numerators and denominators of the fractions in the above expressions and kept the fractions in the form $\frac{\chi^i}{\epsslend\sqrt{\chi^2+\gamma^2}^j}$, as we will next use known expressions for their integrated values.

\subsubsection{Evaluating the inner integrals} \label{Inner_Eval_App}
We can now perform the integrations with respect to $\sdum$, following~\cite{KoensLauga2018}, by using the known values of integrals of the form
\begin{align}
I^{i}_{j}=\int_{-1}^{1}\frac{\chi^i}{\epsslend\sqrt{\chi^2+\gamma^2}^j}\dd \sdum,\end{align}
where $\gamma$ is a constant with respect to $\sdum$, $i,j$ positive integers, and we recall $\sdum=s+\epsslend\chi$. The following leading order expressions are given by \citet{KoensLauga2018}, 
\begin{subequations}
\begin{align}
I^{i=0}_{j=1}&=\log\left(\frac{4(1-s^2)}{\epsslend^2\gamma^2}\right), \qquad\qquad
I^{i=0}_{j=3}=\frac{2}{\gamma^2},  \qquad
I^{i=0}_{j=5}=\frac{4}{3\gamma^4},  \label{Ii0}\\
I^{i=1}_{j=3}&=\frac{2s\epsslend}{s^2-1}, \qquad \qquad \qquad \qquad
I^{i=1}_{j=5}=0,\qquad  \qquad\label{Ii1}\\
I^{i=2}_{j=3}&=\left[\log\left(\frac{4(1-s^2)}{\epsslend^2\gamma^2}\right)-2
\right],  \quad
I^{i=2}_{j=5}=\frac{2}{3\gamma^2}, \qquad\label{Ii2}
\end{align}
\end{subequations}

%{\color{red} SHOW CALCULATION/expansion to illustrate that h.o.ts are of $O(eps^2)$}
and we also evaluate
\begin{align}
I^{i=1}_{j=1}
&=\int_{-1}^{1}\frac{\chi}{\epsslend\sqrt{\chi^2 + \gamma^2}}\dd \sdum=\int_{-(1+s)/\epsslend}^{(1-s)/\epsslend}\frac{\chi}{\sqrt{\chi^2 + \gamma^2}}\dd \chi\nonumber\\
%&=\left[\sqrt{\chi^2 + \gamma^2}\right]_{-(1+s)/\epsslend}^{(1-s)/\epsslend}=\frac{1}{\epsslend}\left[\sqrt{(1-s)^2 + \epsslend^2\gamma^2} - \sqrt{(1+s)^2 + \epsslend^2\gamma^2}\right] 
%\nonumber\\
&=\frac{1}{\epsslend}\left[|1-s|- |1+s| \right]\left[1+O(\epsslend^2)\right]
=-\frac{2s}{\epsslend}\left[1+O(\epsslend^2)\right], \label{Ii1j1}
\end{align}
where we used $|1-s|-|1+s|=(1-s)-(1+s)=-2s$ for $-1<s<1$. 
Using Eqs.~\eqref{Ii0} to \eqref{Ii1} in $\innerexp{\kerone},\innerexp{\kertwo}$ and simplifying gives

%SUMMARY: 
%Inner:
\begin{align}
\int_{-1}^{1}\innerexp{\kerone} \dd \sdum
&=
\crossradius(s)\activity(s,\thetadum)\left[\log\left(\frac{2(1-s^2)}{\epsslend^2 \crossradius^2(s)}\right)-\log\left[1-\cos(\theta-\thetadum)\right]\right]
\nonumber\\
&\quad
-2s
\partial_s\left[\crossradius(s)\activity(s,\thetadum) \right] \left[1+O(\epsslend^2)\right]
\nonumber\\
&\quad
-\epsslend\activity(s,\thetadum) \left[\log\left(\frac{2(1-s^2)}{\epsslend^2 \crossradius^2(s)}\right)-\log\left[1-\cos(\theta-\thetadum)\right]\right]
\crossradius^2(s)\curvature(s)\cos\theta_m(s,\thetadum)
\nonumber\\
&\quad
+\epsslend\crossradius^2(s)\activity(s,\thetadum)
\frac{1}{2}\curvature(s)  \left[\cos\theta_m(s,\thetadum) + \cos\theta_m(s,\theta)\right]\times\nonumber\\&
\qquad\qquad\qquad\qquad\qquad\left[\log\left(\frac{2(1-s^2)}{\epsslend^2 \crossradius^2(s)}\right)-2-\log\left[1-\cos(\theta-\thetadum)\right]\right]
\nonumber\\
&\quad
+ \epsslend\crossradius^2(s)\activity(s,\thetadum)\frac{2s\epsslend}{s^2-1}\ddrhods [\cos(\theta-\thetadum)-1]
\nonumber\\
&\quad
+ O(\epsslend^2),
\\
\int_{-1}^{1}\innerexp{\kertwo}\dd \sdum
&=
-c(s,\thetadum)  \nonumber\\
&\quad
+\frac{1}{2}\epsslend c(s,\thetadum)
\left[\log\left(\frac{2(1-s^2)}{\epsslend^2 \crossradius^2(s)}\right)-\log\left[1-\cos(\theta-\thetadum)\right]\right]
\curvature(s)\crossradius(s)
\cos\theta_m(s,\thetadum) \nonumber\\
&\quad
-\epsslend c(s,\thetadum) 
\frac{1}{2}\crossradius(s)\curvature(s) \left[\cos\theta_m(s,\thetadum) + \cos\theta_m(s,\theta)\right] \nonumber\\
& \quad
+
\epsslend\frac{2s\epsslend}{s^2-1} \crossradius(s) \left[\cos(\theta-\thetadum)-1\right]\partial_s\left[\crossradius(s) c(s,\thetadum)\right] +O(\epsslend^2 c).
\end{align}

\subsection{Matching: common part} \label{MatchingRegion_App} 
In order to obtain $\expouterkernelininner{\kerone}$ and $\expouterkernelininner{\kerone}$, we substitute $\sdum=s+\epsslend\chi$ in the expressions for $\outerexp{\kerone}$ and $\outerexp{\kertwo}$ and expand. Here we show these expansions for  $\bvRo$, $\bvRo/|\bvRo|^2$ and $\Deltaerhorho$, using the Serret-Frenet equations,  %Eqs.~\eqref{SerretFrenetEqnst}-\ref{SerretFrenetEqnsb},   
%$ \dds{\tanhat}= \curvature \norhat(s), \quad  \dds{\norhat}= -\curvature\tanhat(s) + \torsion \binorhat(s),\quad  \dds{\binorhat}=-\torsion \norhat(s)$,
%we substitute $\rc'(s)=\tanhat(s)$, $\rc''(s)=\curvature(s)\norhat(s)$ and $\rc'''(s)=-\curvature^2(s)\tanhat(s)+ \curvature'(s)\norhat + \curvature(s)\torsion(s) \binorhat(s)$
%into the Taylor expansion of the centreline position vector at $\sdum$,
%\begin{align}
%\rc(\sdum)-\rc(s)= (\epsslend\chi)\rc'(s)+ \frac{1}{2} (\epsslend\chi)^2 \rc''(s) + \frac{1}{6}(\epsslend\chi)^3 \rc'''(s) + O(\epsslend^4),
%\end{align}
%to obtain 
%\begin{align} \rc(\sdum) &=\rc(s) + \epsslend\chi\tanhat(s) + \frac{1}{2} (\epsslend\chi)^2\curvature(s) \norhat(s) + O(\epsslend^3)\end{align}
%
%
%\begin{align}
%\rc(\sdum)-\rc(s)&=\epsslend\chi\begin{pmatrix}
%\left[1-\frac{1}{6}(\epsslend\chi)^2\curvature^2(s)\right]\tanhat(s)\\
%+\left[\frac{1}{2}(\epsslend\chi)\curvature(s) + %\frac{1}{6}(\epsslend\chi)^2\curvature'(s)\right]\norhat(s)\\
%+\left[\frac{1}{6}(\epsslend\chi)^2\curvature(s)\torsion(s)\right]\binorhat(s)
%\end{pmatrix} + O(\epsslend^4)
%\end{align}
%Thus
\begin{subequations}
\begin{align}
\bvRo%\equiv&\rc(s)-\rc(\sdum)
=&-\epsslend\chi\bigg[\tanhat(s) + \frac{1}{2}(\epsslend\chi)\curvature(s)\norhat(s)   
\nonumber\\
&\qquad\quad+\frac{(\epsslend\chi)^2}{6} 
\begin{bmatrix}
-\curvature^2(s)\tanhat(s) + \curvature'(s)\norhat(s) + \curvature(s)\torsion(s) \binorhat(s)
\end{bmatrix}
+ O(\epsslend^3)
\bigg],  \label{MatchingRegion_App_Eq1}\\
|\bvRo|
%\epsslend\chi\left|\tanhat(s) + \frac{1}{2}(\epsslend\chi)\curvature(s)\norhat(s)\right| + O(\epsslend^3)=\epsslend\chi\sqrt{1 + \frac{1}{4}(\epsslend\chi)^2\curvature^2(s)} + O(\epsslend^3)\\
=&|\epsslend\chi|\left[1 - \frac{1}{24}(\epsslend\chi)^2\curvature^2(s)+ O(\epsslend^3)\right], \\ %
%\end{align}
%
%The expansion for $\bvRo/|\bvRo|^2$ is
%\begin{align}
%&
%\frac{\bvRo}{|\bvRo|^2}
%\nonumber\\
%&=
%-\frac{1}{(\epsslend\chi)}
%\left[1 - \frac{(\epsslend\chi)^2\curvature^2(s)}{24}+ O(\epsslend^3)\right]^{-2}
%\left[\tanhat(s) + \frac{(\epsslend\chi)\curvature(s)}{2}\norhat(s) %
%+\frac{(\epsslend\chi)^2}{6} \begin{pmatrix}
%-\curvature^2(s)\tanhat(s)\\ + \curvature'(s)\norhat(s)\\ + \curvature(s)\torsion(s) \binorhat(s)
%\end{pmatrix}+ O(\epsslend^3)
%\right] 
%\nonumber\\
%
%&=
%-\frac{1}{(\epsslend\chi)^2}\left[1 - \frac{1}{24}(\epsslend\chi)^2\curvature^2(s)+ O(\epsslend^3)\right]^{-2}\epsslend\chi\left[\tanhat(s) + \frac{1}{2}(\epsslend\chi)\curvature(s)\norhat(s) 
%+\frac{(\epsslend\chi)^2}{6} \begin{pmatrix}
%-\curvature^2(s)\tanhat(s)\\ + \curvature'(s)\norhat(s)\\ + \curvature(s)\torsion(s) \binorhat(s)
%\end{pmatrix}+ O(\epsslend^3)
%\right] 
%\nonumber\\
%
%
%&=
%-\frac{1}{(\epsslend\chi)}\left[1 + \frac{1}{12}(\epsslend\chi)^2\curvature^2(s)+ O(\epsslend^3)\right]\left[\tanhat(s) + \frac{1}{2}(\epsslend\chi)\curvature(s)\norhat(s) 
%+\frac{(\epsslend\chi)^2}{6} \begin{pmatrix}
%-\curvature^2(s)\tanhat(s)\\ + \curvature'(s)\norhat(s)\\ + \curvature(s)\torsion(s) \binorhat(s)
%\end{pmatrix}+ O(\epsslend^3)
%\right]
%\nonumber
%\\
%
%
%&=
%-\frac{1}{(\epsslend\chi)}
%\left[\tanhat(s)
%+ \frac{1}{2}(\epsslend\chi)\curvature(s)\norhat(s) 
%+ \frac{1}{12}(\epsslend\chi)^2\curvature^2(s)\tanhat(s)
%+\frac{(\epsslend\chi)^2}{6} \begin{pmatrix}
%-\curvature^2(s)\tanhat(s)\\ + \curvature'(s)\norhat(s)\\ + \curvature(s)\torsion(s) \binorhat(s)
%\end{pmatrix}+ O(\epsslend^3)
%\right]
%\nonumber\\
%
%
\frac{\bvRo}{|\bvRo|^2}=&
-\frac{1}{(\epsslend\chi)}
\bigg[\tanhat(s)
+ \frac{1}{2}(\epsslend\chi)\curvature(s)\norhat(s) 
\nonumber\\
&\qquad\quad
+\frac{(\epsslend\chi)^2}{6} \begin{bmatrix}
-\frac{\curvature^2(s)}{2}\tanhat(s) + \curvature'(s)\norhat(s) + \curvature(s)\torsion(s) \binorhat(s)
\end{bmatrix}+ O(\epsslend^3)
\bigg], \label{fracbvRobvRotwoexpoutininner}
\\
%\end{align}
%Substituting $\sdum=s+\epsslend\chi$ in $\Deltaerhorho$ and expanding gives
%We first substitute $\sdum=s+\epsslend\chi$ in the term $\frac{\bvRo}{|\bvRo|^2}\cdot\Deltaerhorho$ and expand.
%\begin{align}
%\Deltaerhorho
%&= \crossradius(s)\erho(s,\theta)-\crossradius(\sdum)\erho(\sdum,\thetadum) \nonumber\\
%
%
%\Deltaerhorho
%&=\crossradius(s)\erho(s,\theta)  -\left[\crossradius(s) + \epsslend\chi \crossradius'(s) + \frac{1}{2}(\epsslend\chi)^2\crossradius''(s)\right] \left[\erho(s,\thetadum) - \epsslend\chi\curvature(s) \cos\theta_m(s,\thetadum)\tanhat(s)\right] + O(\epsslend^2) \nonumber\\
%
%
\Deltaerhorho=&\crossradius(s)\left[\erho(s,\theta) - \erho(s,\thetadum)\right]  \nonumber\\& + \epsslend\chi \left[\crossradius(s)\curvature(s) \cos\theta_m(s,\thetadum)\tanhat(s) -\ddrhods\erho(s,\thetadum)\right] + O(\epsslend^2),
\label{Deltaerhorhoexpoutininner}\\
%\end{align}
%and 
%\begin{align}
%\frac{\bvRo}{|\bvRo|^2}&=-\frac{\epsslend\chi}{(\epsslend\chi)^2}\left[\tanhat(s) + \frac{1}{2}(\epsslend\chi)\curvature(s)\norhat(s)+ O(\epsslend^2)\right]\left[1 +\frac{1}{12}(\epsslend\chi)^2\curvature^2(s)\right]  \nonumber\\
%&=-\frac{1}{(\epsslend\chi)}\left[\tanhat(s) + \frac{1}{2}(\epsslend\chi)\curvature(s)\norhat(s)+O(\epsslend^2)\right] %
%\end{align}
%
%Putting Eqs.~\eqref{fracbvRobvRotwoexpoutininner} and \ref{Deltaerhorhoexpoutininner} together we obtain the expansion of 
% $\frac{\bvRo}{|\bvRo|^2}\cdot\Deltaerhorho$,
%\begin{align}
%&
\frac{\bvRo\cdot\Deltaerhorho}{|\bvRo|^2}
%\nonumber\\
%
%&=-\frac{\left[\tanhat(s) + \frac{1}{2}(\epsslend\chi)\curvature(s)\norhat(s)+O(\epsslend^2)\right]}{(\epsslend\chi)} \cdot
%\bigg\{
%\crossradius(s)\left[\erho(s,\theta) - \erho(s,\thetadum)\right]
%\nonumber\\
%&\quad\qquad\qquad\qquad\qquad\qquad\qquad\qquad\qquad
%+ \epsslend\chi \left[\crossradius(s)\curvature(s) \cos\theta_m(s,\thetadum)\tanhat(s) -\crossradius'(s)\erho(s,\thetadum)\right]  + O(\epsslend^2)
%\bigg\}
%\nonumber\\
%
%&=-
%\crossradius(s)\curvature(s)\left[
%\frac{1}{2}\norhat(s)
%\cdot \left[\erho(s,\theta) - \erho(s,\thetadum)\right]+ \cos\theta_m(s,\thetadum)\right] 
%+ O(\epsslend)
%\nonumber\\
=&-\frac{1}{2}\crossradius(s)\curvature(s)\left[\cos\theta_m(s,\theta) + \cos\theta_m(s,\thetadum) \right]+ O(\epsslend).
\end{align}  \label{MatchingRegion_App_Eq1}
\end{subequations}
%Substituting $\sdum=s+\epsslend\chi$ in the expression for $\outerexp{\kerone}$ we obtain
%
%
%\scriptsize
The above expansions, after some calculations lead to \begin{align}
\expouterkernelininner{\kerone}&=\frac{\crossradius(s)\activity(s,\thetadum)}{|\sdum-s|}+\sign{\sdum-s}\partial_s\left[\crossradius(s)\activity(s,\thetadum)\right]
\nonumber\\
&\qquad
  -\frac{1}{2}\epsslend\frac{\crossradius(s)\activity(s,\thetadum) }{|\sdum-s|}\crossradius(s)\curvature(s)\left[
\cos\theta_m(s,\thetadum)-\cos\theta_m(s,\theta)
\right] 
 + O(\epsslend^2) \\
\expouterkernelininner{\kertwo}&=\epsslend^2\frac{\crossradius^2(s)c(s,\thetadum)}{ |\sdum-s|^3}
\left[\cos(\theta-\thetadum)-1\right]\left[1+O(\epsslend
)\right].
 \label{K2_expouterkernelininner}
\end{align}

%\section{Calculation for the Inner integrals} \label{Inner_App}
%%%%%%%%%
%%%%%%%%%
Since $\int_{-1}^{1}\sign{\sdum-s}\dd \sdum=\int_{-1}^{s}(-1)\dd \sdum+\int_{s}^{1}(+1)\dd \sdum=-(1+s)+(1-s)=-2s$ the integrals of the common parts simplify to 
\begin{align}
\int\limits_{-1}^{1}\!\int\limits_{-\pi}^{\pi}&\expouterkernelininner{\kerone}\dd \sdum\dd\thetadum
=
\int_{-1}^{1}\frac{\crossradius(s)}{|\sdum-s|}\int_{-\pi}^{\pi}\activity(s,\thetadum)\dd\thetadum \dd \sdum
+(-2s)\partial_s\int_{-\pi}^{\pi}\left[\crossradius(s)\activity(s,\thetadum)\right] \dd\thetadum
\nonumber\\
& 
-\frac{1}{2}\epsslend\crossradius^2(s)\curvature(s)\int_{-1}^{1}\frac{1}{|\sdum-s|}\int_{-\pi}^{\pi}\activity(s,\thetadum)\left[
\cos\theta_m(s,\thetadum)-\cos\theta_m(s,\theta)
\right]\dd\thetadum \dd \sdum
 + O(\epsslend^2),
\\
&\int\limits_{-1}^{1}\int\limits_{-\pi}^{\pi}\expouterkernelininner{\kertwo}\dd \sdum\dd\thetadum=O(\epsslend^2).
\end{align}

\section{Expansion of the concentration field}
\subsection{Full expression for the expansion of the concentration field} \label{full_sum_appendix}
The full BI equations are approximated by the adding the outer and inner expansions and subtracting from each the common part,
\footnotesize
\begin{align}
&	2\pi c(s,\theta) \approx 
	\int\limits_{-1}^{1}\!\int\limits_{-\pi}^{\pi} 
	\left(
     \innerexp{\kerone} 
    + \innerexp{\kertwo}
    +\outerexp{\kerone}
    + \outerexp{\kertwo} 
    - \expinnerkernelinouter{\kerone}  
    - \expinnerkernelinouter{\kertwo} \right) 
	\dd\thetadum\dd \sdum\nonumber\\
%
%
%
%
%&\textbf{(+Inner1)}\nonumber\\
=&+
\crossradius(s)\int_{-\pi}^{\pi}\activity(s,\thetadum)\left[\log\left(\frac{2(1-s^2)}{\epsslend^2 \crossradius^2(s)}\right)-\log\left[1-\cos(\theta-\thetadum)\right]\right]\dd \thetadum
\nonumber\\
&+\left(-2s\right)\int_{-\pi}^{\pi}
\partial_s\left[\crossradius(s)\activity(s,\thetadum) \right]\dd \thetadum \left[1+O(\epsslend^2)\right]
\nonumber\\
&
- \epsslend\crossradius^2(s)\curvature(s)\int_{-\pi}^{\pi}\activity(s,\thetadum)\cos\theta_m(s,\thetadum)\left[\log\left(\frac{2(1-s^2)}{\epsslend^2 \crossradius^2(s)}\right)-\log\left[1-\cos(\theta-\thetadum)\right]\right]
 \dd \thetadum
\nonumber\\
&+
\frac{1}{2}\epsslend\crossradius^2(s)\curvature(s) \int_{-\pi}^{\pi}\activity(s,\thetadum)\left[\cos\theta_m(s,\thetadum) + \cos\theta_m(s,\theta)\right]\left[\log\left(\frac{2(1-s^2)}{\epsslend^2 \crossradius^2(s)}\right)-2\right]  \dd \thetadum
\nonumber\\
&-
\frac{1}{2}\epsslend\crossradius^2(s)\curvature(s) \int_{-\pi}^{\pi}\activity(s,\thetadum)\left[\cos\theta_m(s,\thetadum) + \cos\theta_m(s,\theta)\right]\log\left[1-\cos(\theta-\thetadum)\right]  \dd \thetadum
\nonumber\\
&+ \epsslend\crossradius^2(s)\frac{2s\epsslend}{s^2-1}\ddrhods \int_{-\pi}^{\pi}\activity(s,\thetadum)[\cos(\theta-\thetadum)-1]\dd \thetadum
\nonumber
\\
%
%&\textbf{(+Inner2)}\nonumber\\
%\int_{-\pi}^{\pi}\int_{-1}^{1}\innerexp{\kertwo}\dd \sdum \dd \thetadum
%=
&
-\int_{-\pi}^{\pi}c(s,\thetadum)\dd \thetadum \nonumber\\
&
+\frac{1}{2} \epsslend \crossradius(s)\curvature(s)\int_{-\pi}^{\pi}c(s,\thetadum) 
\left[\log\left(\frac{2(1-s^2)}{\epsslend^2 \crossradius^2(s)}\right)-\log\left[1-\cos(\theta-\thetadum)\right]\right]
\cos\theta_m(s,\thetadum) 
\dd \thetadum\nonumber\\
&
-\frac{1}{2} \epsslend \crossradius(s)\curvature(s)\int_{-\pi}^{\pi}c(s,\thetadum)
\left[\cos\theta_m(s,\thetadum) + \cos\theta_m(s,\theta)\right] 
\dd \thetadum\nonumber\\
%&\quad \nonumber\\
%
& \quad
+
\epsslend\frac{2s\epsslend}{s^2-1} \crossradius(s) \int_{-\pi}^{\pi}\left[\cos(\theta-\thetadum)-1\right]\partial_s\left[\crossradius(s) c(s,\thetadum)\right] \dd \thetadum
\nonumber\\
%%%
%%%
%%%
%&\textbf{(+Outer1)}\nonumber\\
%\int_{-\pi}^{\pi}\int_{-1}^{1}\outerexp{\kerone} \dd \sdum\dd\thetadum
%
&+	\int_{-1}^{1}\frac{\crossradius(\sdum)}{ |\bvRo|} \int_{-\pi}^{\pi}\activity(\sdum,\thetadum)\dd\thetadum  \dd \sdum
\nonumber\\
&-\epsslend\int_{-1}^{1}\frac{\crossradius(\sdum)}{ |\bvRo|} \int_{-\pi}^{\pi} \activity(\sdum,\thetadum)\left[\crossradius(\sdum)\curvature(\sdum)\cos\dthetaprime+\frac{\bvRo}{|\bvRo|^2}\cdot\Deltaerhorho \right] \dd\thetadum\dd \sdum 
\nonumber\\
%
%&\textbf{(+Outer2)}\nonumber\\
%\int_{-\pi}^{\pi}\int_{-1}^{1}\outerexp{\kertwo}
&+\epsslend\int_{-1}^{1}\frac{\crossradius(\sdum)}{ |\bvRo|^3}
\int_{-\pi}^{\pi}c(\sdum,\thetadum)\left\{ \bvRo\cdot\erho(\sdum,\thetadum)
+O(\epsslend ) \right\}\dd\thetadum\dd \sdum
\nonumber\\
%%%
%%%
%%%
%&\textbf{(-Common part 1)}\nonumber\\
%-\int_{-\pi}^{\pi}\int_{-1}^{1}\expouterkernelininner{\kerone}\dd \sdum\dd\thetadum
&-
\int_{-1}^{1}\frac{\crossradius(s)}{|\sdum-s|}\int_{-\pi}^{\pi}\activity(s,\thetadum)\dd\thetadum \dd \sdum
+2s\partial_s\int_{-\pi}^{\pi}\left[\crossradius(s)\activity(s,\thetadum)\right] \dd\thetadum
\nonumber\\
&\quad 
+\frac{1}{2}\epsslend\crossradius^2(s)\curvature(s)\int_{-1}^{1}\frac{1}{|\sdum-s|}\int_{-\pi}^{\pi}\activity(s,\thetadum)\left[
\cos\theta_m(s,\thetadum)-\cos\theta_m(s,\theta)
\right]\dd\thetadum \dd \sdum
 + O(\epsslend^2). \label{fullsum}
\end{align}

%\subsection{Leading order concentration field}
%To leading order, after neglecting the term  $\epsslend\frac{2s\epsslend}{s^2-1} \crossradius(s) \int_{-\pi}^{\pi}\left[\cos(\theta-\thetadum)-1\right]\partial_s\left[\crossradius(s) c(s,\thetadum)\right] \dd \theta$ in the integral of $\innerexp{\kertwo}$, Eq.~\eqref{fullsum} gives
%\begin{align}
%	2\pi \zerothorder{c}(s,\theta)+\int_{-\pi}^{\pi}\zerothorder{c}(s,\thetadum)\dd \thetadum   \approx & 
%+	\int_{-1}^{1}\frac{\crossradius(\sdum)}{ |\bvRo|} \int_{-\pi}^{\pi}\activity(\sdum,\thetadum)\dd\thetadum  \dd \sdum
%-\int_{-1}^{1} \frac{\crossradius(s)}{|\sdum-s|}\int_{-\pi}^{\pi}\activity(s,\thetadum) \dd\thetadum \dd \sdum \nonumber\\
%&+\crossradius(s)\log\left(\frac{2(1-s^2)}{\epsslend^2 \crossradius^2(s)}\right)\int_{-\pi}^{\pi}\activity(s,\thetadum)\dd \thetadum
%\nonumber\\
%&-\crossradius(s)\int_{-\pi}^{\pi}\activity(s,\thetadum)\log\left[1-\cos(\theta-\thetadum)\right]\dd \thetadum,
%\end{align}
%where the terms $\pm 2s\partial_s\int_{-\pi}^{\pi}\left[\crossradius(s)\activity(s,\thetadum)\right] \dd\thetadum$ cancel out.

\subsection{Leading order concentration field}\label{LeadingOrder_app}
To leading order, after neglecting the term 
$\epsslend\frac{2s\epsslend}{s^2-1} \crossradius(s) \int_{-\pi}^{\pi}\left[\cos(\theta-\thetadum)-1\right]\partial_s\left[\crossradius(s) c(s,\thetadum)\right] \dd \theta$
in the integral of $\innerexp{\kertwo}$, Eq.~\eqref{fullsum} gives
\begin{align}
	2\pi \zerothorder{c}(s,\theta)+\int_{-\pi}^{\pi}\zerothorder{c}(s,\thetadum)\dd \thetadum   \approx & 
+	\int_{-1}^{1}\frac{\crossradius(\sdum)}{ |\bvRo|} \int_{-\pi}^{\pi}\activity(\sdum,\thetadum)\dd\thetadum  \dd \sdum
-\int_{-1}^{1}
\frac{\crossradius(s)}{|\sdum-s|}\int_{-\pi}^{\pi}\activity(s,\thetadum)
\dd\thetadum \dd \sdum \nonumber\\
&
+
\crossradius(s)\log\left(\frac{2(1-s^2)}{\epsslend^2 \crossradius^2(s)}\right)\int_{-\pi}^{\pi}\activity(s,\thetadum)\dd \thetadum
\nonumber\\
&-\crossradius(s)\int_{-\pi}^{\pi}\activity(s,\thetadum)\log\left[1-\cos(\theta-\thetadum)\right]\dd \thetadum,
\label{c0_App}
\end{align}
where the terms $\pm 2s\partial_s\int_{-\pi}^{\pi}\left[\crossradius(s)\activity(s,\thetadum)\right] \dd\thetadum$ cancel out.
{
Integrating this again over $\theta$ gives 
\begin{align}
	4\pi\langle\zerothorder{c}(s)\rangle
= & 
2\pi \left\{
	\int_{-1}^{1} \left[
\frac{\crossradius(\sdum)\langle\activity(\sdum)\rangle}{ |\bvRo(s,\sdum)|} 
-\frac{\crossradius(s)\langle\activity(s)\rangle}{|\sdum-s|} \right]
 \dd \sdum 
+
\crossradius(s)\langle\activity(s)\rangle\log\left(\frac{2(1-s^2)}{\epsslend^2 \crossradius^2(s)}\right)\right\}
\nonumber\\
&-\crossradius(s)\int_{-\pi}^{\pi}\int_{-\pi}^{\pi}\activity(s,\thetadum)\log\left[1-\cos(\theta-\thetadum)\right]\dd \thetadum \dd \theta, \\
= & 
2\pi\left\{	\int_{-1}^{1} \left[
\frac{\crossradius(\sdum)\langle\activity(\sdum)\rangle}{ |\bvRo(s,\sdum)|} 
-\frac{\crossradius(s)\langle\activity(s)\rangle}{|\sdum-s|} \right]
 \dd \sdum 
+
\crossradius(s)\langle\activity(s)\rangle\log\left(\frac{2(1-s^2)}{\epsslend^2 \crossradius^2(s)}\right) \right\}
\nonumber\\
&+2\pi\log(2)\crossradius(s) \int_{-\pi}^{\pi}\activity(s,\thetadum)\dd \thetadum,\end{align}
where we simplified the last term by exchanging the order of the two integrals (over $\dd \thetadum$ and $\dd \theta$) and using   $\int_{-\pi}^{\pi}\log\left[1-\cos(\theta-\thetadum)\right]\dd \thetadum=-2\pi\log(2)$.
Dividing by $2\pi$ and collecting terms we arrive at
\begin{align}
  2\langle\zerothorder{c}(s)\rangle= & 
+	\int_{-1}^{1} \left[
\frac{\crossradius(\sdum)\langle\activity(\sdum)\rangle}{ |\bvRo(s,\sdum)|} 
-\frac{\crossradius(s)\langle\activity(s)\rangle}{|\sdum-s|} \right]
 \dd \sdum 
+
\crossradius(s)\langle\activity(s)\rangle\log\left(\frac{4(1-s^2)}{\epsslend^2 \crossradius^2(s)}\right)
\label{c_zerothorder_theta_av_App} 
\end{align}

Subtracting half of  Eq.~\eqref{c_zerothorder_theta_av_App} from Eq.~\eqref{c0_App} we obtain the leading order slender boundary integral equation in Eq.~\eqref{c_zerothorder}.
}

\subsection{Next Order Concentration field} \label{NextOrder_App}
Going to the next order, 
Eq.~\eqref{fullsum} gives the first order correction to the slender boundary integral formula
%\begin{tcolorbox}[colback=red!5!white,colframe=red!75!black,title=First Order Correction to the Slender BI equation]
\begin{align}
	&2\pi \firstorder{c}(s,\theta) =\nonumber\\
&
- \crossradius^2(s)\curvature(s)\int_{-\pi}^{\pi}\activity(s,\thetadum)\cos\theta_m(s,\thetadum)\left[\log\left(\frac{2(1-s^2)}{\epsslend^2 \crossradius^2(s)}\right)-\log\left[1-\cos(\theta-\thetadum)\right]\right]
 \dd \thetadum
\nonumber\\
&+\frac{1}{2}\crossradius^2(s)\curvature(s)
 \int_{-\pi}^{\pi}\activity(s,\thetadum)\left[\cos\theta_m(s,\thetadum) + \cos\theta_m(s,\theta)\right]\left[\log\left(\frac{2(1-s^2)}{\epsslend^2 \crossradius^2(s)}\right)-2\right]  \dd \thetadum
\nonumber\\
&-\frac{1}{2}\crossradius^2(s)\curvature(s)
 \int_{-\pi}^{\pi}\activity(s,\thetadum)\left[\cos\theta_m(s,\thetadum) + \cos\theta_m(s,\theta)\right]\log\left[1-\cos(\theta-\thetadum)\right]  \dd \thetadum
\nonumber\\
&
-\int_{-\pi}^{\pi}\firstorder{c}(s,\thetadum)\dd \thetadum  
\nonumber\\
&
+\frac{1}{2}  \crossradius(s)\curvature(s)\int_{-\pi}^{\pi}\zerothorder{c}(s,\thetadum)  
\left[\log\left(\frac{2(1-s^2)}{\epsslend^2 \crossradius^2(s)}\right)-\log\left[1-\cos(\theta-\thetadum)\right]\right]
\cos\theta_m(s,\thetadum)  
 \dd \thetadum\nonumber\\
&
\nonumber\\
&
-\frac{1}{2}  \crossradius(s)\curvature(s)\int_{-\pi}^{\pi}\zerothorder{c}(s,\thetadum) 
\left[\cos\theta_m(s,\thetadum) + \cos\theta_m(s,\theta)\right] 
 \dd \thetadum\nonumber\\
&
\nonumber\\
&
-\int_{-1}^{1}\frac{\crossradius(\sdum)}{ |\bvRo|} \int_{-\pi}^{\pi} \activity(\sdum,\thetadum)\left[\crossradius(\sdum)\curvature(\sdum)\cos\dthetaprime+\frac{\bvRo}{|\bvRo|^2}\cdot\Deltaerhorho \right] \dd\thetadum\dd \sdum +  O(\epsslend^2)
,\nonumber\\
&+\int_{-1}^{1}\frac{\crossradius(\sdum)}{ |\bvRo|^3}
\int_{-\pi}^{\pi}\zerothorder{c}(\sdum,\thetadum)\left\{ \bvRo\cdot\erho(\sdum,\thetadum)
+O(\epsslend ) \right\}\dd\thetadum\dd \sdum
\nonumber\\
&
+\frac{1}{2}\crossradius^2(s)\curvature(s)\int_{-1}^{1}\frac{1}{|\sdum-s|}\int_{-\pi}^{\pi}\activity(s,\thetadum)\left[
\cos\theta_m(s,\thetadum)-\cos\theta_m(s,\theta)
\right]\dd\thetadum \dd \sdum.  \label{c1_full}
\end{align}
%\end{tcolorbox}
%
%

In the above expression we have neglected the term $2\epsslend^2\frac{s\crossradius^2(s)}{s^2-1}\ddrhods \int_{-\pi}^{\pi}\activity(s,\thetadum)[\cos(\theta-\thetadum)-1]\dd \thetadum$ in the intregral of $\innerexp{\kerone}$. This is permissible for   prolate spheroidal ends, as the $\crossradius^2(s)$   on the numerator cancels the singulatity in the denominator at $s=\pm1$). 

For axisymmetric activity, $\activity(s,\theta)\equiv\activity(s)$, this simplifies as 
 some of the integrals vanish or simplify, 
for example,
$
\int_{-\pi}^{\pi}\erho(\sdum,\thetadum)
\dd\thetadum=0$  and 
$\int_{-\pi}^{\pi}\cos\theta_m(s,\thetadum)
\dd\thetadum=0$,

\begin{align}
&	2\pi \firstorder{c}(s,\theta) +\langle\firstorder{c}(s)\rangle 
\nonumber\\
&=
+\frac{1}{2} \crossradius^2(s)\curvature(s)\activity(s) 
\int_{-\pi}^{\pi}\cos\theta_m(s,\thetadum)\log\left[1-\cos(\theta-\thetadum)\right]
 \dd \thetadum
\nonumber\\
&\quad
+\frac{1}{2} \crossradius^2(s)\curvature(s)\activity(s)
\cos\theta_m(s,\theta)\left\{2\pi\left[\log\left(\frac{2(1-s^2)}{\epsslend^2 \crossradius^2(s)}\right)-2\right]-\int_{-\pi}^{\pi}\log\left[1-\cos(\theta-\thetadum)\right] \dd \thetadum\right\} 
\nonumber\\
&
 \quad
-\frac{1}{2}  \crossradius(s)\curvature(s)\zerothorder{c}(s)\begin{pmatrix} 2\pi\cos\theta_m(s,\theta) +
\int_{-\pi}^{\pi} \log\left[1-\cos(\theta-\thetadum)\right]
\cos\theta_m(s,\thetadum) \dd \thetadum
\end{pmatrix}\nonumber\\
&\quad
\nonumber\\
&\quad
-2\pi\crossradius(s)\int_{-1}^{1}\crossradius(\sdum)\activity(\sdum)\frac{\bvRo}{|\bvRo|^3}\dd \sdum\cdot \erho(s,\theta) 
,\nonumber\\
&\quad 
-\pi\crossradius^2(s)\curvature(s)\int_{-1}^{1}\frac{1}{|\sdum-s|}\activity(s)\cos\theta_m(s,\theta)
\dd \sdum.
\end{align}
Now, as in \citet{KoensLauga2018} we use that 
\begin{subequations}
\begin{align}
\int_{-\pi}^{\pi}\log\left[1-\cos(\theta-\thetadum)\right]\dd \thetadum&= - 2\pi\log(2),\\
\int_{-\pi}^{\pi}\log\left[1-\cos(\theta-\thetadum)\right]\cos\theta_m(s,\thetadum)\dd \thetadum&= - 2\pi\cos\theta_m(s,\theta),
\end{align}
\end{subequations}
%{\color{red} SHOW CALCULATION}
to get
\begin{align}
&	2\pi \firstorder{c}(s,\theta) +\langle\firstorder{c}(s)\rangle 
\nonumber\\
=&
\quad
+\frac{1}{2} \crossradius^2(s)\curvature(s)\activity(s)
\begin{pmatrix}
- 2\pi
\cos\theta_m(s,\theta)
+
\cos\theta_m(s,\theta)\left\{2\pi\left[\log\left(\frac{2(1-s^2)}{\epsslend^2 \crossradius^2(s)}\right)-2\right]
+ 2\pi\log(2)\right\} 
\end{pmatrix}\nonumber\\
&
 \quad
-\frac{1}{2}  \crossradius(s)\curvature(s)\zerothorder{c}(s)\begin{pmatrix} 2\pi\cos\theta_m(s,\theta) - 2\pi
\cos\theta_m(s,\theta)
\end{pmatrix}\nonumber\\
&\quad
-2\pi\crossradius(s)\int_{-1}^{1}\crossradius(\sdum)\activity(\sdum)\frac{\bvRo}{|\bvRo|^3}\dd \sdum\cdot \erho(s,\theta) 
,\nonumber\\
&\quad 
-\pi\crossradius^2(s)\curvature(s)\int_{-1}^{1}\frac{1}{|\sdum-s|}\activity(s)\cos\theta_m(s,\theta)
\dd \sdum.
\end{align}

{
 \section{End point error estimation}\label{app:errors}
To complement the numerical validation of our results, here we provide some analysis of the end-errors in our theory. In the derivation of SPT, in particular the inner region expansion, we have assumed that 
$\epsslend\dd \crossradius/ \dd s \ll 1$, however for prolate spheroidal filaments $\crossradius(s)= \sqrt{1-s^2}$, and $\dd \crossradius/ \dd s$ diverges close to the ends. Indeed, in the region where $\epsslend \dd \crossradius / \dd s \sim 1/\epsslend^n, n\geq 0$, we must have $s \sim 1/\sqrt{(1+\epsslend^{2(n+1)})}
\rightarrow |s|\sim 1 - \epsslend^{2(n+1)}/2$, which gives the length of the region as $\delta = O(\epsslend^{2(n+1)})$. Not accounting for this divergence at the ends introduces a small error in the concentration calculation over the filament, which we now estimate via scaling arguments on the boundary integral equation. We then discuss the slip velocity in this region, and the impact on swimming velocity.
 
   \subsection{Error analysis for the concentration}
The concentration field over the filament is given by, 
\begin{align} 
2\pi c(s,\theta)=
&  \int\limits_{-l}^{l} \!\int\limits_{-\pi}^{\pi}
\left[
K_1 
+ 
K_2
\right]\, 
\dd \thetadum \dd \sdum , 
\label{diffusionBIgeneral_appendix}
\end{align}
with 
\begin{align}
    K_1 (s, \theta,\sdum, \thetadum) &= \frac{\activity(\sdum,\thetadum)}{|\bvR|} \bigg|\ddsprime{\surf}\times\ddthprime{\surf}\bigg|,
    \qquad
    %\\
    %
    %
    K_2 (s, \theta,\sdum, \thetadum)
    %&
    = - \frac{c(\sdum,\thetadum)  \bvR}{ |\bvR|^3} \cdot\left(\ddsprime{\surf}\times\ddthprime{\surf}\right).  
    \label{surfelementvec_append_scaling}
\end{align} 
Consider the expressions for the surface element and its magnitude
 \begin{align} 
 \ddsprime{\surf}\times\ddthprime{\surf} 
&= \epsslend \crossradius  \bigg\{ \epsslend \ddrhodsprime \tanhat(\sdum)
-\left[1-\epsslend\crossradius(\sdum) \curvature(\sdum) \cos\dthetaprime\right]
\erho(\sdum,\thetadum)      \bigg\},
\label{surfelementvec_nondim_App}
%\nonumber
%\\&
%=\epsslend\crossradius \left\{
%-\erho(\sdum,\thetadum) + \epsslend \left[ \ddrhodsprime\tanhat(\sdum) +\crossradius(\sdum) \curvature(\sdum) \cos\dthetaprime \erho (\sdum,\thetadum)\right]
%\right\} 
\\
\bigg|\ddsprime{\surf}\times\ddthprime{\surf} \bigg|
&= \epsslend \crossradius(\sdum)  \sqrt{ 
\left[1-\epsslend\crossradius(\sdum) \curvature(\sdum) \cos\dthetaprime\right]^2 + \epsslend^2 \left(\ddrhodsprime\right)^2  }.     \label{surfelementscalar_nondim_App}
\end{align}
While $\dd\crossradius/\dd s$ diverges as $s\to \pm 1$, the product $\rho \dd\crossradius/\dd s$ remains $O(1)$. Thus, for  $\epsilon \dd\crossradius/\dd s \sim 1$, $\crossradius \sim \epsslend,$ and the size of the surface element $\sim \epsslend\crossradius\sqrt{O(1)} = O(\epsilon^2)$. Similarly, for $\epsilon \dd\crossradius/\dd s \gg 1$, the size of the surface element $\sim \epsslend \crossradius\sqrt{(\epsilon \dd\crossradius/\dd s)^2} \sim \epsslend^2  \crossradius\, \dd\crossradius/\dd s = O(\epsslend^2)$.

%for most of the filament, $\epsilon \dd\crossradius/\dd s \ll 1, \crossradius \sim 1$, and the size of the surface element is $O(\epsslend)$.  W

Following the argument of \citet{KoensLauga2018}, in the inner expansion, $|\bvR|  = O(\epsslend)$, hence in the region $\delta$ near the ends, scaling Eqs.~\eqref{surfelementvec_append_scaling} gives $K_1 \sim \frac{1}{\epsslend} \epsslend^2 = O(\epsslend)$, $K_2 \sim \frac{c \epsslend}{\epsslend^3}\epsslend^2 = O(1),$ since $c= O(1)$. Thus, returning to the boundary integral equation, the error in the integral from either endpoint $c_{\text{err}}$,
\begin{equation}
c_{\text{err}} \sim \int\limits_{-l}^{-l+\delta} \!\int\limits_{-\pi}^{\pi}{K_1  + K_2} \sim (\epsslend + 1)\delta \sim \delta.
\end{equation}
Thus at worst, the contribution to the concentration along the filament from the end regions is region is $O(\epsslend^2)$, and therefore negligible.

\subsection{The slip velocity}
The slip velocity is given by Eq.~\eqref{vslip_mobility}. Using Eqs.~\eqref{surfelementvec_nondim_App} and \eqref{surfelementscalar_nondim_App}, 
we evaluate the normal  to the filament's surface, pointing out of the filament, as
 \begin{align}
     \no(s,\theta) 
&=
     \frac{ 
\left[1- \epsslend\crossradius(s) \curvature(s) \cos\dtheta   \right]\erho(s,\theta ) - \epsslend   \ddrhods \tanhat(s) 
}{\sqrt{ 
\left[1-\epsslend\crossradius(s) \curvature(s) \cos\dtheta \right]^2 + \epsslend^2 \left(\ddrhods \right)^2 }} \label{normal_filament_full}
 \end{align}
The boundary condition of Eq.~\ref{flux_activity}   
% $\no\cdot\nabla c = -\activity/D$ on the surface of the filament in Eq., 
allows us to  
  write the slip velocity as 
 \begin{align}
	\vslip|_{S} &= \mobility(\bv{x}) \left(\idmat - \no \no \right)\cdot\boldsymbol{\nabla} c \nonumber\\
	%
	%
	%
	%& = \mobility \left(\boldsymbol{\nabla} c - \no \no\cdot\boldsymbol{\nabla} c \right) \nonumber\\
	%
	%
	%
	& = \mobility  \left(\boldsymbol{\nabla} c + \no \activity  \right),  \label{vslip_nBC}
\end{align}
and using the local cylindrical polars (with the axisymmetry axis along $\tanhat(s)$),  we have that   
\begin{align}
    \boldsymbol{\nabla} c &= 
    \tanhat (\tanhat \cdot \boldsymbol{\nabla} c )
    +   \etheta (\etheta \cdot \boldsymbol{\nabla} c )
    +  \erho (\erho \cdot \boldsymbol{\nabla} c ) \\
    &
   =  \tanhat \partial_s c
    +   \etheta \frac{1}{\epsslend \crossradius}\partial_\theta c  
    +  \erho (\erho \cdot \boldsymbol{\nabla} c ),  \label{nablacEq}
\end{align}
where in the last line we used that $\tanhat\cdot\boldsymbol{\nabla}c = \partial_s c, \quad$ $\etheta\cdot\boldsymbol{\nabla}c = \frac{1}{\epsslend \crossradius(s)} \partial_\theta c$.   
Using the boundary condition in Eq.~\ref{flux_activity} ($\no   \cdot \nabla c = -\activity $), together with the expression for $\no$ in  Eq.~\ref{normal_filament_full}, we can  evaluate $\erho\cdot \boldsymbol{\nabla}c$ as  
 \begin{align}
(\erho\cdot \nabla c)  
=  \frac{1}{\left[1- \epsslend\crossradius \curvature  \cos\theta_m   \right]} \left\{  \epsslend   \ddrhods \partial_s c- \activity \sqrt{ 
\left[1-\epsslend\crossradius  \curvature  \cos\theta_m  \right]^2 + \epsslend^2 \left(\ddrhods \right)^2 } \right\}. \label{erhonablac}
 \end{align}
 Thus we calculate the slip velocity by substituting in the expressions for $\boldsymbol{\nabla} c$ (Eq.~\eqref{nablacEq}) and $\no$ (Eq.~\eqref{normal_filament_full}) in Eq.~\eqref{erhonablac}
 \begin{align}
\frac{1}{\mobility}	\vslip 
	%
	%
	%
	%& =     \boldsymbol{\nabla} c + \no \activity /D  \nonumber\\
	%
	%
	%
	&= \tanhat \partial_s c
    +   \etheta \frac{1}{\epsslend \crossradius}\partial_\theta c  
    +  \erho (\erho \cdot \boldsymbol{\nabla} c )
    +
    \no \activity   
    \nonumber\\
    	&= \tanhat \partial_s c
    +   \etheta \frac{1}{\epsslend \crossradius}\partial_\theta c  
    +  \erho \frac{\epsslend   \ddrhods \partial_s c- \activity \sqrt{ 
\left[1-\epsslend\crossradius  \curvature  \cos\theta_m  \right]^2 + \epsslend^2 \left(\ddrhods \right)^2 }}{\left[1- \epsslend\crossradius \curvature  \cos\theta_m   \right]}
    \nonumber\\
    &\qquad+
         \frac{ 
\left[1- \epsslend\crossradius  \curvature \cos\dtheta   \right]\erho  - \epsslend   \ddrhods \tanhat 
}{\sqrt{ 
\left[1-\epsslend\crossradius  \curvature \cos\dtheta \right]^2 + \epsslend^2 \left(\ddrhods \right)^2 }}\activity  
 \end{align}

\subsection{Swimming velocity error analysis} 
  We now proceed to estimate the effect of the regions with large variations in the cross-sectional radius $\crossradius(s)$ in the swimming velocity. We focus our attention to the prolate spheroidal cross-sectional profile $\crossradius(s) = \sqrt{1-s^2}$ 
and show that 
in both regions (see the following subsections $\S\ref{eddrhodsO1}$ and $\S\ref{eddrhodsgg1}$),  the contribution of the  end-point region to the  swimming kinematics is negligible at leading order.
 
Note the form of the normal vector to the filament surface close to the endpoints,
\begin{subequations}
	\begin{align}
	\epsslend \ddrhods &= O(1),\quad \text{then} \quad 
      \no(s,\theta) =
     \frac{ 
\erho(s,\theta) - \epsslend   \ddrhods \tanhat(s)  
 }{\sqrt{ 
1 + \epsslend^2 \left(\ddrhods \right)^2 }} + O(\epsslend), \\
\epsslend \ddrhods &= O(1/\epsslend^n), \quad n\geq1,\quad \text{then} \quad
 \no(s,\theta) = - \sign{\epsslend   \ddrhods} \tanhat   + O(\epsslend^n).
\end{align}
\end{subequations}
At $s=\pm 1, \no = \pm \tanhat$, and the boundary condition for the flux gives us that $\partial_s c = \pm \activity(s= \pm 1)$. Since in the majority of the filament $s\in(-1+\epsslend^2,1-\epsslend^2)$ we have $\partial_s c = O(1)$ (provided one is away from a jump in activity), by continuity we therefore expect $\partial_s  c = O(1)$ in the region $\delta$. Similarly, we will assume that, provided activity is axisymmetric in the region $\delta$, we have $\partial_\theta c = O(\epsslend\crossradius)$.

Since we are interested in the contribution of the end-regions to the swimming kinematics, we estimate $\int_{-\pi}^{\pi} \int_{1-|s|=O(\delta)}  \vslip(s,\theta) \dd \theta \dd s $ where $\delta$ is the arclength extend of the region close to the end-points in each of the cases examined below. For the purposes of the scaling analysis, we assume $\mobility=1$.

   \subsubsection{Region with $\epsslend \ddrhods = O(1)$} \label{eddrhodsO1}
% If $\epsslend \ddrhods = O(1)$, then 
 For $\crossradius(s) = \sqrt{1-s^2}$ this region has size %   $c   \sim  \delta$ and $\partial_s c  \sim \delta/\epsslend$. 
%For $\crossradius(s) = \sqrt{1-s^2}$, 
%{\color{red} 
$\delta = O(\epsslend^2) $ and within it $\crossradius = O(\epsslend)$. 
%}
We can estimate the scalings of the various terms in the surface slip velocity as follows 
\small
\begin{align}
   	&   \int_{-\pi}^{\pi} \int_{1-|s|=O(\delta)}  \vslip(s,\theta) \dd \theta \dd s 	 \nonumber\\
   	&=   \underbrace{
   \int_{-\pi}^{\pi} \int_{1-|s|=O(\delta)} \tanhat(s)	 (\partial_s c)}_{O(\delta \cdot 1) = O(\epsslend^2)}
    +   \underbrace{
    \int_{-\pi}^{\pi} \int_{1-|s|=O(\delta)} 
    \etheta \frac{1}{\epsslend \crossradius}(\partial_\theta c) }_{O(\delta \cdot \frac{1}{\epsslend^2} \cdot \epsslend^2) =O(\epsslend^2) \text{ for axisymm } \activity }  
    \nonumber\\\nonumber\\
    &
    +  \underbrace{    \int_{-\pi}^{\pi} \int_{1-|s|=O(\delta)} \erho \epsslend   \ddrhods (\partial_s c)\left[1    + O(\epsslend^2 )\right]}_{O(\delta\cdot 1) = O(\epsslend^2)} \nonumber\\ 
   & -  \underbrace{\int_{-\pi}^{\pi} \int_{1-|s|=O(\delta)}
    \erho \activity\sqrt{ 
1 + \epsslend^2 \left(\ddrhods \right)^2 } \left[1   + O(\epsslend^2 )\right]}_{O(\delta \cdot 1) = O(\epsslend^2)}
    \nonumber\\\nonumber\\
    &+
\underbrace{\int_{-\pi}^{\pi} \int_{1-|s|=O(\delta)}          \activity \frac{ 
 \erho   - \epsslend   \ddrhods \tanhat  
}{\sqrt{ 
1 + \epsslend^2 \left(\ddrhods \right)^2 }} \left[1   + O(\epsslend^2 )\right] }_{O(\delta \cdot 1) = O(\epsslend^2)} 
\end{align}
  Thus, the regions of size $O(\epsslend^{ 2})$, in which $\epsslend\ddrhods=O(1)$, have an $O(\epsslend^2)$ contribution to the swimming kinematics, which is negligible compared with the leading order swimming kinematics.

 \subsubsection{$\epsslend \ddrhods \sim 1/\epsslend^n, \quad n\geq 1$} \label{eddrhodsgg1}
For $\crossradius(s) = \sqrt{1-s^2}$, the region in which $\epsslend \ddrhods \sim 1/\epsslend^n, n\geq 1$ has  size  
%{\color{red}
$\delta = O( \epsslend^{2n+2} ) $,  within it $\crossradius=O(\epsslend^{n+1})$,
 %},
and hence we can estimate the scalings of the various terms in the surface  slip velocity as follows
%\begin{align}
%   	& \int_{-\pi}^{\pi} \int_{1-|s|=O(\delta)}	\vslip (s,\theta) \nonumber\\\nonumber\\
%   	&=   
%  \underbrace{ \int_{-\pi}^{\pi} \int_{1-|s|=O(\delta)}\tanhat(s)	 (\partial_s c)}_{O(\delta \cdot 1)= O(\epsslend^{2n+2})} \nonumber\\\nonumber\\
%    &\qquad+\underbrace{  \int_{-\pi}^{\pi} \int_{1-|s|=O(\delta)} \etheta \frac{1}{\epsslend \crossradius}(\partial_\theta c) }_{O(\delta \cdot 1) = O(\epsslend^{2n+2})}  
%    \nonumber\\\nonumber\\
%    &\qquad 
%    + \underbrace{\int_{-\pi}^{\pi} \int_{1-|s|=O(\delta)} \erho    \epsslend   \ddrhods (\partial_s c) \left[1+  O(\epsslend^2)\right]}_{
%    O(\delta\frac{1}{\epsslend^{n }} \cdot 1) = O(\epsslend^{n+2})} 
%    \nonumber\\\nonumber\\
%    &\qquad-
%       \underbrace{ \int_{-\pi}^{\pi} \int_{1-|s|=O(\delta)} \erho 
%    \activity
%    \left|\epsslend \ddrhods \right| \left[1   + O(\epsslend^{2n} )\right]
%    \left[1   + O(\epsslend )\right]}_{O( \delta \frac{1}{\epsslend^{n}}) = O(\epsslend^{n+2})} 
%    \nonumber\\
%    &\qquad+
%        \underbrace{\int_{-\pi}^{\pi} \int_{1-|s|=O(\delta)} \left[ 
%   - \sign{\epsslend   \ddrhods} \tanhat  + O(\epsslend^n) \right]
% \left[1   + O(\epsslend^{2n} )\right]\left[1   + O(\epsslend )\right]}_{O(\delta \cdot 1)= O(\epsslend^{2n+2})}   
%\end{align}
%
\begin{align}
   	& \int_{-\pi}^{\pi} \int_{1-|s|=O(\delta)}	\vslip (s,\theta) \nonumber\\\nonumber\\
   	&=   
  \underbrace{ \int_{-\pi}^{\pi} \int_{1-|s|=O(\delta)}\tanhat(s)	 (\partial_s c)}_{O(\delta \cdot 1)= O(\epsslend^{2n+2})} +\underbrace{  \int_{-\pi}^{\pi} \int_{1-|s|=O(\delta)} \etheta \frac{1}{\epsslend \crossradius}(\partial_\theta c) }_{O(\delta \cdot 1) = O(\epsslend^{2n+2})}  
    \nonumber\\\nonumber\\
    &\qquad 
    + \underbrace{\int_{-\pi}^{\pi} \int_{1-|s|=O(\delta)} \erho    \epsslend   \ddrhods (\partial_s c)}_{
    O(\delta\frac{1}{\epsslend^{n }} \cdot 1) = O(\epsslend^{n+2})} -
       \underbrace{ \int_{-\pi}^{\pi} \int_{1-|s|=O(\delta)} \erho 
    \activity
    \left|\epsslend \ddrhods \right| }_{O( \delta \frac{1}{\epsslend^{n}}) = O(\epsslend^{n+2})} 
    \nonumber\\
    &\qquad+
        \underbrace{\int_{-\pi}^{\pi} \int_{1-|s|=O(\delta)}  
   - \activity \,\sign{\epsslend   \ddrhods} \tanhat  + O(\epsslend^n) 
}_{O(\delta \cdot 1)= O(\epsslend^{2n+2})}   
\end{align}
Thus, the regions of size $O(\epsslend^{2n+2})$ in which $\epsslend\ddrhods=O(\frac{1}{\epsslend^{n}})$, have an $O(\epsslend^{n+2})$ contribution to the swimming kinematics, which is negligible compared to the leading order swimming kinematics.}
 
\bibliographystyle{jfm}
\bibliography{phoreticfilamentsrefs}

\end{document}